\documentclass[a4paper,11pt]{article}
\pdfoutput=1 % if your are submitting a pdflatex (i.e. if you have
\usepackage{jheppub} % for details on the use of the package, please
\usepackage[T1]{fontenc} % if needed

\usepackage[utf8]{inputenc}
\usepackage{amssymb}
\usepackage{amsmath}
\usepackage{amsfonts}
\usepackage{graphicx}
\usepackage{color}
\usepackage{xspace}
\usepackage{comment}
\usepackage{hyperref}
\usepackage[normalem]{ulem}
\usepackage[section]{placeins}
\usepackage{afterpage}
\usepackage{slashed}
\usepackage{enumerate}
\usepackage{enumitem}
\usepackage{caption}
\usepackage{subfigure}
\usepackage{float}
\usepackage{ulem}
\usepackage{verbatim}
\usepackage{multirow,rotating}
\usepackage[dvipsnames]{xcolor}
\usepackage[final]{microtype} % 
\usepackage[nameinlink,capitalize,noabbrev]{cleveref}
\crefname{equation}{Eq.}{Eqs.}
\crefname{figure}{Fig.}{Figs.}
\crefname{table}{Table}{Tables}
\usepackage{siunitx}
\usepackage{booktabs}

\definecolor{dcolour}{rgb}{.5, .5, .5}
\def\gsim{\raise0.3ex\hbox{$\;>$\kern-0.75em\raise-1.1ex\hbox{$\sim\;$}}}
\def\lsim{\raise0.3ex\hbox{$\;<$\kern-0.75em\raise-1.1ex\hbox{$\sim\;$}}}
\def\gsim{\raise0.3ex\hbox{$\;>$\kern-0.75em\raise-1.1ex\hbox{$\sim\;$}}}
\def\lsim{\raise0.3ex\hbox{$\;<$\kern-0.75em\raise-1.1ex\hbox{$\sim\;$}}}

\newcommand{\ba}[1]{\begin{eqnarray} \label{(#1)}}
\newcommand{\ea}{\end{eqnarray}}

\newcommand{\iab}{\rm ab^{-1}}
\newcommand{\ifb}{\rm fb^{-1}}
\newcommand{\mltp}{{\mkern-2mu\times\mkern-2mu}}

\newcommand{\met}{\slashed{E}_{\rm T}}

\newcommand{\GeV}{\,{\mathrm{GeV}}}
\newcommand{\TeV}{\,{\mathrm{TeV}}}

\title{\boldmath 
Discovery prospects for photophobic axion-like particles at a 100~TeV proton--proton collider
}

\author[a,b,1]{Zilong Ding\note{Corresponding author.}}
\author[a,1]{, Jiaojiao Feng}
\author[a,1]{, Ying-nan Mao}
\author[a,1]{, Kechen Wang}
\author[a,c,1]{and Yiheng Xiong}

\affiliation[a]{Department of Physics, School of Physics and Mechanics, Wuhan University of Technology,\\430070 Wuhan, Hubei, China}
\affiliation[b]{College of Physical Science and Technology, Central China Normal University, \\ 430079 Wuhan, Hubei, China}
\affiliation[c]{School of Physics, Beihang University, 102206, Beijing, China}

\emailAdd{zilong.d@mails.ccnu.edu.cn}
\emailAdd{jiaojiao.feng@whut.edu.cn}
\emailAdd{ynmao@whut.edu.cn}
\emailAdd{kechen.wang@whut.edu.cn}
\emailAdd{yihenghht@buaa.edu.cn}

\abstract{
We study heavy photophobic axion-like particles (ALPs) in the limit of an effectively vanishing diphoton coupling, $g_{a\gamma\gamma}\simeq 0$, for which diphoton production and decay are suppressed and collider phenomenology is driven by electroweak interactions ($aWW$, $aZ\gamma$, $aZZ$). We perform detector-level searches at a future $\sqrt{s}=$ 100 TeV $pp$ collider (SppC/FCC-hh), with an integrated luminosity of $\mathcal{L} =$ 20 ab$^{-1}$.
We consider $a\to Z\gamma$ and $a\to W^+W^-$ decays. For $pp\to jj\,a$ we include both $s$-channel electroweak exchange and vector boson fusion (VBF)-like topologies, while the tri-$W$ signature arises from associated production $pp\to W^\pm a$ (via $s$-channel exchange) followed by $a\to W^+W^-$. We analyze three final states--$Z\gamma jj$ with $Z\to\ell^+\ell^-$, tri-$W$ ($W^\pm W^\pm W^\mp$) with same-sign dimuons plus jets, and $W^+W^-jj$ with opposite-sign, different-flavor dilepton ($e^\pm\mu^\mp$) plus jets.
Among the two $WW$ final states, the VBF-assisted $jj\,a(\to W^+W^-)$ channel overtakes the purely $s$-channel tri-$W$ mode for $m_a\gtrsim 1~\TeV$, reflecting 100~TeV signal/background kinematic shifts beyond naive energy/luminosity rescaling.
A boosted-decision-tree (BDT) classifier built from kinematic observables provides the final signal--background separation, using detector-level simulations of signal and high-statistics SM backgrounds.
At $\sqrt{s}=100$ TeV and $\mathcal{L} =$ 20 ab$^{-1}$, we present discovery sensitivities to the ALP--$W$ coupling $g_{aWW}$ over $m_a\in[100,\,7000]$ GeV. In parallel, we report model-independent discovery thresholds on $\sigma\times\mathrm{Br}$ for $pp\to jj\,a$ with $a\to Z\gamma$ and $a\to W^+W^-$, as well as for associated production $pp\to W^\pm a$ with $a\to W^+W^-$. Taken together, the three channels deliver broad coverage and mutually reinforcing checks across decay modes and production regimes, strengthening robustness against channel-dependent systematics. The 100~TeV program extends the discovery reach well beyond 14~TeV projections within a general effective field theory (EFT) description that does not assume a specific ultraviolet completion.
}

\begin{document}

\maketitle

\flushbottom

\section{Introduction}
\label{sec:intro}

Axion-like particles (ALPs) are widely used effective descriptions of new pseudo-scalar states that couple to Standard-Model (SM) fields through higher-dimensional operators, and they provide a sharp target for collider searches in scenarios where the relevant degrees of freedom are electroweak gauge bosons.
While the quantum chromodynamics (QCD) axion is tied to the Peccei--Quinn (PQ) solution of the strong-CP problem in QCD~\cite{Peccei:1977hh,Peccei:1977ur,Weinberg:1977ma,Wilczek:1977pj,Kim:1979if,Zhitnitsky:1980tq,Dine:1981rt,Dine:2000cj,Kim:2008hd,Kim:2009xp,Quevillon:2019zrd,Hook:2019qoh}, collider-oriented ALP studies typically treat the mass and effective couplings as largely independent parameters~\cite{Svrcek:2006yi,Irastorza:2018dyq,DiLuzio:2020wdo,Choi:2020rgn}, motivating broad experimental programs across astrophysical, laboratory, and high-energy frontiers~\cite{Mimasu:2014nea,Jaeckel:2015jla,Liu:2017zdh,Bauer:2017ris,Dolan:2017osp,Bauer:2018uxu,Gavela:2019cmq,dEnterria:2021ljz,Agrawal:2021dbo,Zhang:2021sio,Tian:2022rsi,Ghebretinsaea:2022djg,deGiorgi:2022oks,Wang:2022ock,Schafer:2022shi,Galanti:2022ijh,Cheung:2023nzg,Antel:2023hkf,Dev:2023hax,Biswas:2023ksj,Yue:2023mew,Yue:2023mjm,Balkin:2023gya,Esser:2023fdo,Lu:2024fxs,Qiu:2024muo,Cheung:2024qve,Esser:2024pnc,Wang:2024pqa,Marcos:2024yfm,Yang:2024dic,Li:2024zhw,Inan:2025bdw,Yue:2025ksr,Ai:2025cpj,Wang:2025ncc,Bao:2025tqs,Alda:2025uwo,Wang:2025vhy,Bedi:2025hbz,Figliolia:2025dtw,Yue:2025wkn,Esser:2025nmd,Butterworth:2025szb,Bhattacharya:2025hme}.

Over a wide ALP mass range, existing laboratory, astrophysical, and collider searches place stringent bounds on the ALP--photon interaction, making the diphoton coupling $g_{a\gamma\gamma}$ one of the most tightly constrained ALP portals~\cite{CMS:2018erd,Belle-II:2020jti,ATLAS:2020hii,ATLAS:2022abz,BESIII:2022rzz,BESIII:2024hdv,CMS:2024bnt,ParticleDataGroup:2024cfk,ALPlimits}.
This naturally motivates the photophobic limit, in which the effective $a\gamma\gamma$ interaction is suppressed~\cite{Craig:2018kne, Fonseca:2018xzp, Hook:2016mqo} and collider phenomenology is instead driven by electroweak couplings such as $aWW$, $aZ\gamma$, and $aZZ$.

At the LHC, constraints on heavy photophobic ALPs have been widely obtained by reinterpreting analyses optimized for SM targets---including triboson searches at $\sqrt{s}=8~\TeV$~\cite{Craig:2018kne}, vector boson scattering (VBS)-motivated selections interpreted as nonresonant ALP exchange~\cite{Bonilla:2022pxu}, and Run-2 analyses targeting $W^\pm W^\pm W^\mp$ and $Z\gamma$ final states~\cite{Aiko:2024xiv}; see Ref.~\cite{Ding:2024djo} for a recent overview.
Because the signal hypotheses and control strategies were not designed for on-shell ALP production, acceptance and kinematic mismatches can make such recast constraints conservative and non-uniform across parameter space.

To address this, our group has developed a coordinated set of detector-level HL-LHC projections for photophobic ALPs at $\sqrt{s}=14~\TeV$ and $\mathcal{L}=3~\iab$~\cite{Ding:2024djo,Mao:2024kgx,Feng:2025kof}, using mass-dependent multivariate discrimination optimized for each hypothesis.
These studies cover (i) $pp\to jj\,a(\to Z\gamma)$ with $Z\to\ell^+\ell^-$ ($\ell=e,\mu$), including both $s$-channel and vector boson fusion (VBF)-like production, for $m_a\simeq 100$--$4000~\GeV$~\cite{Ding:2024djo};
(ii) a tri-$W$ strategy $pp\to W^\pm X(\to W^+W^-)$ in a same-sign dimuon plus hadronic-$W$ final state, for $m_a\simeq 170$--$3000~\GeV$~\cite{Mao:2024kgx};
and (iii) $pp\to jj\,a(\to W^+W^-)$ in the $e^\pm\mu^\mp+\met$ channel, for $m_a\simeq 170$--$4000~\GeV$~\cite{Feng:2025kof}.

In this work, we turn to a future proton--proton ($pp$) collider with $\sqrt{s}\sim\mathcal{O}(100~\mathrm{TeV})$, such as the Super proton--proton Collider (SppC) or the Future Circular Collider in hadron--hadron mode (FCC-hh)~\cite{CEPC-SPPCStudyGroup:2015esa,Benedikt:2018ofy,CEPCStudyGroup:2018rmc,CEPCStudyGroup:2023quu,FCC:2018vvp,Schulte:2025pgs}.
Importantly, moving from 14 to 100~TeV is not a mere luminosity rescaling: the relevant parton luminosities shift to smaller momentum fractions and electroweak-boson radiation becomes increasingly important, so the relative weight of production topologies and the fiducial kinematics evolve in a non-universal, initial-state-dependent way (see, e.g., Refs.~\cite{Arkani-Hamed:2015vfh,Hinchliffe:2015qma,Mangano:2016jyj,Mangano:2017tke,FCC:2018byv}).
At 100~TeV this reshapes not only the signal but also the dominant SM backgrounds: more boosted and more forward event topologies modify acceptances and kinematic shapes, and hence the optimal analysis strategy.
For photophobic ALPs, a concrete manifestation is an evolving balance between $s$-channel exchange and VBF-like production in $pp\to jj\,a$, together with increasingly boosted $a\to Z\gamma$ and $a\to W^+W^-$ decays.
We therefore perform an end-to-end detector-level study at 100~TeV for both signal and backgrounds, generating high-statistics Monte Carlo samples (especially for the backgrounds) and re-optimizing the multivariate selections at each $m_a$ to quantify how these shape and efficiency effects impact the sensitivity in each channel (Sec.~\ref{sec:results}).
A simple yield rescaling from 14~TeV would miss these kinematic migrations and background-rejection effects.

We study heavy photophobic ALPs in the limit $g_{a\gamma\gamma}\simeq 0$, focusing on ALPs coupled to electroweak field strengths.
We perform dedicated detector-level searches in three complementary channels:
(I) $pp\to jj\,a(\to Z \gamma)$ with $Z\to\ell^+\ell^-$ ($\ell=e,\mu$);
(II) $pp\to W^{\pm}a(\to W^+W^-)$ with a same-sign dimuon signature plus a hadronic decayed $W$ boson;
and (III) $pp\to jj\,a(\to W^+W^-)$ with opposite-sign, different-flavor dilepton $e^\pm\mu^\mp$.
Our analysis incorporates the relevant production topologies---$s$-channel exchange and, where applicable, VBF-like configurations---models signal and backgrounds consistently with detector-level simulations, and employs a boosted decision tree (BDT)-based multivariate analysis (MVA) optimized for each mass hypothesis.
We present projected sensitivities over $m_a=100$--$7000~\GeV$, reporting both discovery reaches on $g_{aWW}$ and model-independent discovery thresholds on the fiducial quantity given by the signal production cross section times branching ratio, $\sigma \times \mathrm{Br}$.
Our main takeaways are a detector-level 100~TeV three-channel sensitivity forecast (mass-optimized BDTs); an explicit $WW$-sector channel crossover beyond naive scaling, where the growing VBF component makes $pp\to jj\,a(\to W^+W^-)$ overtake $pp\to W^\pm a(\to W^+W^-)$ around $m_a\gtrsim 1~\TeV$ (Sec.~\ref{sec:results}); and model-independent discovery thresholds in $\sigma\times\mathrm{Br}$ that enable direct reinterpretation for other resonant hypotheses.
The three final states probe the same photophobic setup from different experimental angles, which strengthens the interpretation by reducing the reliance on any single background model and by helping pinpoint whether any emerging excess prefers $Z\gamma$ or $WW$ decays and which production regime dominates.

This paper is organized as follows.
Sec.~\ref{sec:theory} lays out the theoretical framework.
Sec.~\ref{sec:sig} introduces the signal topologies and the event generation setup.
Sec.~\ref{sec:analysis} presents the analysis strategy and background modeling for each final state, including preselection, ALP-mass reconstruction where applicable, and the MVA approach.
Sec.~\ref{sec:results} shows the main results, followed by the conclusion in Sec.~\ref{sec:conclusion}.
Additional material is provided in the appendices.

\section{Theory Models}
\label{sec:theory}

We adopt an effective-field-theory description of a pseudoscalar axion-like particle (ALP) whose leading interactions are restricted to the electroweak gauge-field strengths.
Keeping the dimension-five operators prior to electroweak symmetry breaking (EWSB), the relevant terms can be written as~\cite{Georgi:1986df}
\begin{equation}
\mathcal{L}_{\rm ALP} \supset
\frac{1}{2}\,\partial_{\mu}a\,\partial^{\mu}a
-\frac{1}{2}\,m_{a}^{2}\,a^{2}
-\frac{c_{\widetilde W}}{f_a}\,a\,W_{\mu\nu}^{b}\,\widetilde W^{b,\,\mu\nu}
-\frac{c_{\widetilde B}}{f_a}\,a\,B_{\mu\nu}\,\widetilde B^{\mu\nu},
\label{eq:L}
\end{equation}
where $a$ and $m_a$ denote the ALP field and its mass, and $f_a$ is the associated decay constant.
Here $W_{\mu\nu}^{b}$ ($b=1,2,3$) and $B_{\mu\nu}$ are the $\mathrm{SU}(2)_{\rm L}$ and $\mathrm{U}(1)_{\rm Y}$ field strengths, and the dual tensor is defined as
$\widetilde X^{\mu\nu}\equiv \tfrac{1}{2}\epsilon^{\mu\nu\alpha\beta}X_{\alpha\beta}$ with $X=W^{b},B$.
In practice, physical observables depend on the ratios $c_{\widetilde W}/f_a$ and $c_{\widetilde B}/f_a$, and we treat $(m_a,\,c_{\widetilde W}/f_a,\,c_{\widetilde B}/f_a)$ as independent inputs at this stage.

After EWSB, rotating to the mass eigenstates $A$, $Z$, and $W^\pm$, the interactions can be organized as
\begin{equation}
\mathcal{L}_{\rm int} \supset
-\frac{1}{4}\,g_{a\gamma\gamma}\,a\,F_{\mu\nu}\,\widetilde F^{\mu\nu}
-\frac{1}{2}\,g_{aZ \gamma}\,a\,Z_{\mu\nu}\,\widetilde F^{\mu\nu}
-\frac{1}{4}\,g_{aZZ}\,a\,Z_{\mu\nu}\,\widetilde Z^{\mu\nu}
-\frac{1}{2}\,g_{aWW}\,a\,W_{\mu\nu}^{+}\,\widetilde W^{-\mu\nu},
\label{eq:L1}
\end{equation}
with the tree-level relations
\begin{align}
g_{a\gamma\gamma} &= \frac{4}{f_a}\,\big(s_\theta^2\,c_{\widetilde W}+c_\theta^2\,c_{\widetilde B}\big), \\
g_{aZZ} &= \frac{4}{f_a}\,\big(c_\theta^2\,c_{\widetilde W}+s_\theta^2\,c_{\widetilde B}\big), \\
g_{aZ \gamma} &= \frac{2}{f_a}\,s_{2\theta}\,\big(c_{\widetilde W}-c_{\widetilde B}\big), \\
g_{aWW} &= \frac{4}{f_a}\,c_{\widetilde W},
\end{align}
where $\theta$ is the weak mixing angle and we use the shorthand
$s_\theta\equiv\sin\theta$, $c_\theta\equiv\cos\theta$, and $s_{2\theta}\equiv\sin 2\theta$ for simplicity.

Motivated by the strict experimental bounds on the diphoton channel, we work in the \emph{photophobic} limit,
\begin{equation}
g_{a\gamma\gamma}=0
\;\;\Longleftrightarrow\;\;
s_\theta^{2}\,c_{\widetilde W}+c_\theta^{2}\,c_{\widetilde B}=0
\;\;\Longleftrightarrow\;\;
c_{\widetilde B}=-t_\theta^{2}\,c_{\widetilde W},
\end{equation}
where $t_\theta\equiv\tan\theta$.
With this condition imposed at tree level, the remaining electroweak couplings are correlated as
\begin{equation}
g_{aZ \gamma}=t_\theta\,g_{aWW},\qquad
g_{aZZ}=(1-t_\theta^{2})\,g_{aWW}.
\end{equation}
Throughout our phenomenological analysis we scan $(m_a,\,g_{aWW})$ and fix $g_{aZ\gamma}$ and $g_{aZZ}$ using the above relations.

Beyond tree level, radiative corrections and renormalization-group running effects can regenerate a small effective $g_{a\gamma\gamma}$ even when the photophobic condition is imposed.
For the collider study presented here, this correction is numerically negligible~\cite{Bauer:2017ris,Craig:2018kne,Bauer:2020jbp,Bonilla:2021ufe,Ding:2024djo}, and we neglect it.

\section{Signal production}
\label{sec:sig}

\begin{figure}[h]
\centering
\includegraphics[width=15cm,height=8cm]{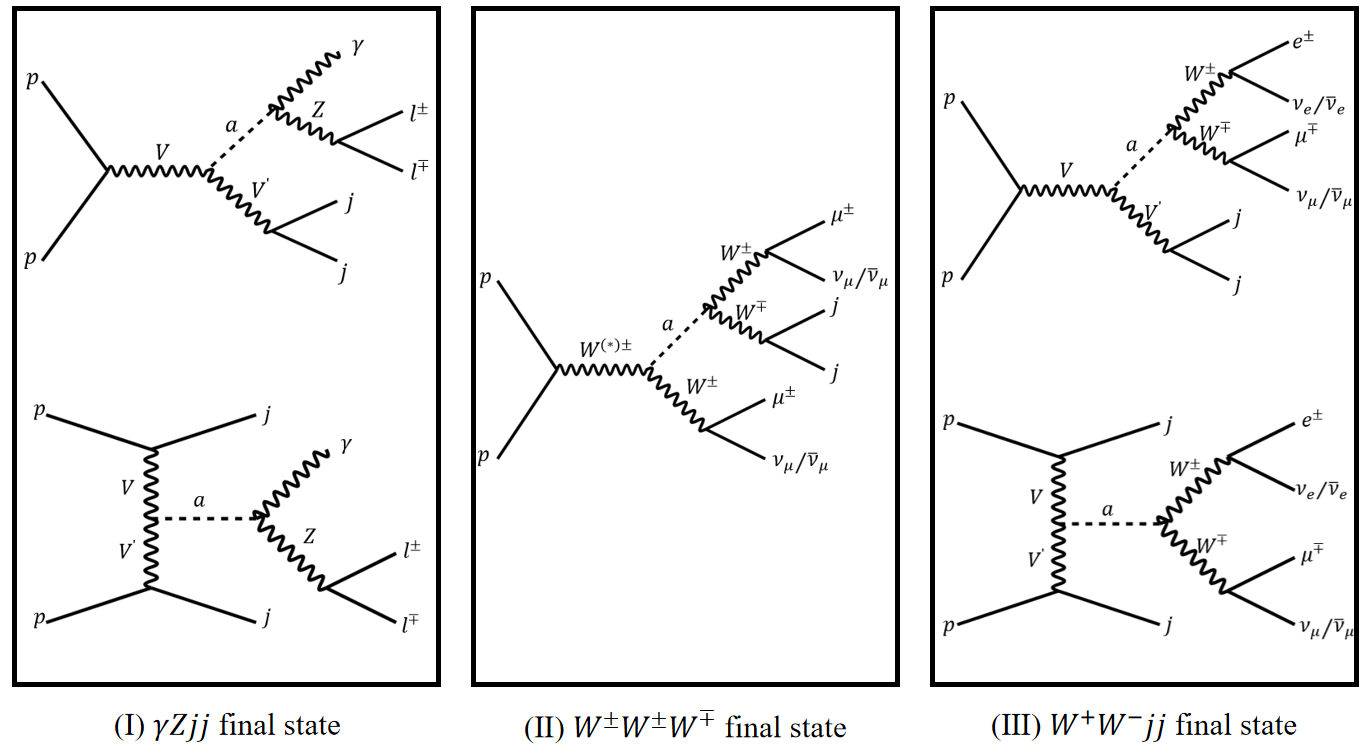}
\caption{
Signal production and decay topologies for the photophobic ALP considered in this work.
(I) $pp \to jj\,a(\to Z \gamma)$ with $Z\to\ell^+\ell^-$ ($\ell=e,\mu$);
(II) $pp \to W^{\pm} a(\to W^+W^-)$, where the same-sign $W$ bosons decay leptonically into muons and neutrinos, $W^\pm\to\mu^\pm \nu_\mu / \bar{\nu}_\mu$, while the remaining $W$ boson decays hadronically, $W^\mp\to jj$;
(III) $pp \to jj\,a(\to W^+W^-)$, where the two $W$ bosons decay leptonically into different-flavor dilepton, $e^\pm\mu^\mp$.
}
\label{fig:signal}
\end{figure}

Although a photophobic ALP does not couple to diphoton at tree level, it still couples to the electroweak diboson states ($aWW$, $aZZ$, and $aZ \gamma$). Heavy ALPs can therefore be produced at colliders via these effective vertices. Fig.~\ref{fig:signal} illustrates the signal topologies considered in this study, namely ALP production in association with two jets at the SppC/FCC-hh, either via an $s$-channel electroweak gauge boson or via vector-boson fusion (VBF). Panel~(I) of Fig.~\ref{fig:signal} corresponds to $pp\to jj\,a(\to Z \gamma)$, where the $Z$ boson decays dileptonically, $Z\to\ell^+\ell^-$ with $\ell=e,\mu$. Panel~(II) shows $pp\to W^{\pm}a(\to W^+W^-)$, where the same-sign $W$ bosons decay leptonically into muons and neutrinos, $W^\pm\to\mu^\pm \nu_\mu / \bar{\nu}_\mu$ 
, while the remaining $W$ boson decays hadronically, $W^\mp\to jj$. Panel~(III) corresponds to $pp\to jj\,a(\to W^+W^-)$, where the two $W$ bosons decay leptonically into different-flavor dilepton, $e^\pm\mu^\mp$, each accompanied by the corresponding neutrino. The diagrams in panel (II) and the upper diagrams of (I) and (III) represent associated production via an $s$-channel gauge boson, while the lower diagrams of (I) and (III) represent VBF production with two forward jets.

To generate the signal events, we implement the ALP model based on the linear Lagrangian~\cite{Brivio:2017ije} in the Universal FeynRules Output (UFO) format~\cite{Degrande:2011ua} and import it into the MadGraph5\_aMC@NLO framework (version 2.6.7)~\cite{Alwall:2014hca} to simulate $pp$ collisions, using the default ``nn23lo1'' parton distribution function (PDF) set~\cite{Ball:2012cx}. Parton showering and hadronization are simulated with PYTHIA~8.3~\cite{Bierlich:2022pfr}, which also handles the decays of unstable SM particles. Detector effects are modeled with Delphes~3.4.2~\cite{deFavereau:2013fsa} using the CMS configuration card. Jets are reconstructed with FastJet~\cite{Cacciari:2011ma} employing the anti-$k_t$ algorithm~\cite{Cacciari:2008gp} with radius parameter $R=0.4$.

For signal~(I), we generate $pp\to jj\, a(\to Z \gamma)$ events with MadGraph, and allow $Z\to\ell^{+}\ell^{-}$ with $\ell=e,\mu,\tau$ at the generator level, while in the analysis we restrict to final states with $\ell=e,\mu$. We generate samples at representative ALP masses
$m_a \in$ \{100, 165, 400, 700, 1000, 1500, 2000, 4000, 5000, 7000\} GeV.
For each $m_a$, the coupling is fixed to the benchmark value $g_{aWW}=1~\mathrm{TeV}^{-1}$, and at least $10^6$ events are generated.

For signal~(II), we generate $pp\to W^{\pm}a(\to W^+W^-)$ events with MadGraph. We generate samples at representative ALP masses
$m_a \in$ \{170, 200, 300, 400, 600, 900, 1400, 1800, 2200, 3000, 5000, 10000, 20000\} GeV.
For each mass point, the coupling is fixed to $g_{aWW}=1~\mathrm{TeV}^{-1}$, and at least $10^6$ events are generated.

For signal~(III), the hard-scattering process $pp\to jj\,a$ is generated with MadGraph, while the subsequent decays $a\to W^+W^-$ and $W^+\to \ell^+\nu$, $W^-\to \ell^-\bar\nu$ ($\ell=e,\mu$), yielding $a\to W^+(\to \ell^+\nu)\,W^-(\to \ell^-\bar\nu)\,jj$, are simulated with PYTHIA~8.3. We generate signal samples at representative ALP masses
$m_a \in$ \{170, 185, 200, 225, 250, 350, 750, 1500, 2500, 4000, 5500, 7000\} GeV.
At the HL-LHC, we generate $1.0$ and $0.305$ million events for the benchmark points $m_a=200$ and $225~\mathrm{GeV}$, respectively; $0.3$ million events for each point with $m_a=170,\,185$ and $250~\mathrm{GeV}$; and $0.275$ million events for each point with $m_a=350,\,750,\,1500,\,2500,\,4000,\,5500$ and $7000~\mathrm{GeV}$.

\begin{figure}[h]
\centering
\includegraphics[width=12cm,height=8cm]{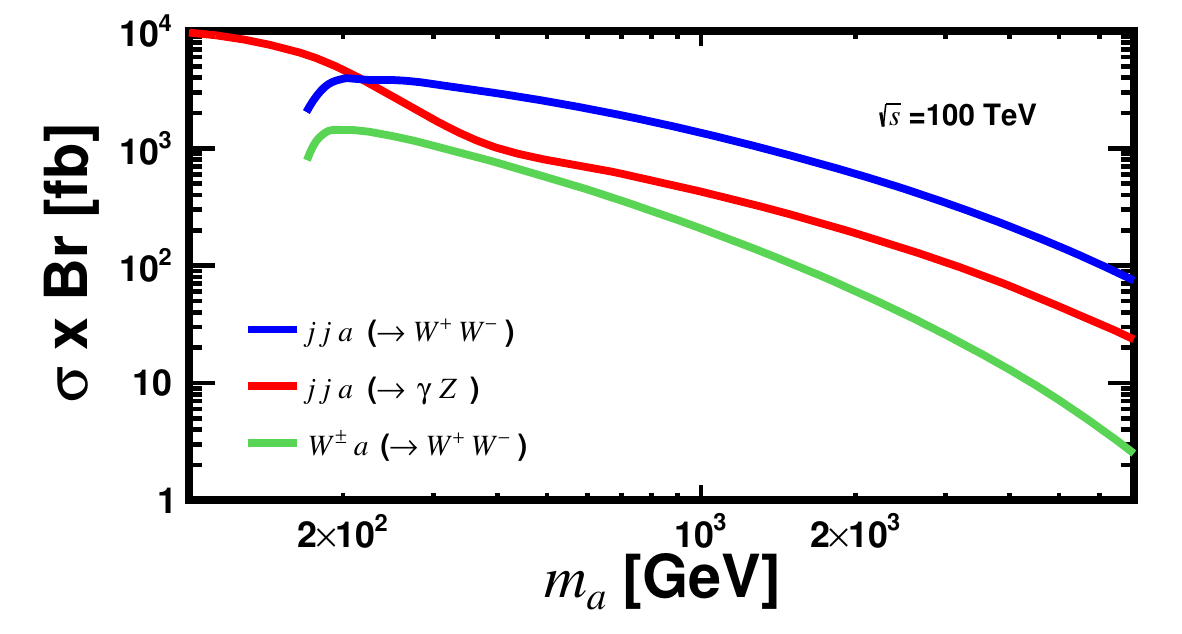}
\caption{
Assuming benchmark couplings $g_{a\gamma\gamma}=0$ and $g_{aWW}=1~\mathrm{TeV}^{-1}$ at the SppC/FCC-hh with $\sqrt{s}=100$~TeV,
the production cross section times branching ratio, $\sigma(pp\to jj\,a)\times\mathrm{Br}(a\to Z \gamma)$, is shown as a function of the ALP mass $m_a$ in the range $100$--$7000$~GeV, while $\sigma(pp\to W^{\pm}a)\times\mathrm{Br}(a\to W^+ W^-)$ and $\sigma(pp\to jj\,a)\times\mathrm{Br}(a\to W^+ W^-)$ are shown as functions of $m_a$ in the range $170$--$7000$~GeV. The red, green, and blue curves correspond to $jj\,a(\to Z \gamma)$, $W^{\pm}a(\to W^+ W^-)$, and $jj\,a(\to W^+ W^-)$, respectively.
}
\label{fig:cro}
\end{figure}

In Fig.~\ref{fig:cro}, we present the rates $\sigma\times\mathrm{Br}$ for the three signal processes as functions of $m_a$ at $\sqrt{s}=100$~TeV, for $g_{a\gamma\gamma}=0$ and $g_{aWW}=1~\mathrm{TeV}^{-1}$. 
All curves fall with increasing $m_a$, reflecting the rapidly decreasing parton luminosities at large momentum fractions. 
For $jj\,a(\to Z\gamma)$ (red), a pronounced suppression around $m_a\sim200$~GeV is mainly due to the opening of competing two-body diboson channels, which quickly depletes $\mathrm{Br}(a\to Z\gamma)$ over $m_a\simeq160$--250~GeV~\cite{Ding:2024djo}. 
For the $W^+W^-$ modes (blue and green), $\mathrm{Br}(a\to W^+W^-)$ turns on rapidly above the threshold $m_a\simeq2m_W$ (e.g.\ $\sim49\%$ at $m_a=185~\mathrm{GeV}$~\cite{Aiko:2023trb,Ding:2024djo}) and then varies only mildly at higher masses; together with the falling production cross sections, this yields a broad maximum in $\sigma\times\mathrm{Br}$ just above threshold, around $m_a\sim190$--200~GeV. 
For the associated-production channel $pp\to W^{\pm}a(\to W^+ W^-)$ (green), the point shown at $m_a=7000~\mathrm{GeV}$ is obtained by a smooth fit to the simulated mass points (including samples generated at higher masses).

\section{Backgrounds and data analysis} 
\label{sec:analysis} 

\subsection{$Z \gamma j j$ final state}
\label{subsec:gammaZjj_bkg_ana}

\subsubsection{Background processes}
\label{subsubsec:gammaZjj_SMbg}

Since the final state contains one photon, an opposite-sign same-flavor lepton pair, and two jets, 
we consider six relevant SM background processes:
$ZW\gamma$, $Z(\to \ell^+\ell^-)\,\gamma jj$, $ZZ\gamma$, $W^+W^-\gamma$, $W(\to \ell\nu)\,\gamma jj$, and $t\bar t\,\gamma$, where $\ell=e,\mu,\tau$ denotes a charged lepton and $\nu$ denotes an (anti)neutrino. 
We label them as ``B1–B6'' throughout.
When a $Z$ boson decays to $\ell^+\ell^-$ and the accompanying heavy $W/Z$ boson decays hadronically, B1–B3 yield the same visible final state as the signal and are thus irreducible. 
B4–B5 can contribute via misidentified leptons in the reconstruction. 
If both $W$ bosons in the final state decay leptonically, B6 also contributes.

Background events are generated with the same tool chain as the signal (MadGraph + PYTHIA + Delphes \cite{Alwall:2014hca,Ball:2012cx,Bierlich:2022pfr,deFavereau:2013fsa}). Event yields for both signal and backgrounds are normalized using the production cross sections reported by MadGraph5\_aMC@NLO. 
To reduce statistical uncertainties, we produce as many background events as allowed by our computing resources.
In total we generate 10.0 million $ZW\gamma$, 22.0 million $Z(\to \ell^+\ell^-)\,\gamma jj$, 10.0 million $ZZ\gamma$, 10.0 million $W^+W^-\gamma$, $1.0\mltp10^{9}$ $W(\to \ell\nu)\,\gamma jj$, and 18.0 million $t\bar t\,\gamma$ events. 
Due to their large production cross sections, B2 and B5 dominate the total background.
To improve generating efficiency for B2 and B5, we force the $Z/W$ bosons to decay leptonically at the generator level, i.e. $Z\to \ell^+\ell^-$ and $W\to \ell\nu$ with $\ell=e,\mu,\tau$.

\subsubsection{Preselection}
\label{subsubsec:gammaZjj_presel}

Final-state objects (photons, leptons, and jets) are ordered by transverse momentum and labeled $O_i$ with $O=\gamma,\ell,j$ and $i=1,2,\ldots$.
We apply the following preselection criteria as the first selection step to target the desired final state and suppress backgrounds:
\begin{enumerate}[label*=(\roman*)]
\item At least one photon, $N(\gamma)\ge 1$, with $p_T(\gamma_1)>20$~GeV.
\item Exactly two same-flavor, opposite-sign leptons ($\ell=e,\mu$), i.e. $N(\ell)=2$ and an $\ell^+\ell^-$ pair, with $p_T(\ell_{1,2})>20$~GeV.
\item At least two jets, $N(j)\ge 2$, with $p_T(j_{1,2})>50$~GeV.
\item No $b$-tagged or $\tau$-tagged jets: $N(j_b)=0$ and $N(j_\tau)=0$.
\end{enumerate}

\begin{table}[h]
\centering
\begin{tabular}{ccccccc}
\hline
\hline
\multicolumn{2}{c}{SppC/FCC-hh} & initial & (i) & (ii) & (iii) & (iv)  \\
\hline
signal & $jj\, a(\to Z \gamma(\to \ell^+ \ell^-))$ & $1.2\mltp10^{6}$ & $1.1 \mltp 10^{6}$ & $4.3 \mltp 10^{5}$ & $3.7 \mltp 10^{5}$  & $2.8 \mltp 10^{5}$ \\ 
\multicolumn{2}{c}{total background} & $7.6\mltp10^{10}$ & $1.7 \mltp 10^{10}$ & $2.1 \mltp 10^{8}$ & $7.5 \mltp 10^{7}$  & $4.8 \mltp 10^{7}$ \\ 
\hline
B1 & $ZW \gamma$ & $5.2\mltp10^{7}$ & $2.1 \mltp 10^{7}$ & $4.8 \mltp 10^{5}$ & $3.2 \mltp 10^{5}$  & $3.1 \mltp 10^{4}$ \\  
B2 & $Z(\to \ell^+ \ell^-) \gamma jj$ & $6.6\mltp10^{9}$ & $1.0 \mltp 10^{9}$ & $1.7 \mltp 10^{8}$ & $5.1 \mltp 10^{7}$  & $4.5 \mltp 10^{7}$ \\ 
B3 & $ZZ \gamma$ & $2.0\mltp10^{7}$ & $6.1 \mltp 10^{6}$ & $3.6 \mltp 10^{5}$ & $7.7 \mltp 10^{4}$  & $5.2 \mltp 10^{4}$ \\ 
B4 & $W^+ W^- \gamma$ & $1.4\mltp10^{8}$ & $4.1 \mltp 10^{7}$ & $1.1 \mltp 10^{6}$ & $6.8 \mltp 10^{4}$  & $5.4 \mltp 10^{4}$ \\ 
B5 & $W(\to \ell\nu) \gamma jj$ & $6.6\mltp10^{10}$ & $1.5 \mltp 10^{10}$ & $1.8 \mltp 10^{6}$ & $1.0\mltp 10^{6}$  & $5.9 \mltp 10^{5}$ \\ 
B6 & $t\bar{t} \gamma$ & $3.4\mltp10^{9}$ & $1.4 \mltp 10^{9}$ & $3.2\mltp 10^{7}$ & $2.3 \mltp 10^{7}$  & $1.9 \mltp 10^{6}$ \\ 
\hline
\hline
\end{tabular}
\caption{
Event yields for the signal with benchmark $m_a=700$~GeV and for each background process after applying preselection criteria (i)–(iv) sequentially. 
Numbers correspond to the SppC/FCC-hh with $\sqrt{s}=100$~TeV and $\mathcal{L}=20~\iab$.
}
\label{tab:gammazjjpreselection}
\end{table}

The expected number of events, $N_{\rm exp}$, is computed as
\begin{equation}
N_{\rm exp}=\sigma_{\rm pro} \times \mathcal{L} \times\epsilon_{\rm sel},
\label{eqn:expected_eventsnumber}
\end{equation}
where $\sigma_{\rm pro}$ is the production cross section of the process, $\mathcal{L}$ is the integrated luminosity, and $\epsilon_{\rm sel}$ is the selection efficiency obtained from our analysis. 
Table~\ref{tab:gammazjjpreselection} shows event yields for the signal with benchmark $m_a=700$~GeV and for each background process after applying preselection criteria (i)–(iv) sequentially. 
Numbers correspond to the SppC/FCC-hh with $\sqrt{s}=100$~TeV and $\mathcal{L}=20~\iab$.

\begin{table*}[h]
\centering 
\scalebox{0.9}{
\begin{tabular}{ccccccc}
\hline
\hline
$m_a$ [GeV] & 100 & 165  & 400 & 700 & 1000 & 1500 \\
&$4.34\mltp10^{-2}$ &$1.27\mltp10^{-1}$ &$20.2\mltp10^{-1}$ &$2.29\mltp10^{-1}$&$2.43\mltp10^{-1}$&$2.56\mltp10^{-1}$ \\
\hline
$m_a$ [GeV]  & 2000 & 4000 & 5000 & 7000 \\
 &$2.63\mltp10^{-1}$&$2.76\mltp10^{-1}$&$2.79\mltp10^{-1}$&$2.83\mltp10^{-1}$\\
\hline
background & B1 & B2 & B3 & B4 & B5 & B6 \\
& $ZW \gamma$ & $Z(\to \ell^+ \ell^-) \gamma jj$ & $ZZ \gamma$& $W^+ W^- \gamma$ & $W(\to \ell\nu) \gamma jj$ & $t\bar{t} \gamma$ \\
 & $5.96\mltp10^{-4}$ & $6.88\mltp10^{-3}$ & $2.60\mltp10^{-3}$ & $3.86\mltp10^{-4}$ & $8.94\mltp10^{-6}$ & $5.59\mltp10^{-4}$  \\
 \hline
\hline
\end{tabular}
}
\caption{
Preselection efficiencies for the signal with representative ALP masses and for the background processes at the SppC/FCC-hh with $\sqrt{s}=100$~TeV. 
}
\label{tab:preselection}
\end{table*}

Table~\ref{tab:preselection} shows the selection efficiencies after applying criteria (i)–(iv) for representative $m_a$ values and for all background processes at $\sqrt{s}=100$~TeV.
One observes that B5 is strongly rejected, with a preselection efficiency of $8.94\mltp10^{-6}$.
Efficiencies for the other backgrounds are typically $\sim10^{-3}$–$10^{-4}$. It is worth noting that, for the signal, the preselection efficiency at $m_a=100$~GeV is $4.34\mltp10^{-2}$, smaller than for other masses. This is because $m_a$ is close to the $Z$-boson mass, yielding softer photons. 

\subsubsection{Reconstruction of the ALP mass}
\label{subsubsec:gammaZjj_ALPmass}

\begin{figure}[h]
\centering
\includegraphics[width=7.3cm,height=5cm]{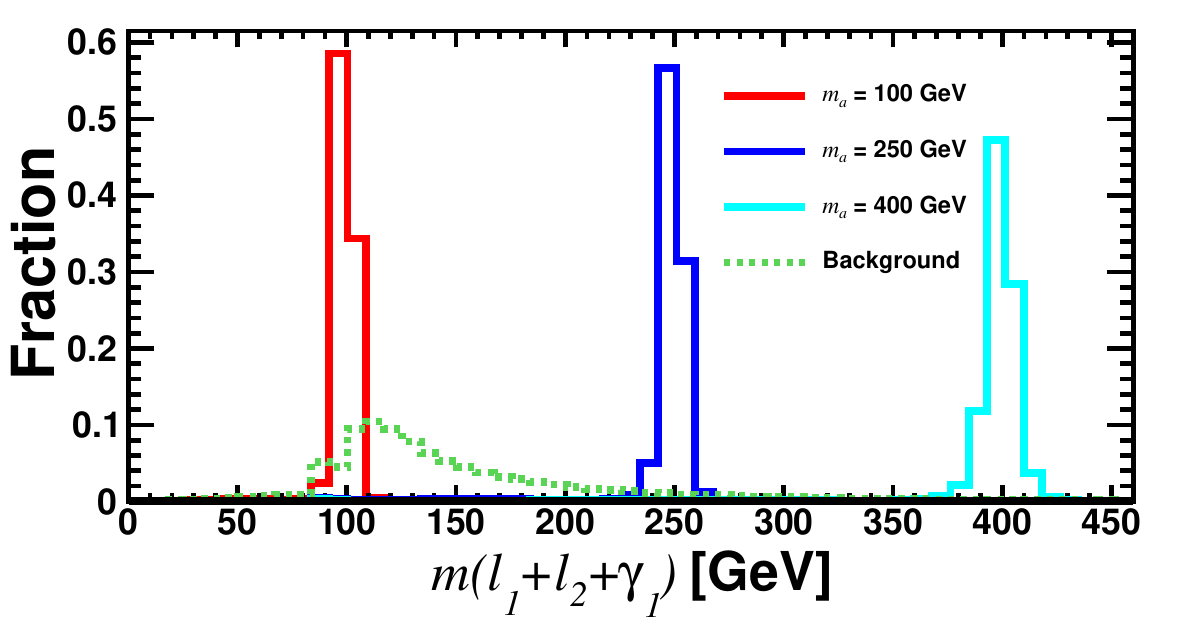}\,\,\,\,\,
\includegraphics[width=7.3cm,height=5cm]{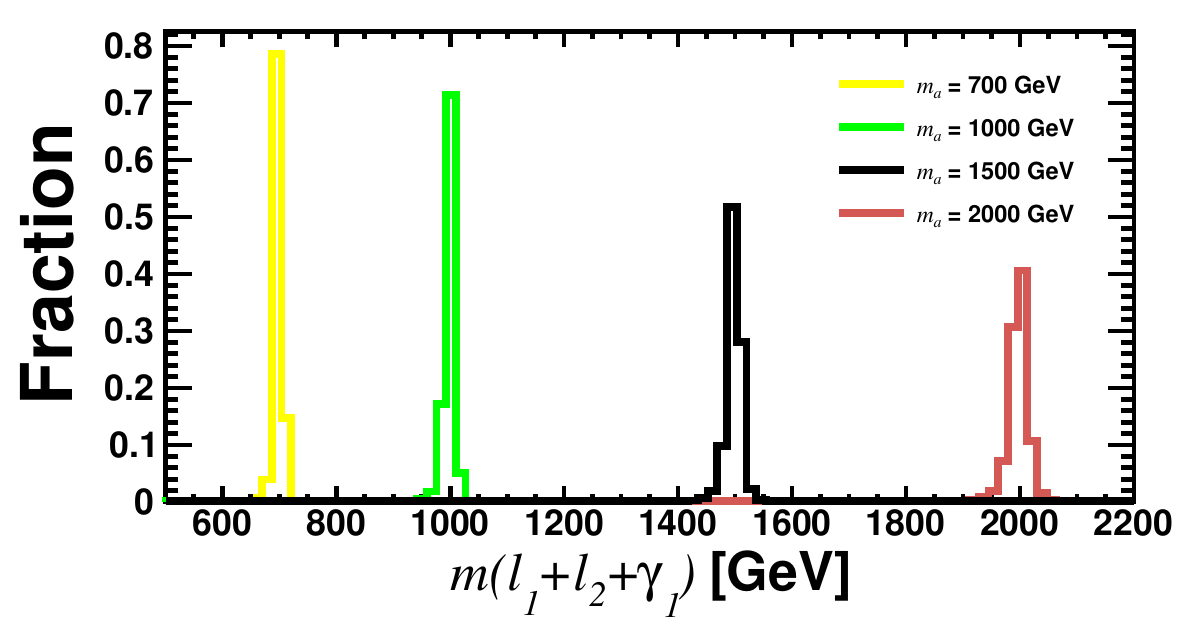}
\includegraphics[width=7.3cm,height=5cm]{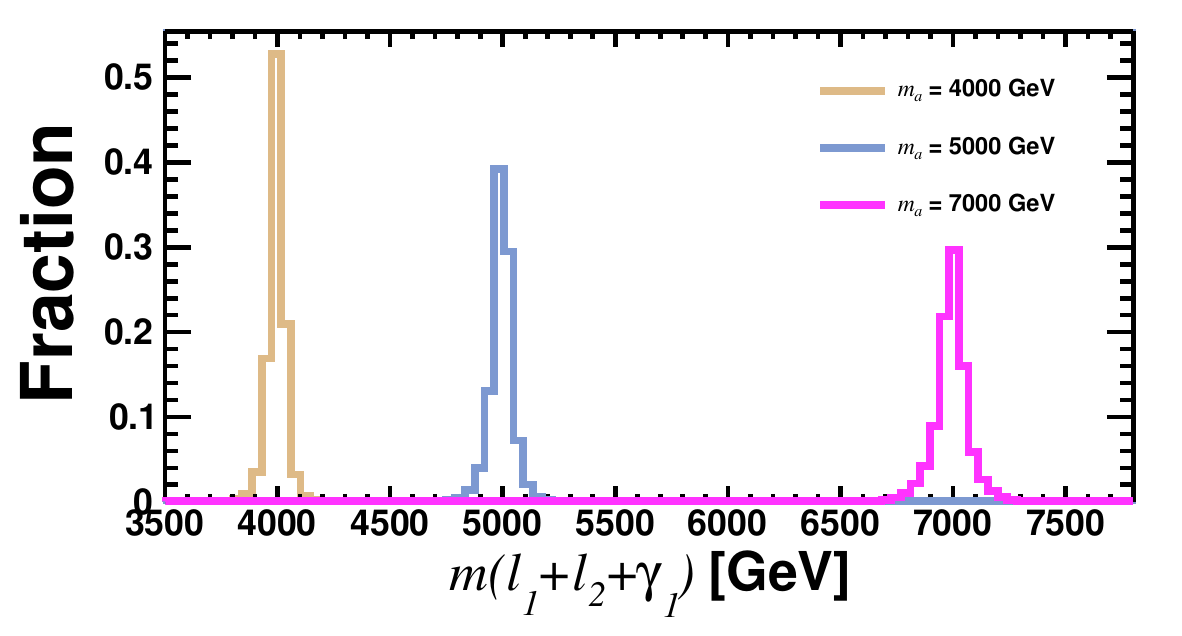}
\caption{
Distributions of the invariant mass of the system formed by the two leading leptons and the leading photon, $m(\ell_1+\ell_2+\gamma_1)$, 
for the total background (dotted) and for signal benchmarks with various $m_a$, after preselection at $\sqrt{s}=100$~TeV.
}
\label{fig:ma1}
\end{figure}

Since the signal contains $a\to Z \gamma$ with $Z\to\ell^+\ell^-$, we reconstruct $m_a$ from the invariant mass of the system formed by the two leading leptons and the leading photon, denoted as $m(\ell_1+\ell_2+\gamma_1)$. 
Fig.~\ref{fig:ma1} shows the corresponding $m(\ell_1+\ell_2+\gamma_1)$ distributions for the total background (dotted) and for signal benchmarks with various $m_a$, after preselection at $\sqrt{s}=100$~TeV. 
One observes that the signal peaks sharply near the input value of $m_a$ throughout the range 100--7000~GeV, whereas the total background exhibits a broad structure around $\sim130$~GeV. 
Thus, the invariant mass $m(\ell_1+\ell_2+\gamma_1)$ provides a reliable estimator of the ALP mass and a powerful discriminator for $m_a\gtrsim200$~GeV.

\subsubsection{Multivariate analysis}
\label{subsubsec:gammaZjj_mva}

After preselection, to further suppress backgrounds, we train a BDT using TMVA with its default settings~\cite{TMVA:2007ngy}. 
The input observables are organized as follows:

\begin{enumerate}[label*=(\Alph*)]
\item Kinematics of the leading objects: 
$E(O)$, $p_T(O)$, $\eta(O)$, and $\phi(O)$ for $O=\gamma_1,\ell_1,\ell_2,j_1,j_2$.
\item Missing-momentum observables: $\met$ and its azimuthal angle $\phi(\met)$.
\item Angular separations $\Delta R\equiv\sqrt{(\Delta\eta)^2+(\Delta\phi)^2}$: 
$\Delta R(j_1,j_2)$ and $\Delta R(\gamma_1,\ell_1+\ell_2)$, where “$\ell_1+\ell_2$” denotes the dilepton system.
\item Invariant masses of combined systems: $m(j_1+j_2)$ and $m(\ell_1+\ell_2+\gamma_1)$.
\item Over all dijet pairs, the minimal angular separation and the corresponding invariant mass: 
$\Delta R(j,j')_{\rm min}$ and $m(j+j')_{{\rm min}\,\Delta R}$.
\item Over all dijet pairs, the maximal pseudorapidity separation and the corresponding invariant mass:
$\Delta\eta(j,j')_{\rm max}$ and $m(j+j')_{{\rm max}\,\Delta\eta}$.
\end{enumerate}

Appendix~\ref{app:gammaZjj_obs} presents distributions of representative observables for the signal with benchmark $m_a=700$~GeV and for the six background processes at the SppC/FCC-hh with $\sqrt{s}=100$~TeV, after the preselection criteria are applied. As expected, $m(\ell\ell\gamma)$ is a key discriminator across all masses. Given that the VBF contribution becomes increasingly important at larger $m_a$, we include variables that capture its topology (e.g., the dijet pair with maximal $|\Delta\eta|$ and its invariant mass). For the $s$-channel topology, we instead use the dijet pair with minimal $\Delta R$ and its invariant mass.

\begin{figure}[h]
\centering
\includegraphics[width=12cm,height=8cm]{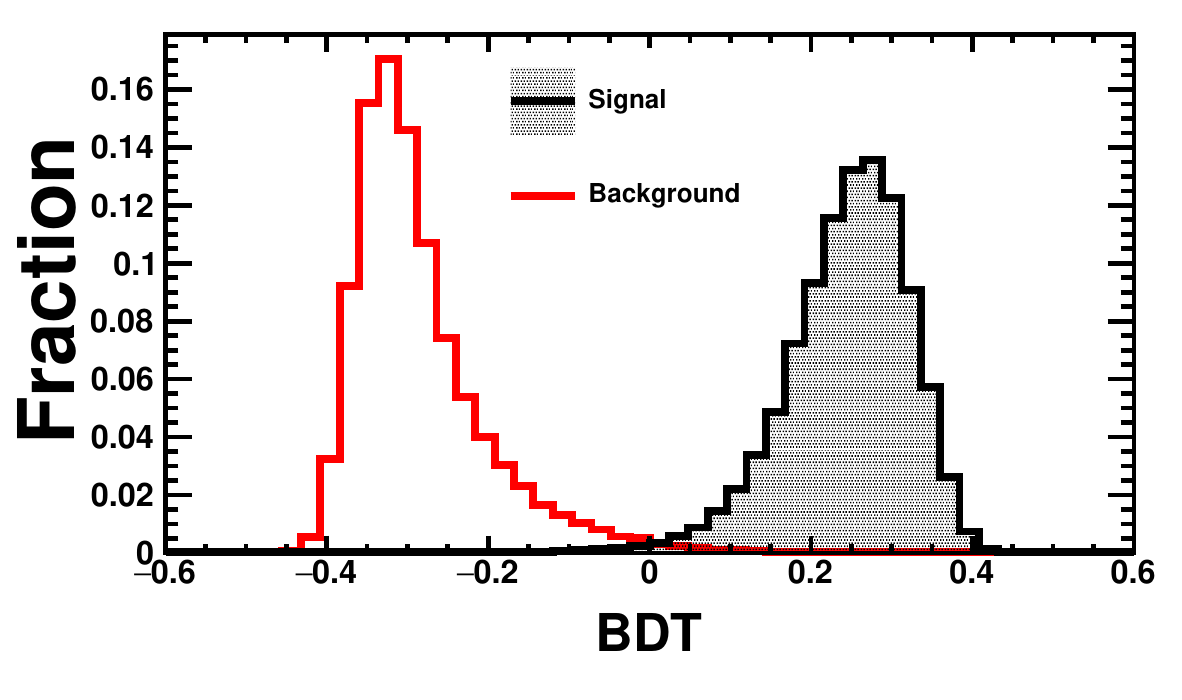}
\caption{
Distributions of BDT responses for the benchmark signal with $m_a=700$~GeV (black, shaded) and the total SM background (red) at the SppC/FCC-hh with $\sqrt{s}=100$~TeV.
}
\label{fig:BDTbenchHad}
\end{figure}

Fig.~\ref{fig:BDTbenchHad} shows the BDT response distributions for the benchmark signal with $m_a=700$~GeV (black, shaded) and for the total SM background (red). Appendix~\ref{app:gammaZjj_BDT} presents the BDT response distributions for the signal and for the six background processes at the SppC/FCC-hh with $\sqrt{s}=100$~TeV, for different $m_a$ values. As $m_a$ increases, the BDT separation between signal and background improves, and for $m_a\gtrsim700$~GeV the overlap becomes small.

The BDT threshold is optimized independently at each $m_a$ to maximize the statistical significance~\cite{Cowan:2011},
\begin{equation}
\sigma_{\rm stat}=\sqrt{2\left[\left(N_s+N_b\right)\ln\!\left(1+\frac{N_s}{N_b}\right)-N_s\right]},
\label{eqn:statSgf}
\end{equation}
where $N_s$ ($N_b$) denotes the number of signal (background) events after preselection and the BDT requirement~\cite{Cowan:2011, ATLAS:2020yaz, Bhattiprolu:2020mwi}.
The selection efficiencies of the optimized BDT requirement for both signal and background processes at the SppC/FCC-hh with $\sqrt{s}=100$~TeV, for different $m_a$ values, are summarized in Appendix~\ref{app:gammaZjj_efficiency}.

\subsection{$W^\pm W^\pm W^\mp$ final state} 
\label{subsec:WWW_bkg_ana}

\subsubsection{Background processes}
\label{subsubsec:WWW_SMbg}

The signal process is $pp \to W^{\pm} a(\to W^+W^-)$, where the same-sign $W$ bosons decay leptonically into muons and neutrinos, $W^\pm\to\mu^\pm\nu_\mu$ (including the charge-conjugate mode), while the remaining $W$ boson decays hadronically, $W^\mp\to jj$; cf.\ panel~(II) of Fig.~\ref{fig:signal}. The resulting final state contains two same-sign muons, a dijet system, and moderate missing transverse energy. The relevant SM processes that can mimic this signature are:
\begin{enumerate}[label*=(\alph*)]
\item single-boson production with two jets: $W^\pm jj$ and $Zjj$;
\item diboson production with two jets: $W^\pm W^\pm jj$, $W^+W^-jj$, $W^\pm Zjj$, and $ZZjj$;
\item triboson production: $W^\pm W^\pm W^\mp$;
\item top-quark pair production: $t\bar{t}$.
\end{enumerate}

After the $W$-boson decays, the processes $W^\pm W^\pm jj$ and $W^\pm W^\pm W^\mp$ yield the same visible final state as the signal and therefore constitute irreducible backgrounds, despite their relatively small production cross sections. The $W^\pm jj$ process, with a much larger production cross section, contributes via misidentified leptons. Charge mismeasurement of final-state leptons can also give rise to backgrounds from $Zjj$, $W^+W^-jj$, and $t\bar{t}$. Furthermore, the $W^\pm Zjj$ process becomes relevant when one lepton is not reconstructed (a lost lepton), and $ZZjj$ can contribute through either lost leptons or charge mismeasurement.

\subsubsection{Preselection}
\label{subsubsec:WWW_presel}

We apply the following preselection criteria before performing the multivariate analysis. Final-state muons and jets are ordered by transverse momentum and labeled as $O_i$ ($i=1,2,\ldots$) with $O=\mu,j$:
\begin{enumerate}[label*=(\roman*)]
\item Exactly two muons, i.e.\ $N(\mu)=2$.
\item Both muons must satisfy $p_{\rm T}(\mu)>10$~GeV.
\item The two muons must have the same electric charge, i.e.\ $\mu^\pm\mu^\pm$.
\item Each jet must satisfy $p_{\rm T}(j)>30$~GeV.
\item At least two jets, $N(j)\ge2$, and no $b$-tagged jets, i.e.\ $N(j_b)=0$.
\end{enumerate}

\begin{table}[h]
\centering
\begin{tabular}{ccccccc}
\hline
\hline
& initial  & (i)--(ii) & (iii) & (iv)--(v) \\ 	
\hline
$W^\pm a(\rightarrow W^\pm W^\mp)$ & $4.90 \mltp 10^{6}$ & $1.31 \mltp 10^{5}$ & $4.35 \mltp 10^{4}$ & $2.48 \mltp 10^{4}$  \\
\hline
$W^{\pm}(\rightarrow\mu^{\pm}\nu_{\mu})\,jj$ & $1.75 \mltp 10^{12}$ & $2.82 \mltp 10^{8}$ & $1.23 \mltp 10^{8}$ & $5.20 \mltp 10^{7}$ \\
$Z(\rightarrow\mu^{+}\mu^{-})\,jj$ & $1.73 \mltp 10^{11}$ & $1.12 \mltp 10^{11}$ & $3.86 \mltp 10^{6}$ & $1.48 \mltp 10^{6}$  \\
$W^\pm W^\pm jj$ & $1.21 \mltp 10^{8}$ & $1.12 \mltp 10^{6}$ & $1.11 \mltp 10^{6}$ & $7.50 \mltp 10^{5}$  \\
$W^\pm(\rightarrow\mu^{\pm}\nu_{\mu})\,W^\mp jj$ & $6.39 \mltp 10^{9}$ & $5.04 \mltp 10^{8}$ & $5.91 \mltp 10^{5}$ & $3.45 \mltp 10^{5}$  \\
$W^\pm Zjj$ & $1.79 \mltp 10^{10}$ & $3.32 \mltp 10^{8}$ & $9.23 \mltp 10^{6}$ & $5.62 \mltp 10^{6}$ \\
$ZZjj$ & $3.38 \mltp 10^{9}$ & $1.38 \mltp 10^{8}$ & $2.04 \mltp 10^{5}$ & $1.24 \mltp 10^{5}$ \\
$W^\pm W^\pm W^\mp$ & $3.16 \mltp 10^{7}$ & $7.20 \mltp 10^{5}$ & $2.36 \mltp 10^{5}$ & $1.22 \mltp 10^{5}$ \\
$t\bar{t}$ & $4.94 \mltp 10^{11}$ & $4.38 \mltp 10^{9}$ & $5.05 \mltp 10^{7}$ & $1.27 \mltp 10^{7}$ \\
\hline
\hline
\end{tabular}
\caption{
Expected number of events for the signal with benchmark mass $m_a=900$~GeV and coupling $g_{aWW}=1~\mathrm{TeV}^{-1}$, and for the background processes, after applying preselection criteria (i)–(v) sequentially at the SppC/FCC-hh with $\sqrt{s}=100$~TeV and $\mathcal{L}=20~\iab$.
}
\label{tab:Crsc}
\end{table}

To illustrate the impact of each preselection criterion, Table~\ref{tab:Crsc} shows the expected number of events for the signal with benchmark $m_a=900$~GeV and for each background process after sequentially applying criteria (i)–(v) at the SppC/FCC-hh with $\sqrt{s}=100$~TeV and integrated luminosity $\mathcal{L}=20~\iab$. Because of its large production cross section and relatively high selection efficiency, the $W^\pm jj$ background dominates and accounts for about 95\% of the total background after preselection.
As shown in Table~\ref{tab:Crsc}, after applying all preselection criteria, the number of background events for most processes is reduced by at least a factor of $10^{4}$, although the total background still exceeds the expected signal yield.

\subsubsection{Multivariate analysis}
\label{subsubsec:WWW_mva}

After the preselection, we employ the TMVA package~\cite{TMVA:2007ngy} to perform a MVA using a BDT with default settings. Events that pass the preselection are evaluated using the following kinematic observables:

\begin{enumerate}[label=(\Alph*)]
\item Missing transverse energy and its azimuthal angle: $\met$ and $\phi(\met)$.
\item The momentum components ($p_x,p_y,p_z$) and energy ($E$) of the two leading jets and two leading muons: \\
$p_x(j_1)$, $p_y(j_1)$, $p_z(j_1)$, $E(j_1)$; $p_x(j_2)$, $p_y(j_2)$, $p_z(j_2)$, $E(j_2)$; \\
$p_x(\mu_1)$, $p_y(\mu_1)$, $p_z(\mu_1)$, $E(\mu_1)$; $p_x(\mu_2)$, $p_y(\mu_2)$, $p_z(\mu_2)$, $E(\mu_2)$.
\item The number of charged tracks ($N_{\rm track}$) and the hadronic-to-electromagnetic energy ratio ($R_{\rm E}$) for the two leading jets: \\
$N_{\rm track}(j_1)$, $N_{\rm track}(j_2)$; $R_{\rm E}(j_1)$, $R_{\rm E}(j_2)$ (typically $R_{\rm E}>1$ for jets).
\item Angular separation and invariant mass of the two leading jets, and the jet multiplicity: \\
$\Delta R(j_1,j_2)$, $m(j_1+j_2)$, $N(j)$.
\item Muon-isolation observables: \\
(i) the scalar $p_{\rm T}$ sum of all other objects within a cone $R=0.4$ around the muon, excluding the muon itself: $p^{\rm iso}_{\rm T}(\mu_1)$, $p^{\rm iso}_{\rm T}(\mu_2)$, and their maximum, $p^{\rm iso}_{\rm T,\,\max}(\mu)$; \\
(ii) the ratio of the transverse energy in a $3\times3$ grid surrounding the muon to the muon transverse momentum (the ``etrat'' variable in Ref.~\cite{lhcoFormat}), which takes values between 0 and 0.99: $R_{\rm grid}(\mu_1)$, $R_{\rm grid}(\mu_2)$; \\
(iii) the minimal separation between each muon and any jet, 
%the minimal separation between each muon and any jet: for each muon we compute $\Delta R$ to every jet and then take the minimal value, 
$\Delta R_{\rm min}(\mu,j)$. Well-isolated muons have small $p^{\rm iso}_{\rm T}$ and $R_{\rm grid}$ but large $\Delta R_{\rm min}(\mu,j)$.
\item Hadronic $W$ reconstruction ($jj_{_W}$): among all dijet pairs, we select the pair with invariant mass closest to $80$~GeV and label the jets by $p_{\rm T}$ as $j_{_{W1}}$ and $j_{_{W2}}$. The input observables are \\
$p_x(j_{_{W1}})$, $p_y(j_{_{W1}})$, $p_z(j_{_{W1}})$, $E(j_{_{W1}})$; \\
$p_x(j_{_{W2}})$, $p_y(j_{_{W2}})$, $p_z(j_{_{W2}})$, $E(j_{_{W2}})$; \\
$N_{\rm track}(j_{_{W1}})$, $N_{\rm track}(j_{_{W2}})$; $R_{\rm E}(j_{_{W1}})$, $R_{\rm E}(j_{_{W2}})$; \\
$\Delta R(j_{_{W1}},j_{_{W2}})$, $m(j_{_{W1}}+j_{_{W2}})$.
\item Transverse masses $m_{\rm T}$~\cite{Han:2005mu}: $m_{\rm T}(\mu_1+\met)$ and $m_{\rm T}(\mu_2+\met)$. \\
Here, the transverse mass $m_{\mathrm T}$ of a visible system and the missing transverse momentum is defined as
$
m_{\mathrm T}^2
= \big(E_{\mathrm T}^{\rm vis} + \met \big)^2
 - \bigl|\vec{p}_{\mathrm T}^{\,\rm vis} + \vec{p}_{\mathrm T}^{\,\rm miss}\bigr|^2 ,
$
where $E_{\mathrm T}^{\rm vis} = \sqrt{|\vec{p}_{\mathrm T}^{\,\rm vis}|^2 + (m^{\rm vis})^2}$ is the transverse energy of the visible system and 
$\met = |\vec{p}_{\mathrm T}^{\,\rm miss}|$ assumes a massless invisible system.
\item Separations between the hadronic $W$ candidate and each muon: we define $\mu_a$ ($\mu_b$) as the muon with the smaller (larger) $\Delta R$ to the hadronic $W$ candidate $jj_{_W}$. The input observables are $\Delta R(j_{_{W1}}+j_{_{W2}},\mu_a)$ and $\Delta R(j_{_{W1}}+j_{_{W2}},\mu_b)$.
\item ALP-mass observables: $m(j_{_{W1}}+j_{_{W2}}+\mu_a)$, $m(j_{_{W1}}+j_{_{W2}}+\mu_b)$, and $m_{\rm T}(j_{_{W1}}+j_{_{W2}}+\mu_a+\met)$, $m_{\rm T}(j_{_{W1}}+j_{_{W2}}+\mu_b+\met)$.
\item Off-shell $W^{(*)}$ reconstruction: $m_{\rm T}(j_{_{W1}}+j_{_{W2}}+\mu_a+\mu_b+\met)$.
\end{enumerate}

Details of $N_{\rm track}(j_i)$, $R_{\rm E}(j_i)$, $R_{\rm grid}(\mu_i)$, and $p^{\rm iso}_{\rm T}(\mu)$ can be found in Ref.~\cite{lhcoFormat}. 
The hadronic $W$ candidate $jj_{_W}$ is reconstructed as described above. 
Appendix~\ref{app:WWW_observables} presents distributions of representative observables for the signal and background processes at the SppC/FCC-hh with $\sqrt{s}=100$~TeV, after applying the preselection criteria.
For $m_a\lesssim500$~GeV, the endpoints of $m(j_{_{W1}}+j_{_{W2}}+\mu_a)$ and $m_{\rm T}(j_{_{W1}}+j_{_{W2}}+\mu_a+\met)$ track $m_a$ closely. For larger $m_a$, the endpoint correlation of $m_{\rm T}(j_{_{W1}}+j_{_{W2}}+\mu_a+\met)$ with $m_a$ persists but becomes less pronounced.

\begin{figure}[h]
\centering
\includegraphics[width=7.3cm,height=5cm]{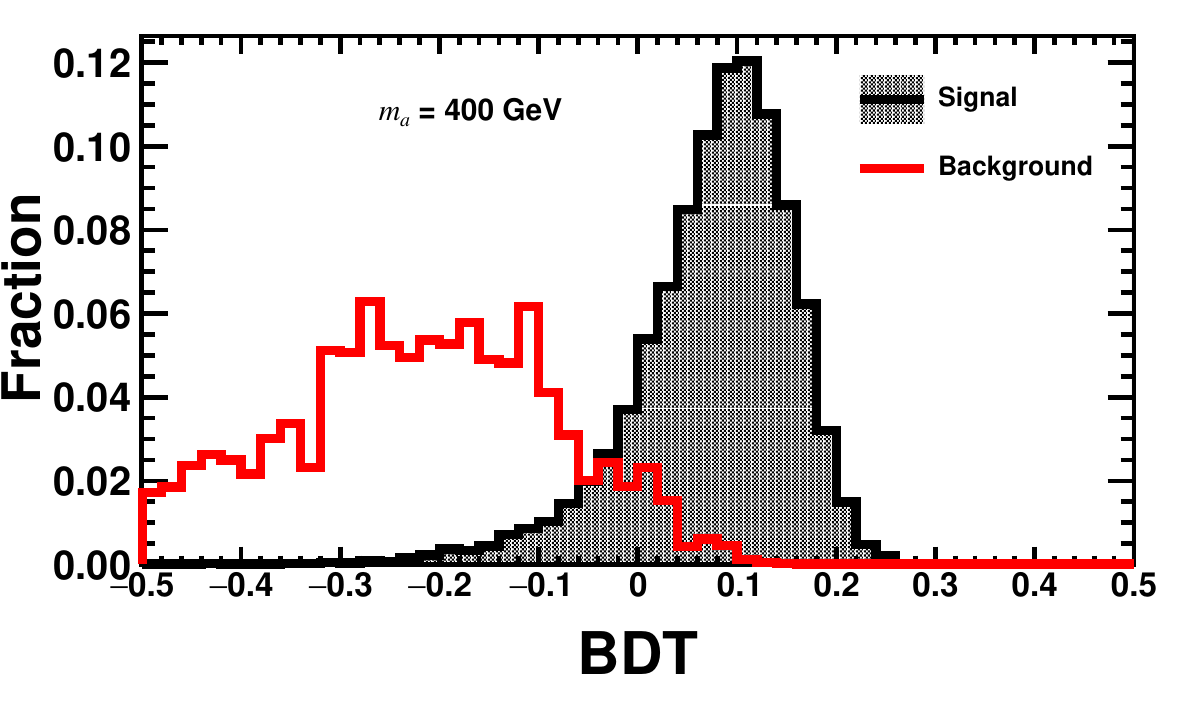}\,\,\,\,\,
\includegraphics[width=7.3cm,height=5cm]{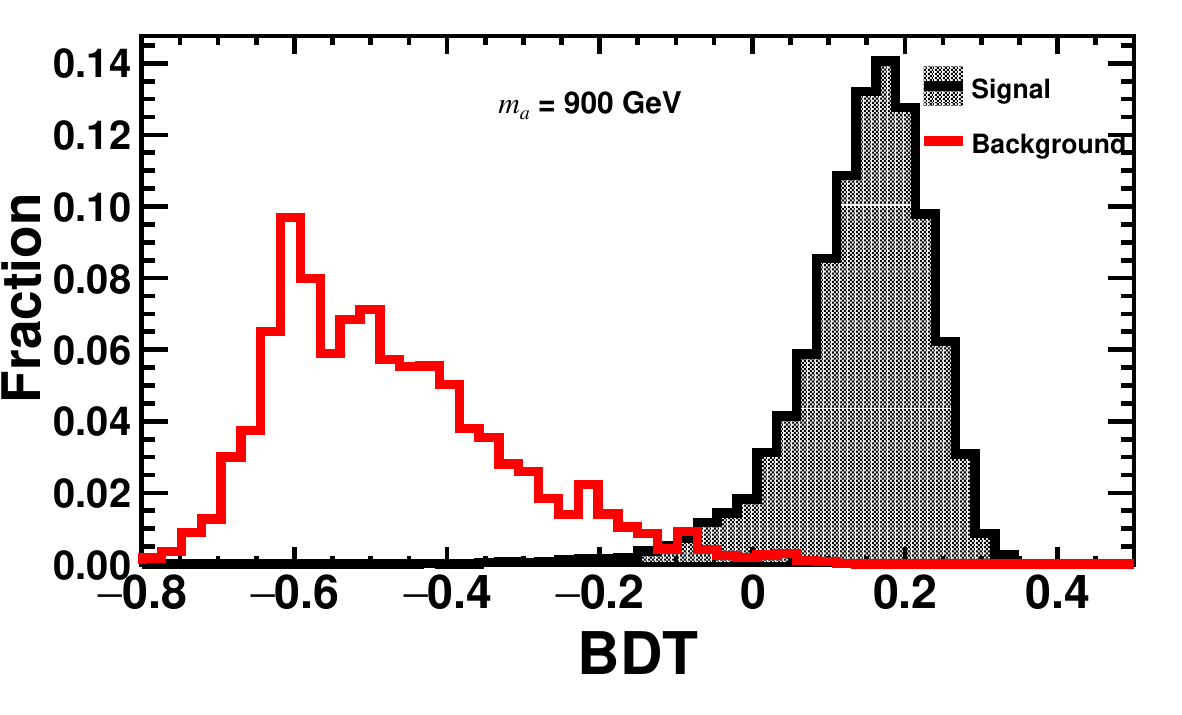}
\caption{
BDT response distributions for the total SM background and the signal $W^\pm a(\to W^\pm W^\mp)$ at the SppC/FCC-hh with $\sqrt{s}=100$~TeV, for $m_a=400$~GeV (left) and $m_a=900$~GeV (right).
}
\label{fig:BDT_400_900}
\end{figure}

Fig.~\ref{fig:BDT_400_900} shows the BDT response distributions for the total SM background and the signal $W^\pm a(\to W^\pm W^\mp)$ at the SppC/FCC-hh with $\sqrt{s}=100$~TeV, for $m_a=400$~GeV (left) and $m_a=900$~GeV (right).
Appendix~\ref{app:WWW_BDT} further shows the BDT response distributions for the signal and background processes at the SppC/FCC-hh with $\sqrt{s}=100$~TeV, after applying the preselection criteria, for representative values of $m_a$.

After the preselection, the BDT threshold is optimized independently at each $m_a$ to maximize the statistical significance defined in Eq.~(\ref{eqn:statSgf}).
Appendix~\ref{app:WWW_Sel_Eff} summarizes the preselection and BDT selection efficiencies for the signal and background processes at the SppC/FCC-hh with $\sqrt{s}=100$~TeV for various $m_a$ benchmark points, where a ``$-$'' entry indicates that the yield is negligible for $\mathcal{L}=20~\iab$.

\subsection{$W^+W^- jj$ final state}
\label{subsec:WWjj_bkg_ana}

\subsubsection{Background processes}
\label{subsubsec:WWjj_SMbg}

The signal targets two \emph{oppositely charged, different-flavour} leptons, at least two jets, and moderate missing transverse momentum, i.e.\ $e^\pm\mu^\mp+jj+\slashed{E}_T$.
In the $e\mu$ channel, Drell--Yan contamination is strongly suppressed. The dominant backgrounds are well known to include diboson and top-quark production, as well as processes with nonprompt or misidentified leptons.
Here, ``nonprompt" denotes leptons from heavy-flavour decays in jets, hadron misidentification, or (for electrons) photon conversions.
We group them by physics origin and by reducibility at detector level.

\paragraph{Irreducible (same visible final state)}
\begin{enumerate}[label*=(\alph*)]
  \item \textbf{$W^+W^-jj$ (QCD+EW).} Inclusive $WW$ production in association with jets, dominated by $q\bar q\!\to WW$ with additional QCD radiation~\cite{CMS:WW136emu2024}. 
  After leptonic $W$ decays, the visible final state coincides with the signal ($e^\pm\mu^\mp+jj+\slashed{E}_T$).
  \item \textbf{$t\bar t$ (dileptonic).} When both $W$ bosons decay leptonically, $t\bar t\!\to\!\ell^+\nu\,\ell^-\bar\nu\,b\bar b$ can be indistinguishable from the signal up to global kinematics if $b$ jets are not correctly tagged; it is therefore effectively irreducible in this channel.
\end{enumerate}

\paragraph{Reducible (lost/misidentified objects)}
\begin{enumerate}[label*=(\alph*),resume]
  \item \textbf{$WZjj$, $ZZjj$.} 
  They contribute either when one prompt lepton is not reconstructed or fails identification/isolation, or when $Z\!\to\!\tau\tau$ produces an $e\mu$ pair.
  \item \textbf{$W^\pm jj$.} Contributes via one prompt lepton from the $W$ and one nonprompt or misidentified lepton (heavy-flavour decays, hadron misidentification, photon conversions).
  \item \textbf{$Zjj$.} This background is strongly suppressed in the $e\mu$ channel but remains nonzero through a misidentified lepton, a lost lepton plus a nonprompt lepton, or via $Z\!\to\!\tau\tau\!\to\!e\mu+X$. 
\end{enumerate}

The $pp \to W^+ W^- jj$ background is generated at matrix-element level with MadGraph5\_aMC@NLO, while the subsequent decays $W \to \ell^\pm \nu_\ell$ ($\ell = e,\mu$) and parton showering are modeled with PYTHIA~8.3. 
The $t\bar{t}$, $WZjj$, and $ZZjj$ samples are likewise produced with MadGraph5\_aMC@NLO and interfaced to PYTHIA~8.3, which handles all kinematically allowed $W$- and $Z$-boson decays. 
Single-boson processes $W^\pm(\to \ell^\pm \nu_\ell)jj$ and $Z(\to \ell^+\ell^-)jj$ with $\ell = e,\mu$ are generated fully differentially in MadGraph5\_aMC@NLO, keeping the leptonic $W$ and $Z$ decays explicit in the matrix element.

\subsubsection{Preselection}
\label{subsubsec:WWjj_presel}

We target final states with two oppositely charged, different-flavour leptons, at least two jets, and moderate missing transverse momentum. Throughout, objects are ordered by transverse momentum, and the leading (subleading) jet is denoted by $j_1$ ($j_2$). The following preselection is applied to select the $e^\pm\mu^\mp+jj+\met$ final state and suppress backgrounds at the first stage:
\begin{enumerate}[label*=(\roman*)]
\item Exactly two leptons: $N(\ell)=2$;
\item Lepton transverse momenta: $p_T(\ell)\ge 10~\mathrm{GeV}$ for both leptons;
\item Different flavour: exactly one electron and one muon, $N(e)=1$ and $N(\mu)=1$;
\item Opposite charge: $Q(e)+Q(\mu)=0$;
\item Jet multiplicity: at least two jets, $N(j)\ge 2$;
\item Leading-jet thresholds: $p_T(j_1)\ge 30~\mathrm{GeV}$ and $p_T(j_2)\ge 30~\mathrm{GeV}$;
\item Top-quark suppression: zero $b$-tagged jets, $N(j_b)=0$.
\end{enumerate}
This opposite-sign $e\mu$ requirement strongly reduces Drell--Yan contamination, while the $b$-jet veto suppresses top-quark backgrounds, in line with LHC $WW\to e^\pm\nu\mu^\mp \bar{\nu}$ measurements~\cite{CMS:WW136emu2024}. 

\begin{table}[h]
\centering
\scalebox{0.71}{
\begin{tabular}{cccccccccc}
\hline
\hline
\multicolumn{1}{c}{100~TeV $pp$} & initial & (i) & (ii) & (iii) & (iv) & (v) & (vi) & (vii) \\
\hline
signal & $2.86\mltp10^{8}$ & $1.89\mltp10^{8}$ & $1.75\mltp10^{8}$ & $8.74\mltp10^7$ & $8.74\mltp10^7$ & $8.41\mltp10^7$ & $8.27\mltp10^7$ & $6.11\mltp10^7$\\ 
\hline
$W^+(\to \ell^+ \nu)\,W^-(\to \ell^- \bar{\nu})\,jj$ & $1.38\mltp10^{9}$ & $8.23\mltp10^{8}$ & $7.35\mltp10^{8}$ & $3.63\mltp10^8$ & $3.63\mltp10^8$ & $2.71\mltp10^8$ & $2.38\mltp10^8$ & $1.87\mltp10^8$\\ 
$WZ\,jj$ & $1.85\mltp10^{10}$ & $1.34\mltp10^{9}$ & $6.27\mltp10^{8}$ & $5.05\mltp10^7$ & $2.83\mltp10^7$ & $2.48\mltp10^7$ & $2.29\mltp10^7$ & $1.70\mltp10^7$\\ 
$ZZ\,jj$ & $3.46\mltp10^{9}$ & $3.23\mltp10^{8}$ & $2.37\mltp10^{8}$ & $3.76\mltp10^6$ & $3.11\mltp10^6$ & $2.78\mltp10^6$ & $2.61\mltp10^6$ & $1.80\mltp10^6$\\
$W^\pm(\to \ell^\pm \nu_\ell/\bar{\nu}_\ell)\,jj$ & $3.50\mltp10^{12}$ & $3.35\mltp10^{11}$ & $8.77\mltp10^{8}$ & $4.30\mltp10^8$ & $2.34\mltp10^8$ & $1.53\mltp10^8$ & $1.22\mltp10^8$ & $8.31\mltp10^7$\\
$Z(\to \ell^+\ell^-)\,jj$ & $3.51\mltp10^{11}$ & $2.06\mltp10^{11}$ & $1.85\mltp10^{11}$  & $2.28\mltp10^{7}$ & $1.18\mltp10^{7}$ & $9.30\mltp10^{6}$ &  $8.03\mltp10^{6}$ & $5.64\mltp10^{6}$ \\
$t\bar{t}$ & $4.93\mltp10^{11}$ & $4.23\mltp10^{10}$ & $1.42\mltp10^{10}$ & $7.03\mltp10^9$ & $6.97\mltp10^9$ & $6.45\mltp10^9$ & $6.11\mltp10^9$ & $6.36\mltp10^8$\\
\hline
\hline
\end{tabular}
}
\caption{
Event yields for the signal with benchmark $m_a=750$~GeV and for the background processes after sequentially applying preselection criteria (i)–(vii). Numbers correspond to a 100~TeV $pp$ collider with $\mathcal{L}=20~\iab$.
}
\label{tab:presel}
\end{table}

Table~\ref{tab:presel} shows the event yields after sequentially applying preselection criteria (i)–(vii) for the signal with benchmark $m_a=750$~GeV and for the background processes at the SppC/FCC-hh with $\sqrt{s}=100$~TeV and $\mathcal{L}=20~\iab$.
With the opposite-sign $e\mu$ requirement, the large inclusive $Wjj$ and $Zjj$ rates are greatly reduced, while $W^+W^-jj$ and $t\bar t$ remain the leading backgrounds in the $e\mu$ channel~\cite{CMS:WW136emu2024}.

\begin{table*}[h]
\centering 
\scalebox{0.75}{
\begin{tabular}{cccccccccccc}
\hline
\hline
$m_a$ [GeV] & 170 & 185 & 200 & 225 & 250 & 350 \\
& $1.94\mltp10^{-1}$ & $1.96\mltp10^{-1}$ & $1.96\mltp10^{-1}$ & $1.98\mltp10^{-1}$ & $1.98\mltp10^{-1}$ & $2.02\mltp10^{-1}$ \\
\hline
$m_a$ [GeV] & 750 & 1500 & 2500 & 4000 & 5500 & 7000 \\
& $2.14\mltp10^{-1}$ & $2.19\mltp10^{-1}$ & $2.25\mltp10^{-1}$ & $2.27\mltp10^{-1}$ & $2.27\mltp10^{-1}$ & $2.28\mltp10^{-1}$ \\
\hline
background & $W^+(\to \ell^+ \nu)\,W^-(\to \ell^- \bar{\nu})\,jj$ & $WZ\,jj$ & $ZZ\,jj$ & $W^\pm(\to \ell^\pm \nu_\ell)\,jj$ & $Z(\to \ell^+\ell^-)\,jj$ & $t\bar{t}$ \\
& $1.36\mltp10^{-1}$ & $9.19\mltp10^{-4}$ & $5.19\mltp10^{-4}$ & $2.37\mltp10^{-5}$ & $1.61\mltp10^{-5}$ & $1.29\mltp10^{-3}$ \\
\hline
\hline
\end{tabular}
}
\caption{
Preselection efficiencies for the signals with different $m_a$ values and for the background processes at a 100~TeV $pp$ collider. 
}
\label{tab:preselection14TeV}
\end{table*}

Table~\ref{tab:preselection14TeV} summarizes the preselection efficiencies for the signals with different $m_a$ values and for the background processes at a 100~TeV $pp$ collider. 
As shown in this table, the $Wjj$ and $Zjj$ processes have the lowest preselection efficiencies because of the tight opposite-sign $e\mu$ requirement. 
Diboson $WZjj$ and $ZZjj$ remain as subleading reducible backgrounds, primarily through lost leptons or $\tau$ decays, while $W^+W^-jj$ and $t\bar t$ are effectively irreducible up to global kinematics; the latter is also strongly reduced by the $b$-tag veto.

\subsubsection{Multivariate analysis}
\label{subsubsec:WWjj_mva}

After preselection, we employ a MVA to further discriminate signal from background. The input observables are:
\begin{enumerate}[label*=(\roman*)]
\item Object kinematics: energy, transverse momentum, pseudorapidity, and azimuthal angle of the final-state objects,
$E(O)$, $p_T(O)$, $\eta(O)$, $\phi(O)$ with $O=\mu,\,e,\,j_1,\,j_2$.
\item Missing transverse momentum (magnitude and azimuthal angle): $\met$ and $\phi(\met)$.
\item Among all jet pairs, we identify the pair with the smallest angular separation, characterized by $\Delta R (j, j')_{\rm min}$ and the corresponding dijet mass $m(j+j')_{ {\rm min}\,\Delta R}$.
\item Among all jet pairs, we also select the pair with the largest pseudorapidity separation, $\Delta \eta(j, j')_{\rm max}$, and its dijet mass $m(j + j')_{ {\rm max}\,\Delta \eta}$.
\item The invariant mass of the two leading jets and the jet multiplicity: $m(j_{1}+j_{2})$ and $N(j)$.
\item Angular separations between key objects:
$\Delta R(j_1,j_2)$, $\Delta\eta(j_1,j_2)$, $\Delta R(e,\mu)$, $\Delta R(\mu,j_1)$, $\Delta R(\mu,j_2)$, $\Delta R(e,j_1)$, $\Delta R(e,j_2)$.
\item Transverse masses of leptons and $\met$: $m_{\rm T}(\mu + \met)$, $m_{\rm T}(e + \met)$, and $m_{\rm T}(\mu + e + \met)$. 
\end{enumerate}

Observables (iv)–(vi) exploit the VBF-like topology, characterized by two forward tagging jets, large $\Delta\eta_{jj}$ and high $m_{jj}$ with reduced central hadronic activity, while (iii) captures $s$-channel topologies where jets tend to be closer in phase space. This choice follows standard LHC practices for VBF and dijet tagging.
Observables (vii) are related to the kinematics of the decay $a\to W^+W^-\to e^\pm\mu^\mp\nu\bar{\nu}$ in the $e^\pm\mu^\mp+\met$ final state.
For completeness, Appendix~\ref{app:WWjj_observables} displays distributions of representative observables for the signal and the six background processes at the SppC/FCC-hh with $\sqrt{s}=100$~TeV, assuming the benchmark $m_a=750$~GeV.

\begin{figure}[htbp]
\centering
\subfigure{
\includegraphics[width=7.3cm,height=5cm]{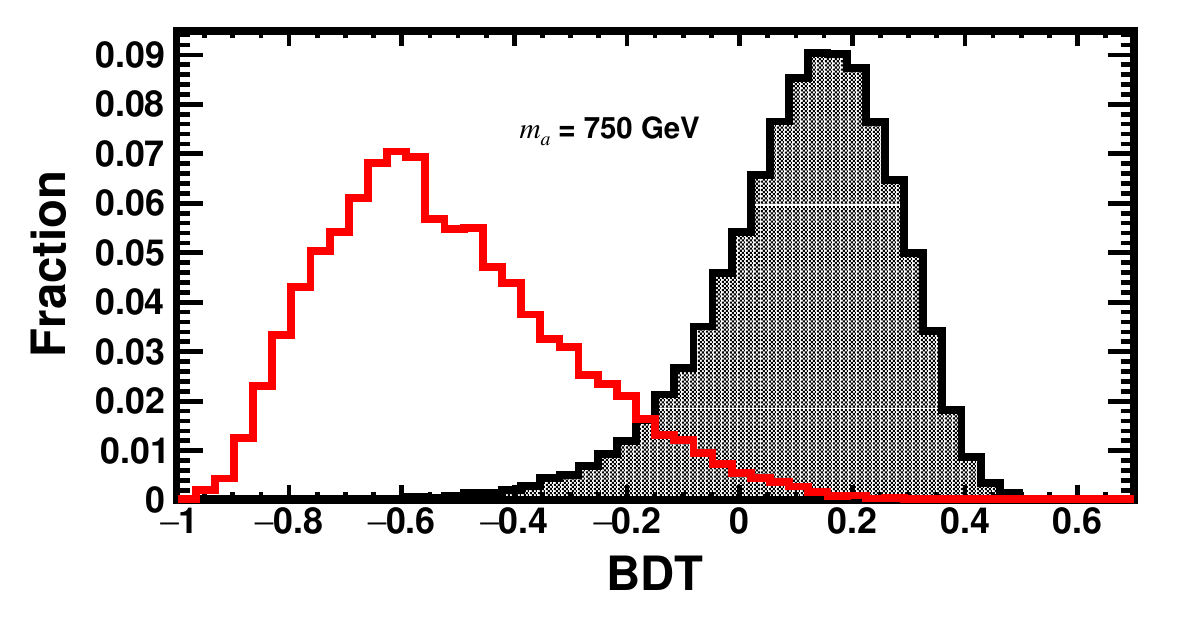}
\includegraphics[width=7.3cm,height=5cm]{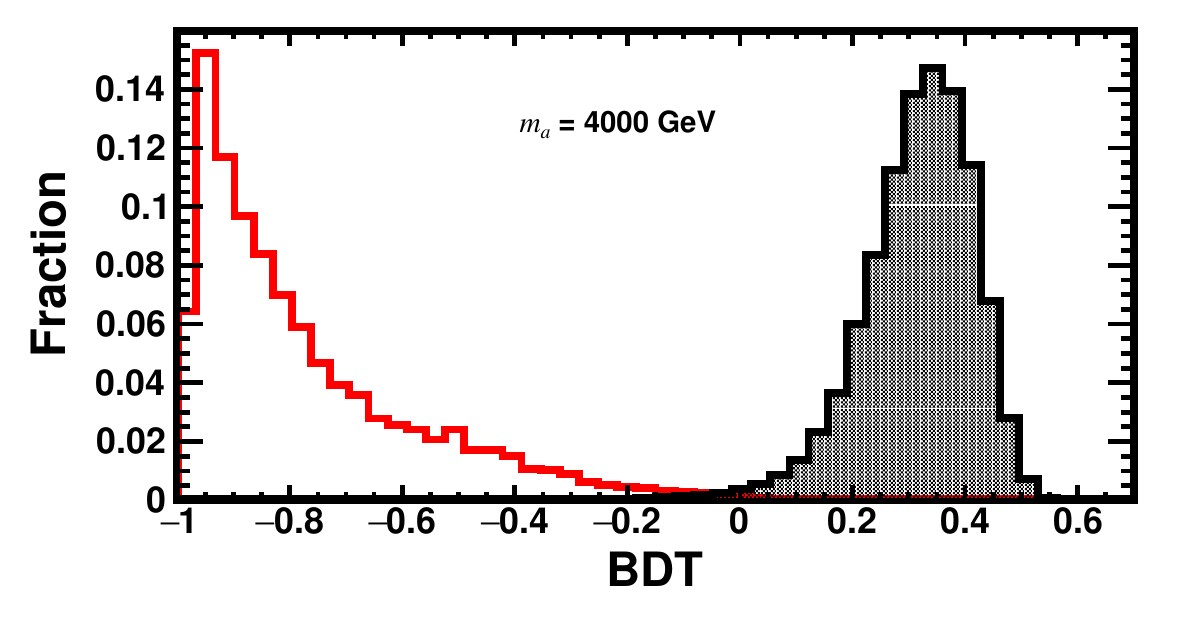}
}
\caption{
Distributions of BDT responses for the signal (black, shaded) and the total background (red) at $\sqrt{s}=100$~TeV, assuming $m_a=750$~GeV (left) and $m_a=4000$~GeV (right).
}
\label{fig:multivariate analysis}
\end{figure}

We implement the MVA using a BDT in the \textsc{TMVA} package integrated into \textsc{ROOT}, adopting the default BDT configuration for classification~\cite{TMVA:2007ngy}. 
This setup is designed to maximize background rejection while retaining high signal efficiency. 
Fig.~\ref{fig:multivariate analysis} shows the distributions of BDT responses for the signal and the total background at $\sqrt{s}=100$~TeV, assuming benchmarks $m_a=750$~GeV (left) and $m_a=4000$~GeV (right).
Appendix~\ref{app:WWjj_BDT} shows BDT response distributions after applying the preselection criteria for the signal and background processes at the SppC/FCC-hh with $\sqrt{s}=100$~TeV for representative $m_a$ values.
As $m_a$ increases, the underlying kinematics becomes more boosted and the separation between signal and background improves, as seen from the shift of the signal (background) distribution toward higher (lower) BDT scores.

Similar to previous analyses, the BDT threshold is then optimized independently at each $m_a$ to maximize the statistical significance defined in Eq.~(\ref{eqn:statSgf}).
Appendix~\ref{app:WWjj_Sel_Eff} summarizes the selection efficiencies of the BDT cut for both signal and background processes at the SppC/FCC-hh with $\sqrt{s}=100$~TeV, assuming different ALP masses. 
Despite their large inclusive cross sections, the $W^\pm(\to\ell^\pm\nu)jj$ and $Z(\to\ell^+\ell^-)jj$ backgrounds are almost completely removed by the opposite-sign $e\mu$ requirement and the multivariate selection, so they are omitted from the appendix table for brevity.

\section{Results}
\label{sec:results}

Based on the preceding analyses, we present the discovery sensitivities for the signal at the SppC/FCC-hh with center-of-mass energy $\sqrt{s}=100$~TeV and integrated luminosity $\mathcal{L}=20~\iab$. Results are shown over the ALP mass range $m_a \in [100,7000]$~GeV, both in terms of the coupling $g_{aWW}$ and in terms of the model-independent fiducial quantity given by the signal production cross section times branching ratio, $\sigma \times \mathrm{Br}$.

\subsection{Sensitivities on $g_{_{aWW}}$}

\begin{figure}[h]
\centering
\includegraphics[width=12cm, height=8cm]{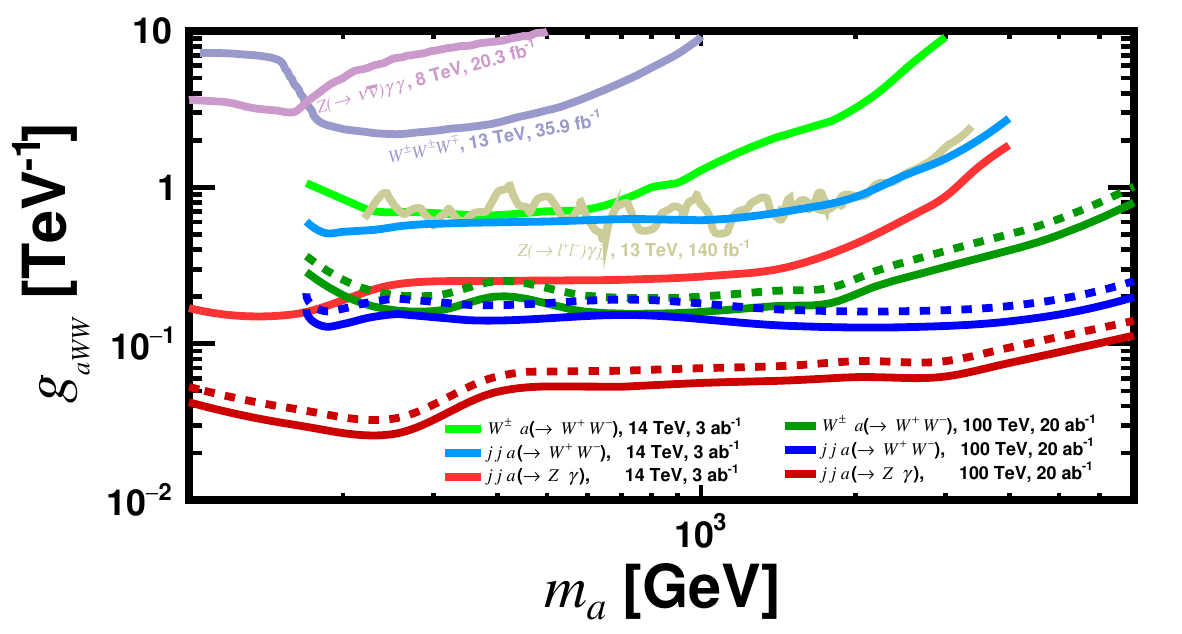}
\caption{
Projected discovery reach on the $aWW$ coupling, $g_{aWW}$, for photophobic ALPs as a function of the ALP mass $m_a$ in the range 100--7000~GeV at the SppC/FCC-hh with $\sqrt{s}=100~\mathrm{TeV}$ and $\mathcal{L}=20~\mathrm{ab}^{-1}$. Solid (dashed) curves indicate $2\sigma$ ($5\sigma$) statistical significances. The projection assumes $g_{a\gamma\gamma}=0$ and a dedicated BDT optimization at each mass point; significances are statistical only.
For comparison, we overlay existing 95\%~CL limits recast from the triboson channel $pp \to Z(\to \nu\bar{\nu})\,\gamma\gamma$ at $\sqrt{s}=8$~TeV with 20.3~$\ifb$~\cite{Craig:2018kne}, and from Run-2 analyses at $\sqrt{s}=13$~TeV: $pp\to W^\pm W^\pm W^\mp$ with 35.9~$\ifb$ and $pp\to Z(\to\ell^+\ell^-)\gamma jj$ with 140~$\ifb$~\cite{Aiko:2024xiv}. Also shown are HL-LHC ($\sqrt{s}=14~\mathrm{TeV}$, $\mathcal{L}=3~\mathrm{ab}^{-1}$) $2\sigma$ projections for $pp\to jj\,a(\to Z \gamma)$~\cite{Ding:2024djo}, $pp\to W^\pm a(\to W^+W^-)$~\cite{Mao:2024kgx}, and $pp\to jj\,a(\to W^+W^-)$~\cite{Feng:2025kof}.
}
\label{fig:sensALP}
\end{figure}

In Fig.~\ref{fig:sensALP}, we summarize the projected discovery reach on the
photophobic ALP--gauge coupling $g_{aWW}$ as a function of the ALP mass
$m_a$. The three sets of red, green and blue curves correspond to the signal topologies introduced in Fig.~\ref{fig:signal} at the SppC/FCC-hh
($\sqrt{s}=100~\mathrm{TeV}$, $\mathcal{L}=20~\mathrm{ab}^{-1}$): 
(I) $pp\to jj\,a(\to Z \gamma)$, 
(II) $pp\to W^\pm a(\to W^+W^-)$, 
and (III) $pp\to jj\,a(\to W^+W^-)$. 
For each topology we show both $2\sigma$ and $5\sigma$ discovery contours in
the $(m_a,g_{aWW})$ plane, obtained after the channel-specific BDT
optimization described in Secs.~\ref{subsec:gammaZjj_bkg_ana},
\ref{subsec:WWW_bkg_ana}, and~\ref{subsec:WWjj_bkg_ana}, and imposing the
photophobic condition $g_{a\gamma\gamma}=0$.  
The corresponding $2\sigma$ HL-LHC projections at
$\sqrt{s}=14~\mathrm{TeV}$ and $\mathcal{L}=3~\mathrm{ab}^{-1}$ from
Refs.~\cite{Ding:2024djo,Mao:2024kgx,Feng:2025kof} are overlaid for direct
comparison with the $100$~TeV reach.
In addition, the figure also displays, for reference, existing $95\%$~CL limits obtained by recasting LHC measurements: the Run-1 triboson channel
$pp\to Z(\to\nu\bar{\nu})\gamma\gamma$ at $\sqrt{s}=8$~TeV with
$20.3~\mathrm{fb}^{-1}$~\cite{Craig:2018kne}, and the Run-2 measurements of
$pp\to W^\pm W^\pm W^\mp$ (35.9~fb$^{-1}$) and
$pp\to Z(\to\ell^+\ell^-)\gamma jj$ (140~fb$^{-1}$) at
$\sqrt{s}=13$~TeV~\cite{Aiko:2024xiv}. 

Comparing the three 100~TeV curves in Fig.~\ref{fig:sensALP}, one finds a clear hierarchy among the signal topologies.  
Over the full mass range, the $jj\,a(\to Z\gamma)$ channel~(I) provides the best reach in $g_{aWW}$.  
This final state combines a fully reconstructible $m(\ell^+ + \ell^- + \gamma)$ resonance, excellent photon and lepton energy resolution, and relatively small irreducible backgrounds, which together permit tight BDT selections with only modest signal loss.
Among the two $WW$ modes, the $jj\,a(\to W^+W^-)$ channel~(III) overtakes the $W^\pm a(\to W^+W^-)$ channel~(II) once $m_a\gtrsim 1~\TeV$.  
The origin is twofold.  
First, channel~(III) receives contributions from both $s$-channel vector-boson exchange and genuine VBF topologies, so at high mass it benefits from the rapid growth of the electroweak vector-boson luminosity and the large phase space for two energetic forward jets.  
By contrast, $pp\to W^\pm a$ in channel~(II) proceeds purely through an $s$-channel configuration and its cross section drops more steeply with $m_a$.  
Second, both $WW$ channels are intrinsically less clean than $Z\gamma$: leptonic $W$ decays introduce sizeable genuine $\slashed{E}_T$, large SM $WW$ and $t\bar t$ backgrounds, and no fully reconstructible resonance.  
Even though $\mathrm{Br}(a\to W^+W^-)$ exceeds $\mathrm{Br}(a\to Z\gamma)$ in the photophobic limit, these effects weaken the net reach in $g_{aWW}$.
In summary, at 100~TeV the $jj\,a(\to Z\gamma)$ channel~(I) sets the benchmark sensitivity; the $jj\,a(\to W^+W^-)$ channel~(III) provides the next-best, and increasingly competitive, reach at high masses thanks to its VBF component; and the $W^\pm a(\to W^+W^-)$ channel~(II), while statistically less powerful especially at high masses, probes the same coupling in a complementary production mode and final state.

A comparison between the SppC/FCC-hh ($\sqrt{s}=100$~TeV, $\mathcal{L}=20~\mathrm{ab}^{-1}$) projections and the HL-LHC ($\sqrt{s}=14$~TeV, $\mathcal{L}=3~\mathrm{ab}^{-1}$) results for the same three final states highlights the combined impact of higher energy and luminosity. In all channels the gain is modest at low $m_a$, where production is dominated by relatively soft partons already abundant at the LHC. Once $m_a$ enters the TeV regime, however, the much larger partonic centre-of-mass energy and enhanced electroweak boson luminosities at 100~TeV increase the signal cross sections by orders of magnitude. This translates into roughly an order-of-magnitude improvement in the reachable $g_{aWW}$ values compared to the HL-LHC, and extends the mass reach by several TeV. The effect is most pronounced in the VBF-assisted channels~(I) and~(III), whose cross sections grow with energy, while the purely $s$-channel tri--$W$ mode~(II) benefits mainly from the larger luminosity but still gains significantly in the multi-TeV mass range.

Despite their weaker raw sensitivity, the tri--$W$ channel~(II) and the $jj\,a(\to W^+W^-)$ channel~(III) provide essential information that is not redundant with the $jj\,a(\to Z\gamma)$ channel~(I). Channel~(II) probes $pp\to W^\pm a$ in a purely $s$-channel configuration with a very clean same-sign dilepton signature; its background composition, trigger strategy and dominant systematics are qualitatively different from those of the $Z\gamma jj$ analysis. Channel~(III), by contrast, probes the same $aWW$ vertex in a mixed $s$+VBF configuration, where two forward/backward tagging jets and large $|\Delta\eta_{jj}|$ make it particularly sensitive to the Lorentz and CP structure of the $aWW$ operator. Both $WW$ channels directly test the $aWW$ coupling via $a\to W^+W^-$, whereas $jj\,a(\to Z\gamma)$ constrains $aZ\gamma$. A combined analysis of all three therefore overconstrains the photophobic relations $g_{aZ\gamma}=t_\theta\,g_{aWW}$ and $g_{aZZ}=(1-t_\theta^2)\,g_{aWW}$, provides powerful cross-checks against channel–specific systematics, and offers a robust basis for global fits. In this sense, the three topologies are not merely alternative search modes but together form a coherent programme to test the gauge structure and high-energy behaviour of photophobic ALP scenarios.

\subsection{Sensitivities on production cross sections}

\begin{figure}[h]
\centering
\includegraphics[width=12cm,height=8cm]{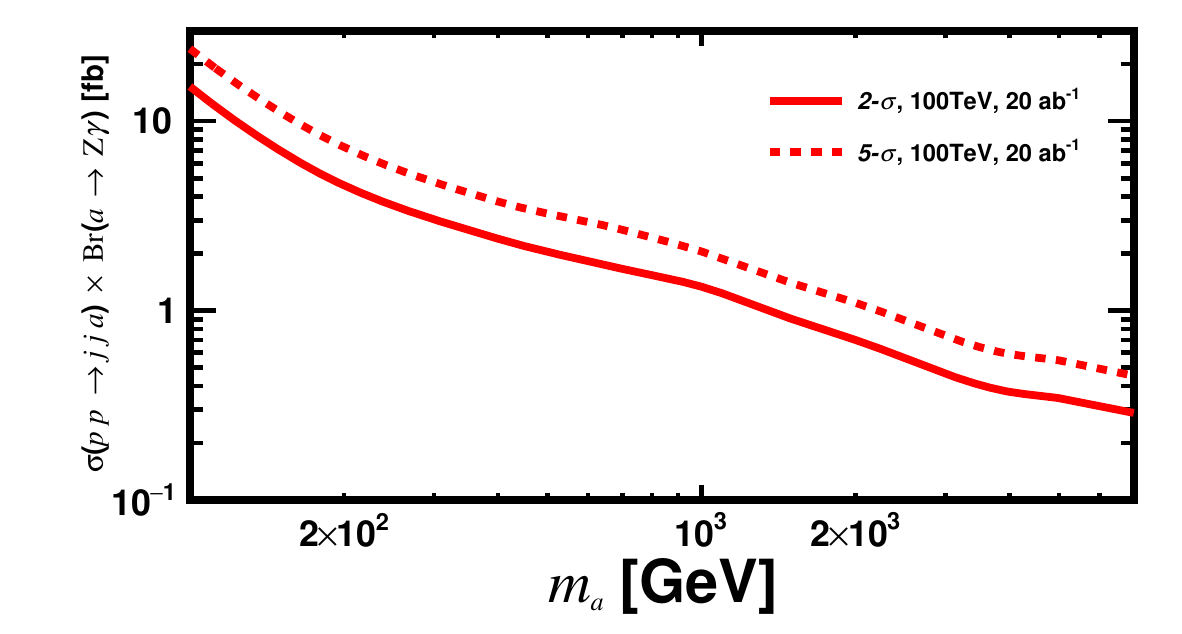}
\caption{
Discovery sensitivities on the production cross section $\sigma(pp \to jj\,a)$ times the branching ratio $\mathrm{Br}(a \to Z \gamma)$ as a function of $m_a$ from 100 to 7000~GeV at the SppC/FCC-hh with $\sqrt{s}=100$~TeV and $\mathcal{L}=20~\iab$.
The solid (dashed) curves correspond to 2$\sigma$ (5$\sigma$) significance.
}
\label{fig:sensGeneral}
\end{figure}

\begin{figure}[h]
\centering
\includegraphics[width=12cm, height=8cm]{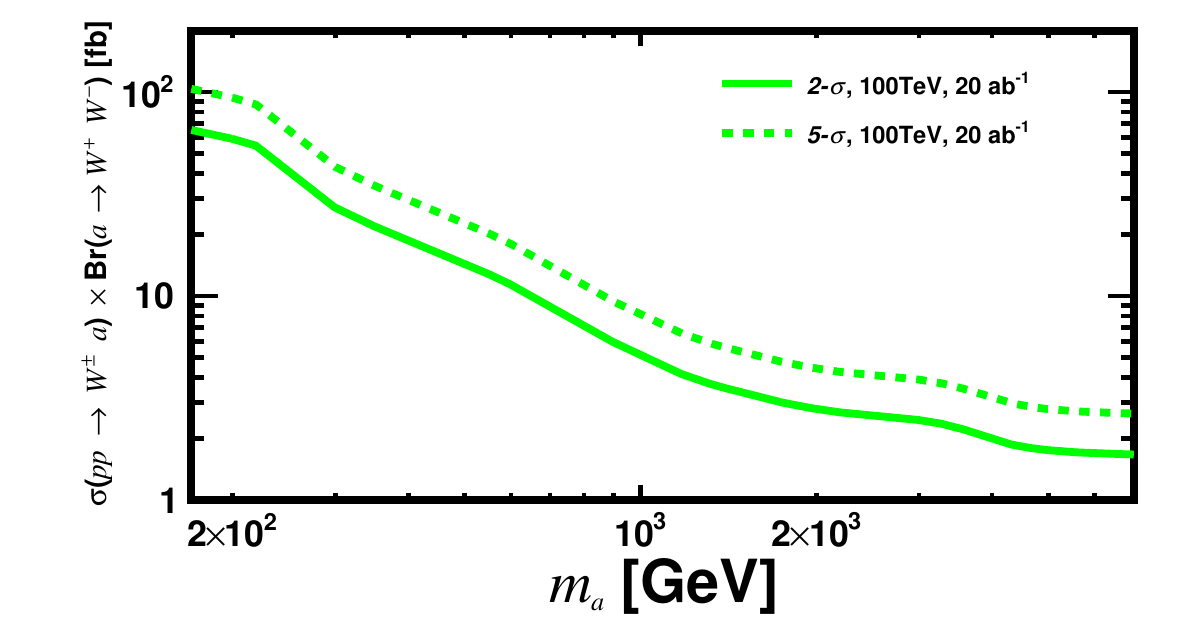}
\caption{
Discovery sensitivities on $\sigma(pp \to W^{\pm} a)$ times $\mathrm{Br}(a \to W^{+}W^{-})$
in the mass range 170--7000~GeV at the SppC/FCC-hh with $\sqrt{s}=100$~TeV and $\mathcal{L}=20~\iab$.
The solid (dashed) curves correspond to 2$\sigma$ (5$\sigma$) significance.
}
\label{fig:sensSigmaBr_Wa}
\end{figure}

\begin{figure}[h]
\centering
\includegraphics[width=12cm, height=8cm]{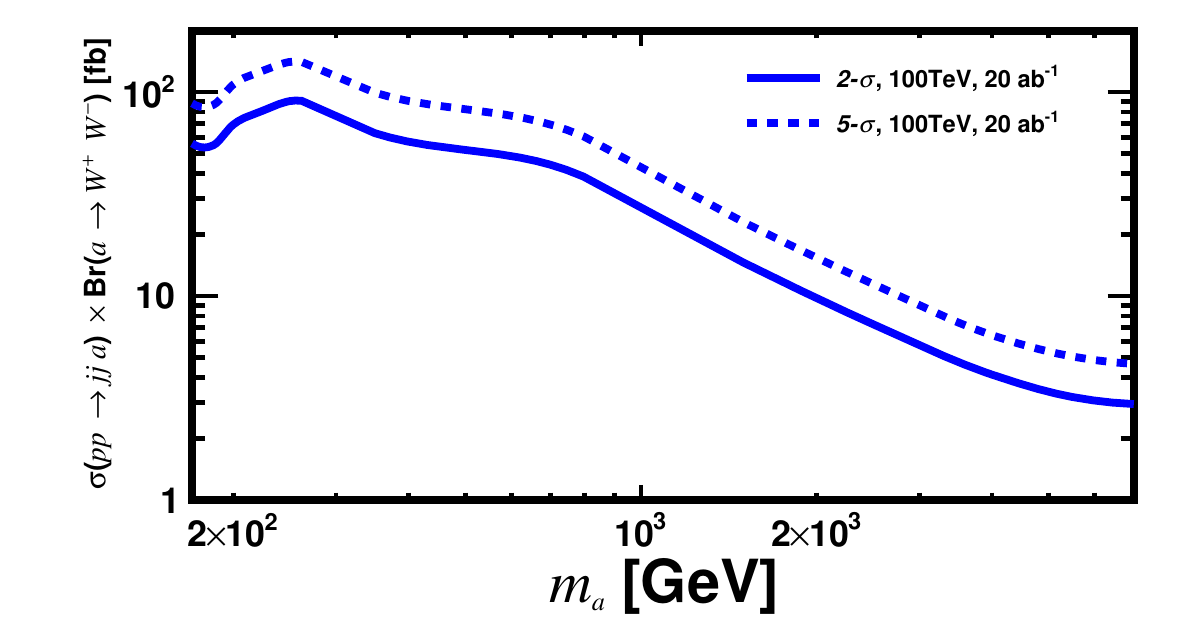}
\caption{
Discovery sensitivities on $\sigma(pp \to jj\,a)$ times $\mathrm{Br}(a \to W^{+}W^{-})$
in the mass range 170--7000~GeV at the SppC/FCC-hh with $\sqrt{s}=100$~TeV and $\mathcal{L}=20~\iab$.
The solid (dashed) curves correspond to 2$\sigma$ (5$\sigma$) significance.
}
\label{fig:sensSigmaBr_jj}
\end{figure}

To facilitate reinterpretations beyond the photophobic benchmark, we also present model-independent discovery thresholds on the inclusive rate
$\sigma\times\mathrm{Br}$ for each signal topology at the SppC/FCC-hh ($\sqrt{s}=100~\mathrm{TeV}$, $\mathcal{L}=20~\mathrm{ab}^{-1}$).
Figs.~\ref{fig:sensGeneral}--\ref{fig:sensSigmaBr_jj} show the minimum $\sigma\times\mathrm{Br}$ required for a $2\sigma$ (solid) or $5\sigma$ (dashed) excess after the full preselection and BDT-based analysis. In all three channels, the reach generally improves (smaller $\sigma\times\mathrm{Br}$) as $m_a$ increases, because the selected SM background falls rapidly in the high-mass region and the signal kinematics become more distinctive, enabling tighter multivariate selections at nearly constant signal efficiency. 

In the $jj \, a(\to Z\gamma)$ channel (I), Fig.~\ref{fig:sensGeneral} reports the discovery reach on $\sigma(pp\to jj\,a)\times\mathrm{Br}(a\to Z\gamma)$. The required rate drops from ${\cal O}(10)~\mathrm{fb}$ at $m_a\sim100~\mathrm{GeV}$ to the sub-fb level toward the multi-TeV regime. This strong performance is driven by the clean experimental signature and, crucially, by the fully reconstructible $m(\ell^+ + \ell^- + \gamma)$ resonance, which effectively localizes the search to a narrow mass region where the continuum background is both small and rapidly falling.

For the associated-production $W^\pm a(\to W^+W^-)$ channel~(II), Fig.~\ref{fig:sensSigmaBr_Wa} shows the reach on $\sigma(pp\to W^\pm a)\times\mathrm{Br}(a\to W^+W^-)$. The sensitivity improves from ${\cal O}(10^2)~\mathrm{fb}$ near threshold to ${\cal O}(1)~\mathrm{fb}$ at multi-TeV masses. Although this channel does not admit a fully reconstructed ALP mass peak due to neutrinos, it benefits from a very rare same-sign dilepton signature, which suppresses many SM backgrounds already at the object-selection level; the remaining backgrounds also become increasingly phase-space limited at large $m_a$, yielding good high-mass reach in $\sigma\times\mathrm{Br}$.

Finally, for the $jj\,a(\to W^+W^-)$ channel~(III),  Fig.~\ref{fig:sensSigmaBr_jj} presents the reach on $\sigma(pp\to jj\,a)\times\mathrm{Br}(a\to W^+W^-)$. At low masses the required rate is comparatively larger, reflecting the intrinsically less constrained kinematics of dileptonic $WW$ decays (genuine $\met$ and no narrow fully reconstructible resonance) and the sizeable irreducible backgrounds from SM $WW$ and $t\bar t$. The reach nonetheless strengthens rapidly with increasing $m_a$, reaching a few-fb level in the multi-TeV regime, consistent with both the falling background and the increasingly boosted/VBF-like event structure that enhances multivariate separation power.

These $\sigma\times\mathrm{Br}$ curves can be directly reused to constrain other new-physics scenarios that populate the same reconstructed final states and have comparable acceptances. Concretely, for a model predicting a narrow resonance $X$ in one of the three topologies of Fig.~\ref{fig:signal}, discovery is expected once $\sigma(pp\to X+\cdots)\times\mathrm{Br}(X\to\text{final state})$ exceeds the corresponding curve, up to an acceptance/efficiency correction if the kinematics differ significantly from those of the benchmark ALP simulation.

\section{Conclusion}
\label{sec:conclusion}

Searches for axion-like particles (ALPs) at colliders are often driven by diphoton signatures, which tightly constrain the effective ALP--photon coupling $g_{a\gamma\gamma}$. This motivates ``photophobic'' scenarios in which the diphoton coupling is suppressed, while electroweak couplings remain sizable. In this work we therefore set $g_{a\gamma\gamma}=0$ and study heavy photophobic ALPs at a future 100~TeV proton--proton collider (SppC/FCC-hh) with an integrated luminosity of $\mathcal{L}=20~\mathrm{ab}^{-1}$.

We consider ALP production through electroweak boson exchange, including both $s$-channel and VBF-like topologies when applicable, and analyze three complementary discovery channels (cf.\ Fig.~\ref{fig:signal}): (I) $pp\to jj\,a(\to Z\gamma)$ with $Z\to\ell^+\ell^-$ ($\ell=e,\mu$); (II) $pp\to W^\pm a(\to W^+W^-)$ with a same-sign dimuon signature plus a hadronic $W$; and (III) $pp\to jj\,a(\to W^+W^-)$ with opposite-sign, different-flavor dilepton $e^\pm\mu^\mp$ in the final state. Signal and background events are generated with MadGraph5\_aMC@NLO and showered with PYTHIA~8, followed by a Delphes detector simulation. Each analysis applies a dedicated preselection and then a TMVA-based BDT using channel-specific kinematic observables; for every ALP-mass hypothesis, the BDT threshold is optimized to maximize the statistical discovery significance.

Our main results are the projected discovery reaches in the $(m_a,g_{aWW})$ plane and the corresponding model-independent thresholds in $\sigma\times\mathrm{Br}$. 
At $\sqrt{s}=100$~TeV, the $jj\,a(\to Z\gamma)$ channel~(I) gives the strongest reach across the full mass range, driven by a fully reconstructible $m(\ell^+ + \ell^- + \gamma)$ resonance and comparatively small irreducible backgrounds. 
The two $WW$ modes are intrinsically less clean---neutrinos induce genuine $\met$ and prevent a narrow mass reconstruction, and SM $WW$ and $t\bar t$ backgrounds are sizable---but they are far from auxiliary. 
Channel~(II), $pp\to W^\pm a(\to W^+W^-)$, is a \emph{pure $s$-channel} probe with a very clean same-sign dilepton signature and largely independent background composition and systematics, providing a robust cross-check in a qualitatively different production environment. 
Channel~(III), $pp\to jj\,a(\to W^+W^-)$, receives both $s$-channel and VBF-like contributions; at high $m_a$ it becomes increasingly competitive thanks to the forward-tagging-jet topology, the growing importance of vector-boson luminosities at 100~TeV, and its sensitivity to the Lorentz/CP structure of the $aWW$ operator through dijet correlations. 
Taken together, 
the three channels can provide an overconstrained test of the photophobic ALP hypothesis in the presence of a signal: 
channel (I) maximizes discovery potential, while channels (II) and (III) directly interrogate $a\to W^+W^-$ in complementary production regimes, enabling stringent internal consistency tests of the electroweak coupling pattern and providing complementary validation and, if an excess emerges, enabling characterization of a potential signal.

Finally, by presenting the discovery thresholds on $\sigma(pp\to jj\,a)\times\mathrm{Br}(a\to Z\gamma)$, $\sigma(pp\to W^\pm a)\times\mathrm{Br}(a\to W^+W^-)$, and $\sigma(pp\to jj\,a)\times\mathrm{Br}(a\to W^+W^-)$, we enable straightforward reinterpretations for other new-physics models that populate the same reconstructed final states. 
Overall, our detector-level study of both signal and backgrounds shows that a 100~TeV collider would substantially extend the discovery reach for photophobic ALPs into the multi-TeV mass regime and improve upon HL-LHC projections. The improvement is driven not by luminosity alone, but by enhanced electroweak-boson luminosities and the associated shift toward more forward and boosted, VBF-like topologies at high masses, which reshapes acceptances and background rejection.
A combined program across the three signal channels (I)--(III) then provides both sensitivity and resilience against channel-specific limitations.

\appendix

%\newpage
\section{Details of the $Z \gamma jj$ analyses}

\subsection{Distributions of representative observables}
\label{app:gammaZjj_obs}	 

\begin{figure}[h]
\centering
\subfigure{
\includegraphics[width=7.3cm, height=4.7cm]{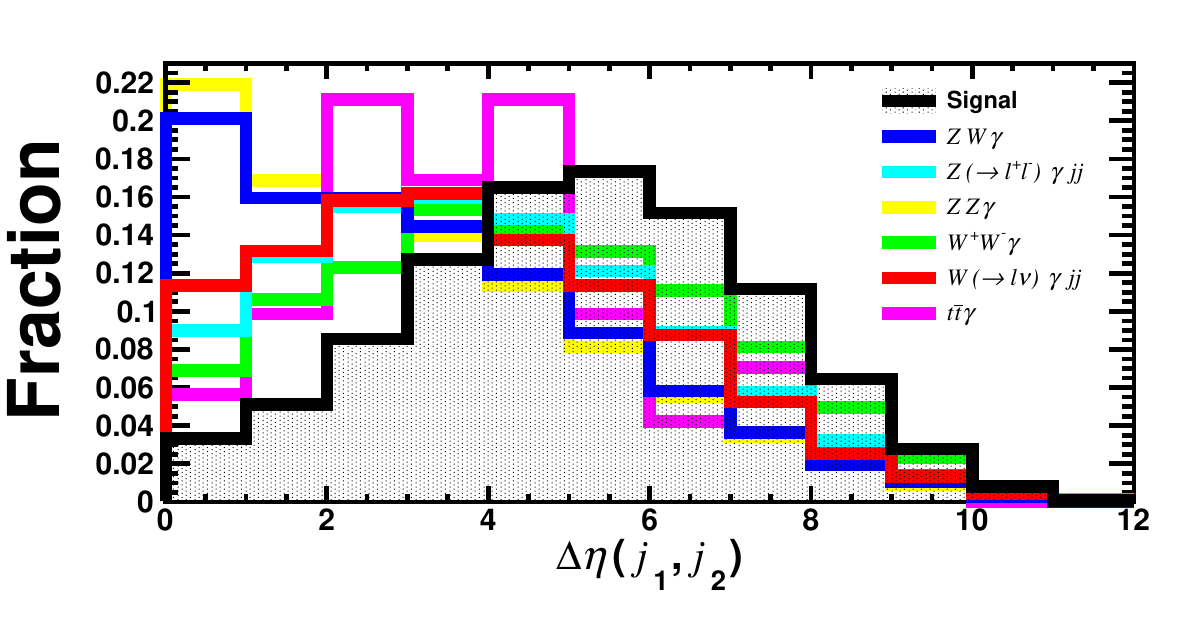}\,\,\,\,\,
\includegraphics[width=7.3cm, height=4.7cm]{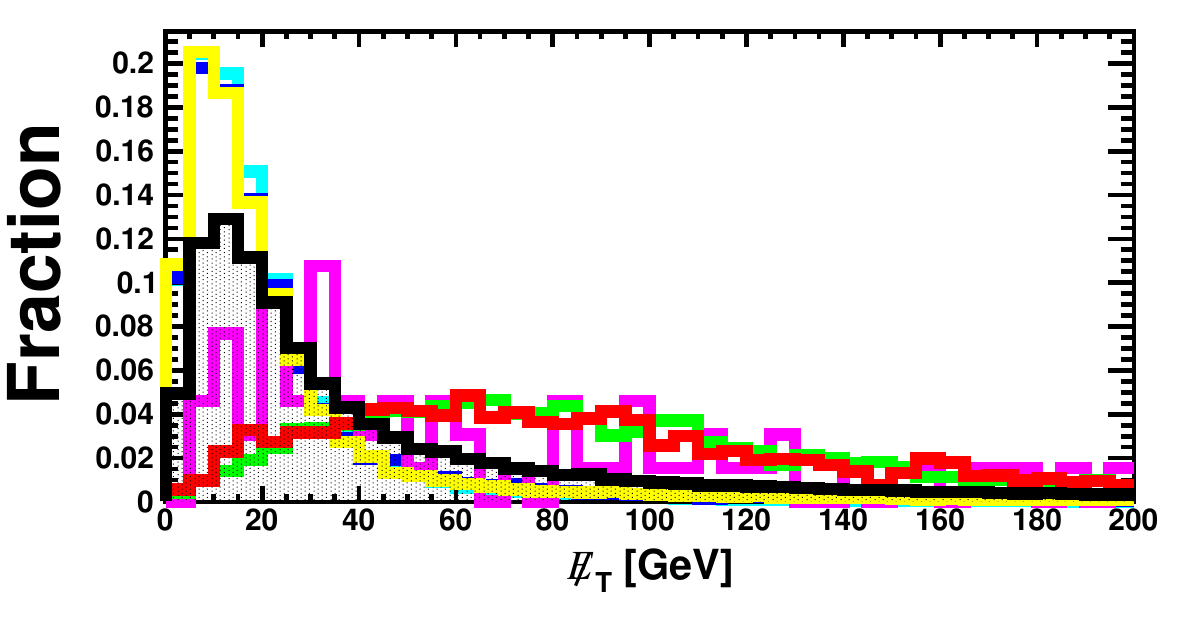}
}
\end{figure}
\addtocounter{figure}{-1}
\vspace{-1.10cm}
\begin{figure}[H]
\centering
\addtocounter{figure}{1}
\subfigure{
\includegraphics[width=7.3cm, height=4.7cm]{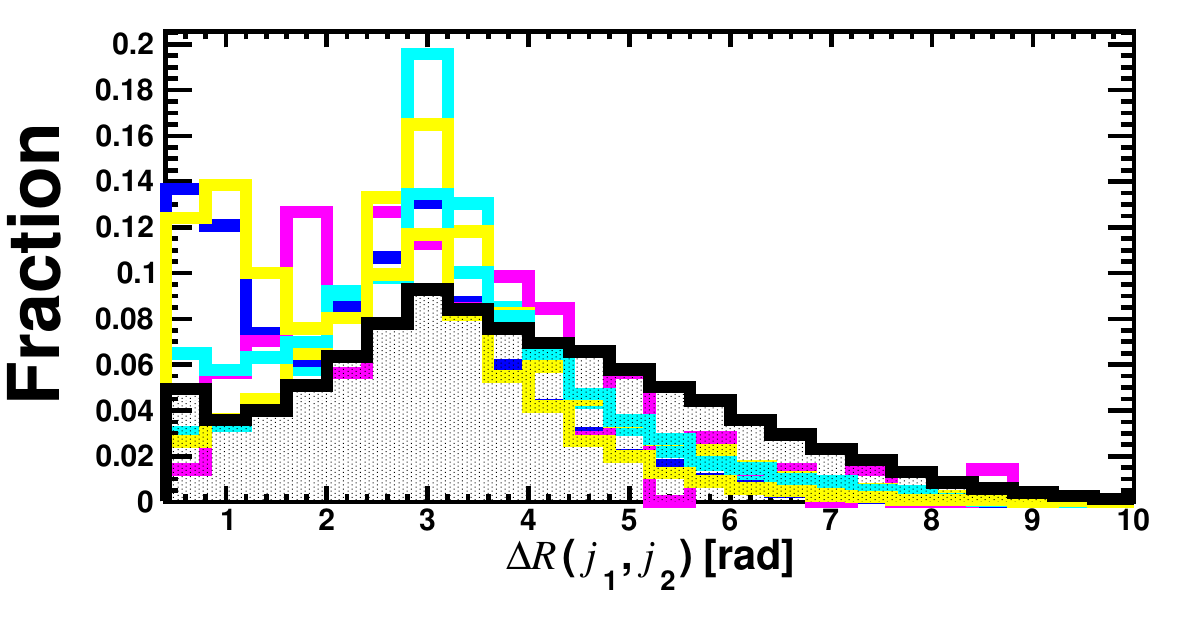}\,\,\,\,\,
\includegraphics[width=7.3cm, height=4.7cm]{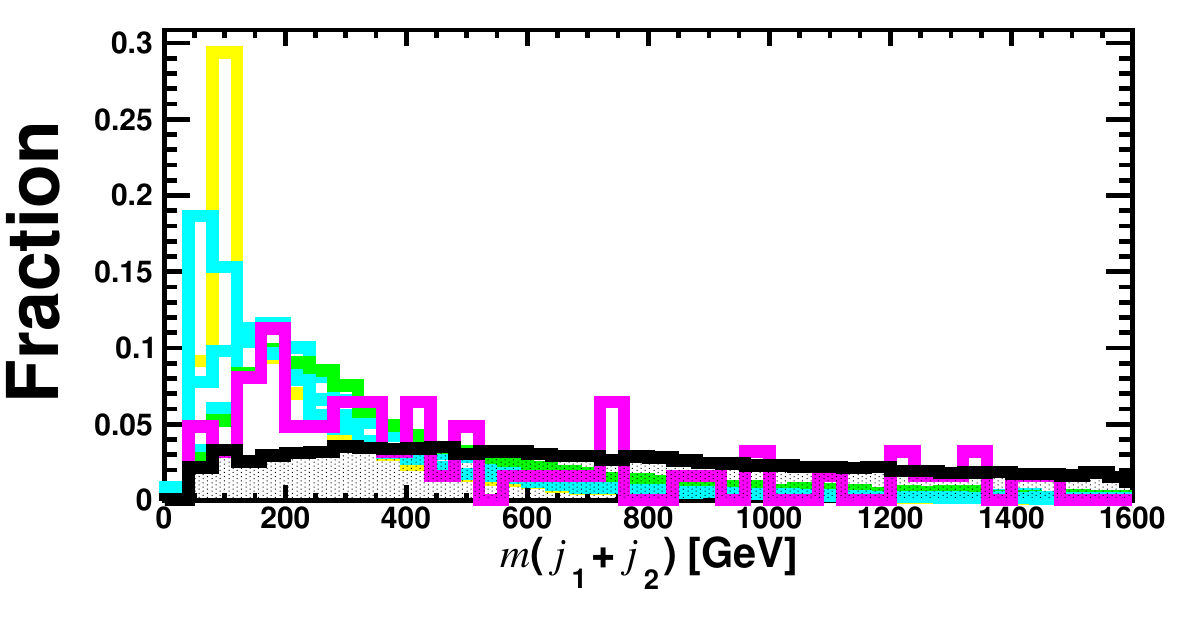}
}
\end{figure}
\vspace{-1.10cm}
\begin{figure}[H] 
\centering
\addtocounter{figure}{-1}
\subfigure{
\includegraphics[width=7.3cm, height=4.7cm]{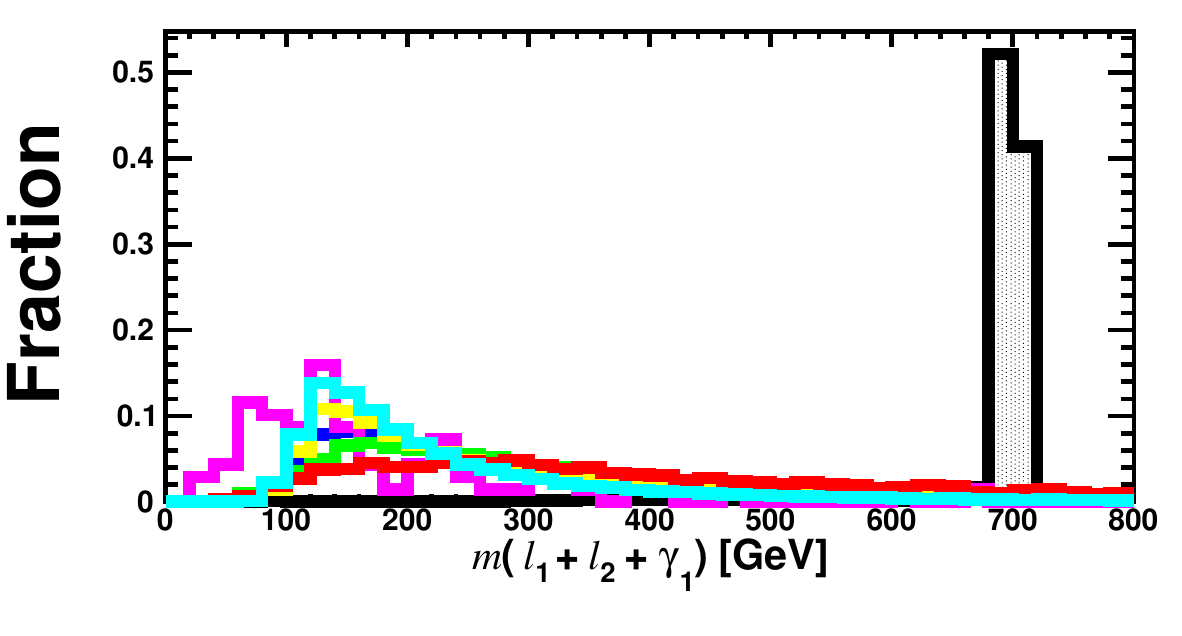}\,\,\,\,\,
\includegraphics[width=7.3cm, height=4.7cm]{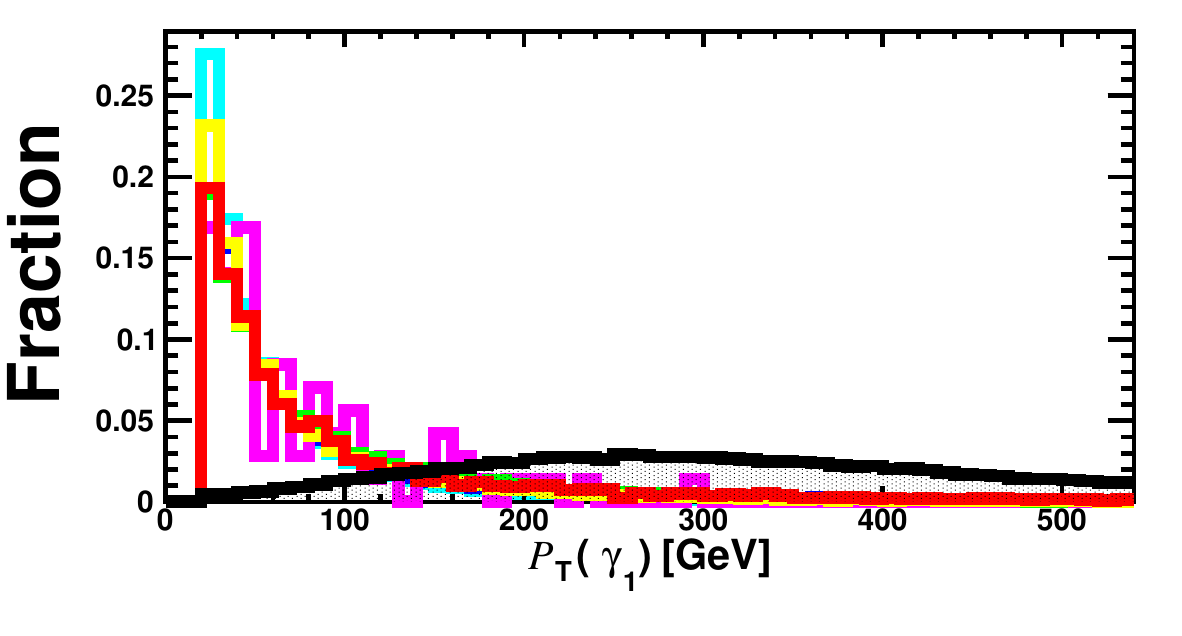}
}
\end{figure}
\vspace{-1.10cm}
\begin{figure}[H] 
\centering
\addtocounter{figure}{1}
\subfigure{
\includegraphics[width=7.3cm, height=4.7cm]{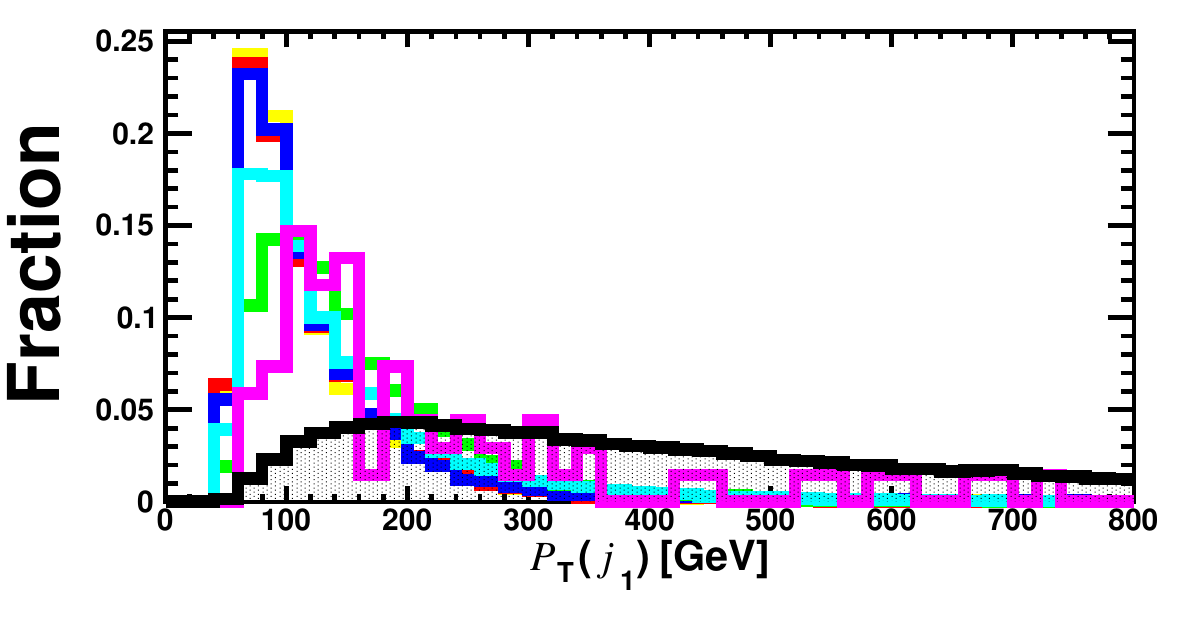}\,\,\,\,\,
\includegraphics[width=7.3cm, height=4.7cm]{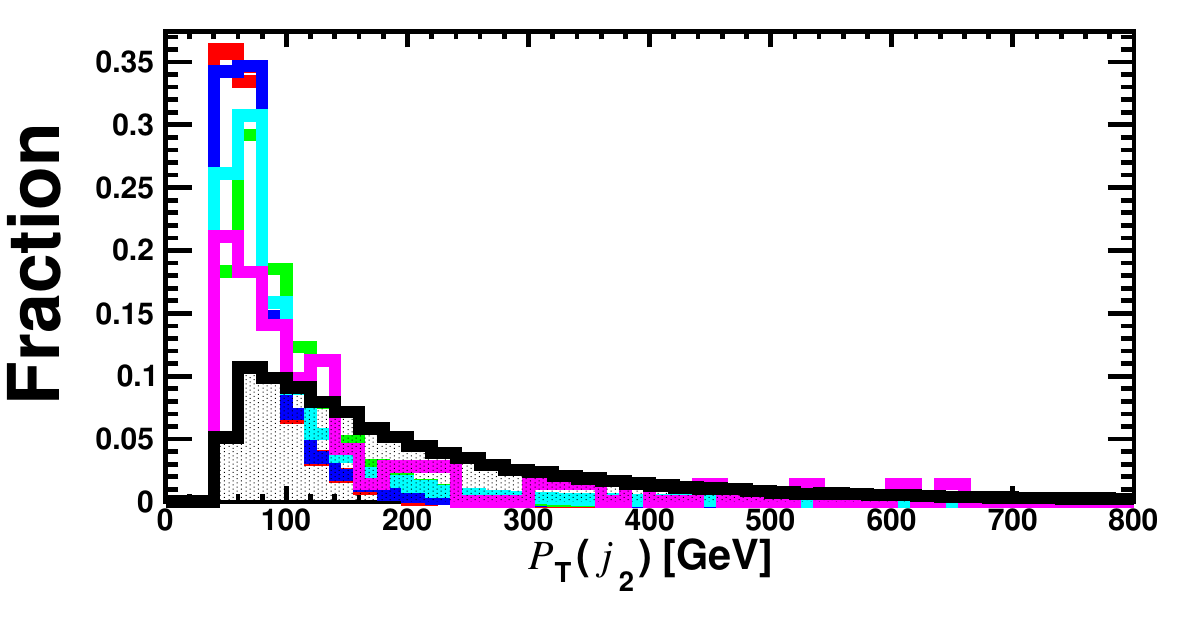}
}
\end{figure}
\vspace{-1.10cm}
\begin{figure}[H] 
\centering
%\addtocounter{figure}{-1}
\subfigure{
\includegraphics[width=7.3cm, height=4.7cm]{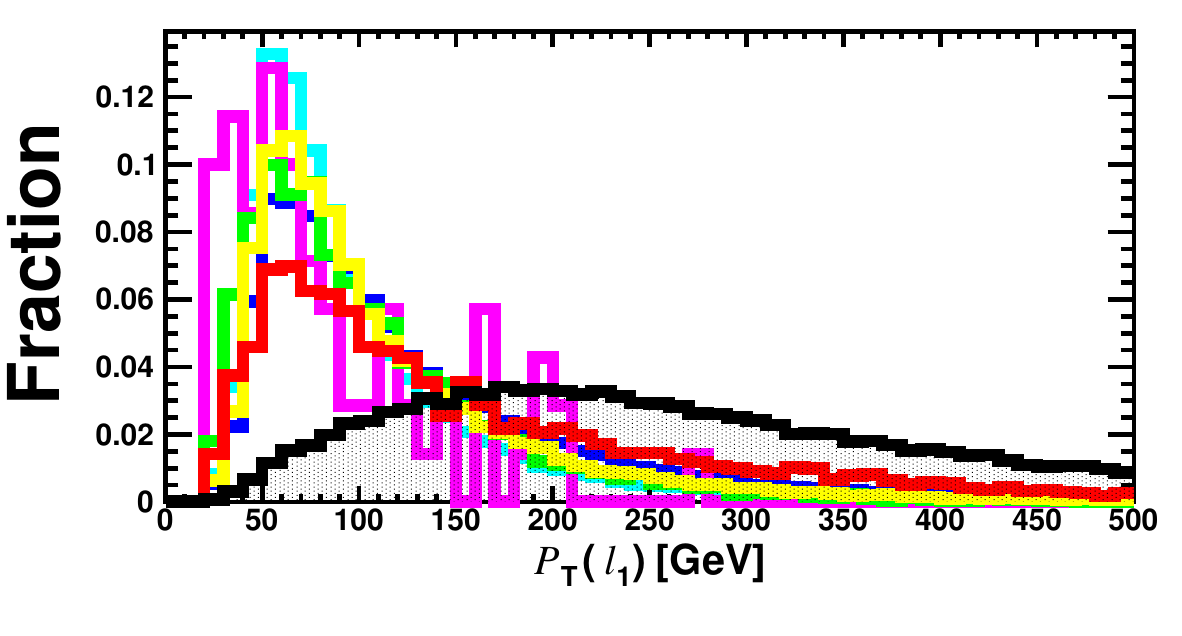}\,\,\,\,\,
\includegraphics[width=7.3cm, height=4.7cm]{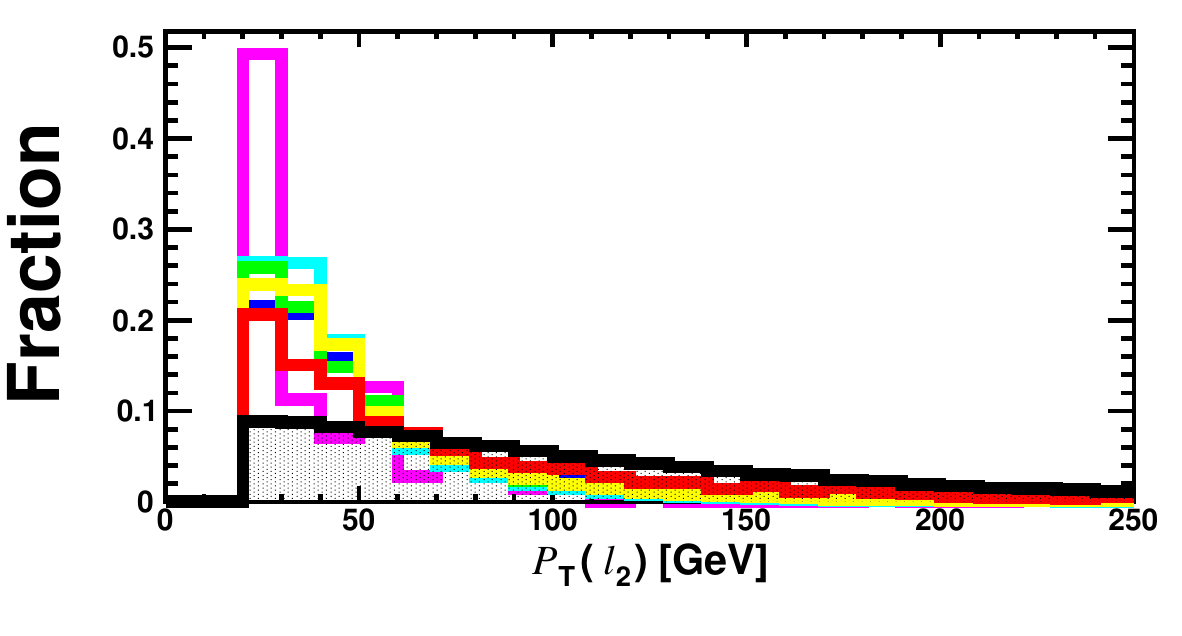}
}
\caption{
Distributions of representative observables for the signal with benchmark $m_a=700$~GeV and for the six background processes at the SppC/FCC-hh with $\sqrt{s}=100$~TeV, after the preselection criteria are applied.
}
\label{fig:obs}
\end{figure}

%\newpage
\subsection{Distributions of BDT responses}
\label{app:gammaZjj_BDT}

\begin{figure}[H]
\centering
\subfigure
{
\includegraphics[width=7.3cm, height=4.7cm]{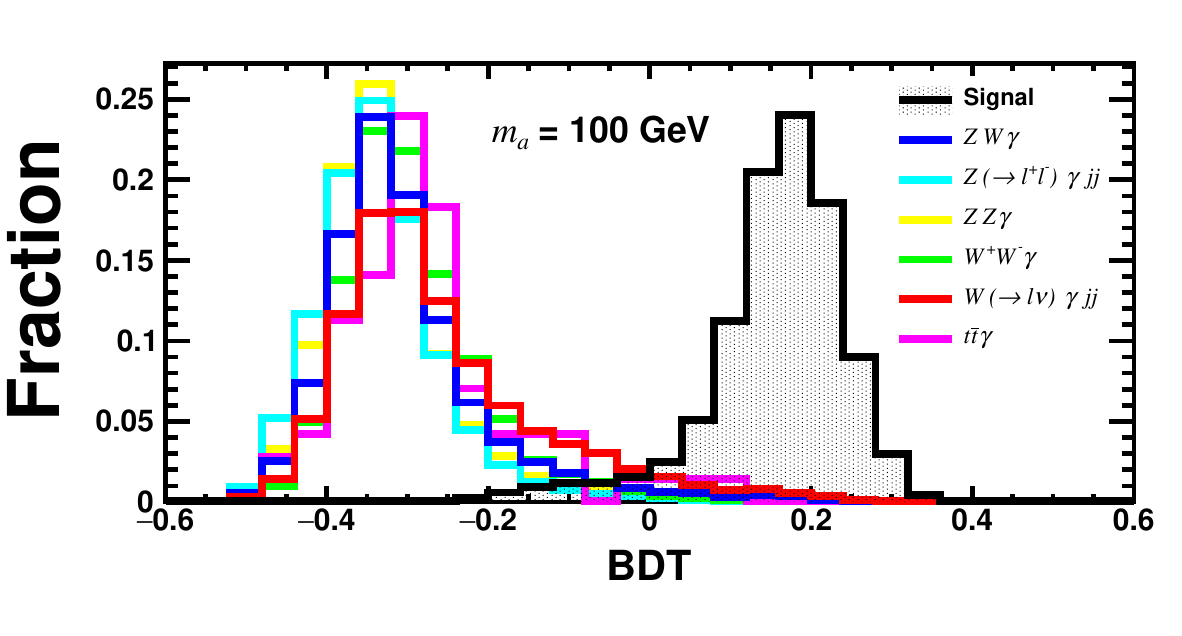}\,\,\,\,\,
\includegraphics[width=7.3cm, height=4.7cm]{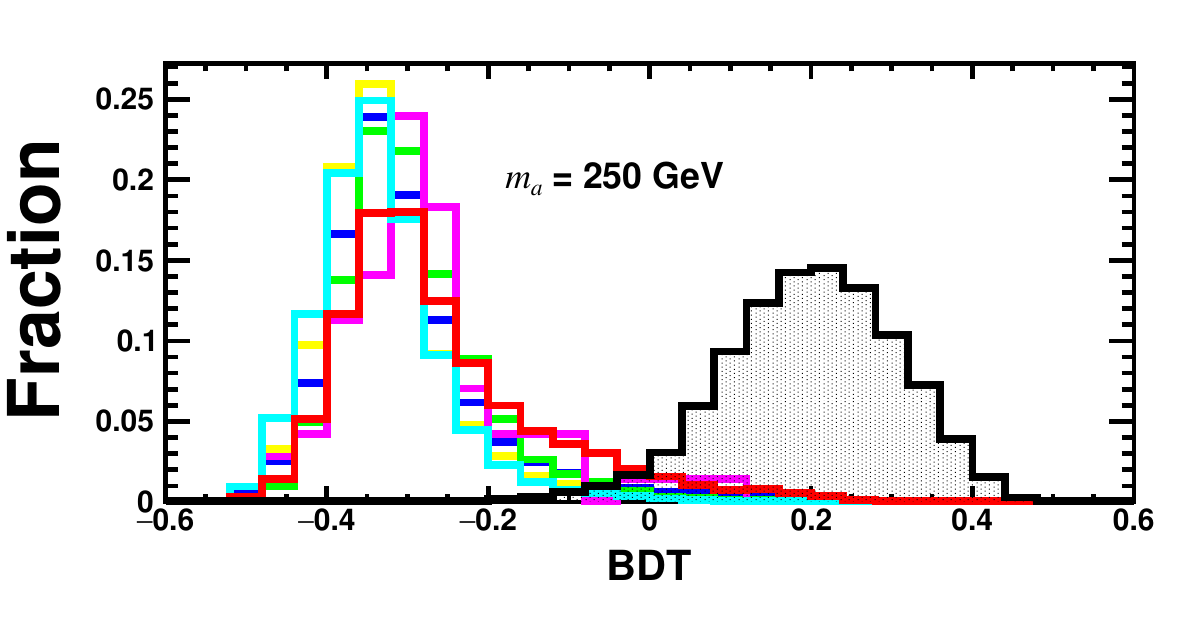}
}
\end{figure}
\vspace{-1.10cm}
\begin{figure}[H]
\centering
\subfigure
{
\includegraphics[width=7.3cm, height=4.7cm]{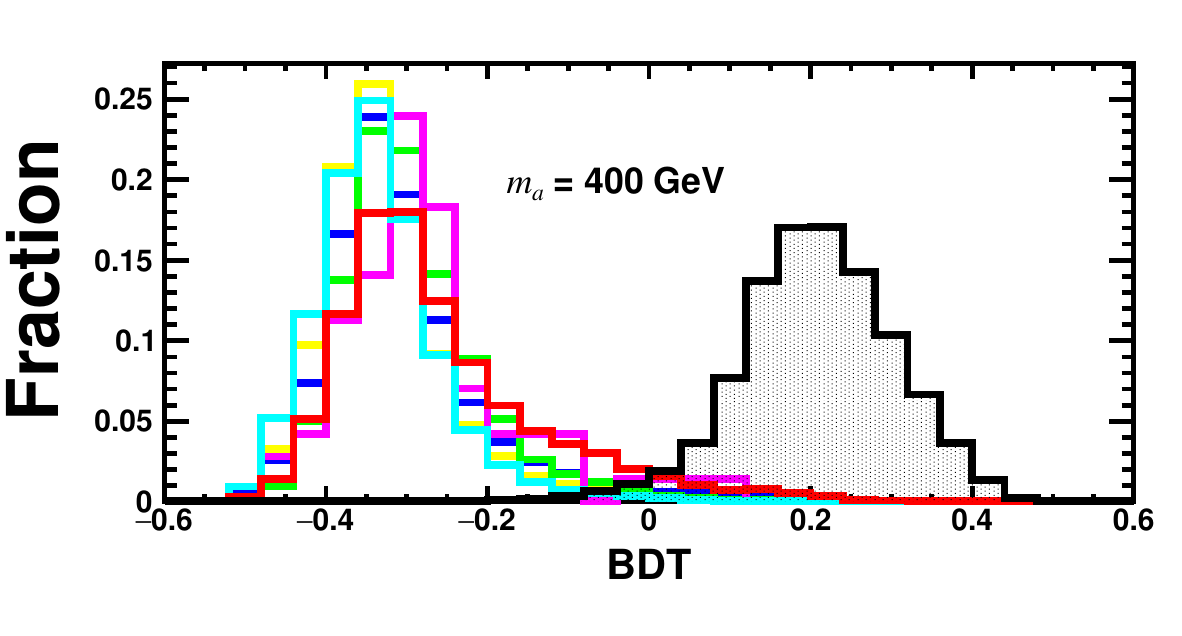}\,\,\,\,\,
\includegraphics[width=7.3cm, height=4.7cm]{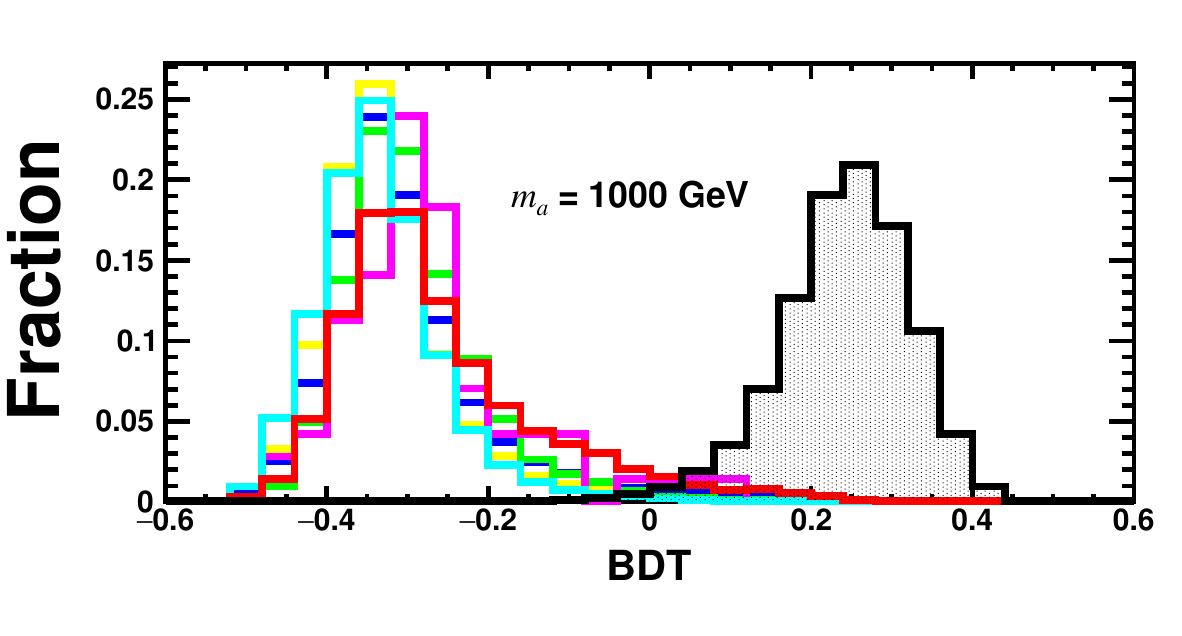}
}
\end{figure}
\vspace{-1.10cm}
\begin{figure}[H]
\centering
\subfigure
{
\includegraphics[width=7.3cm, height=4.7cm]{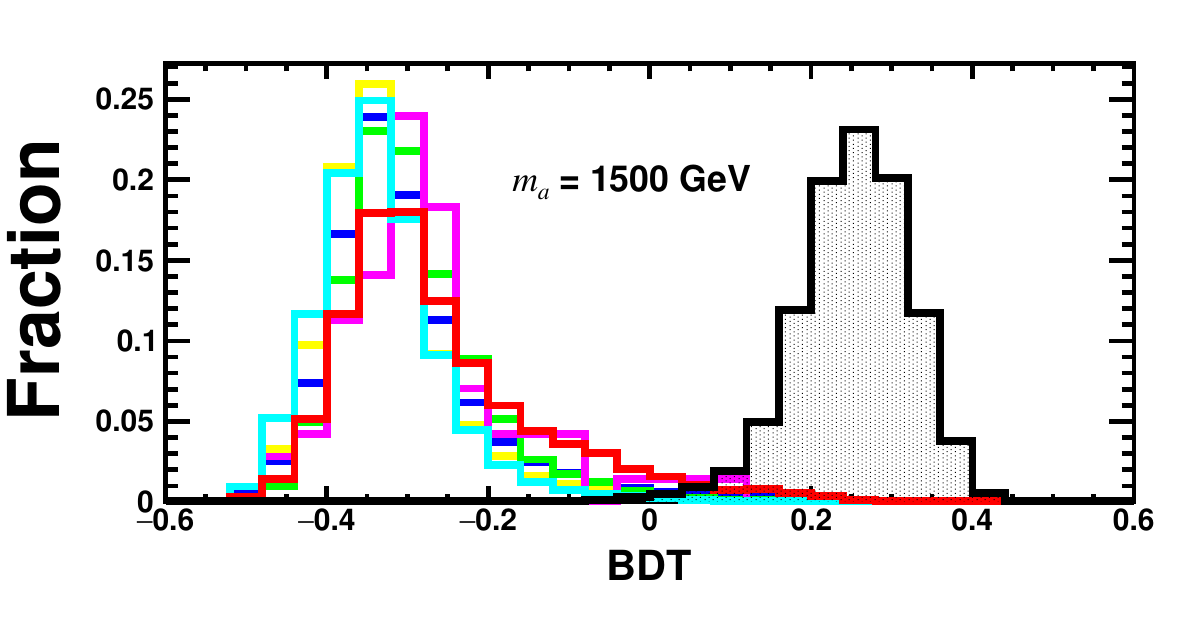}\,\,\,\,\,
\includegraphics[width=7.3cm, height=4.7cm]{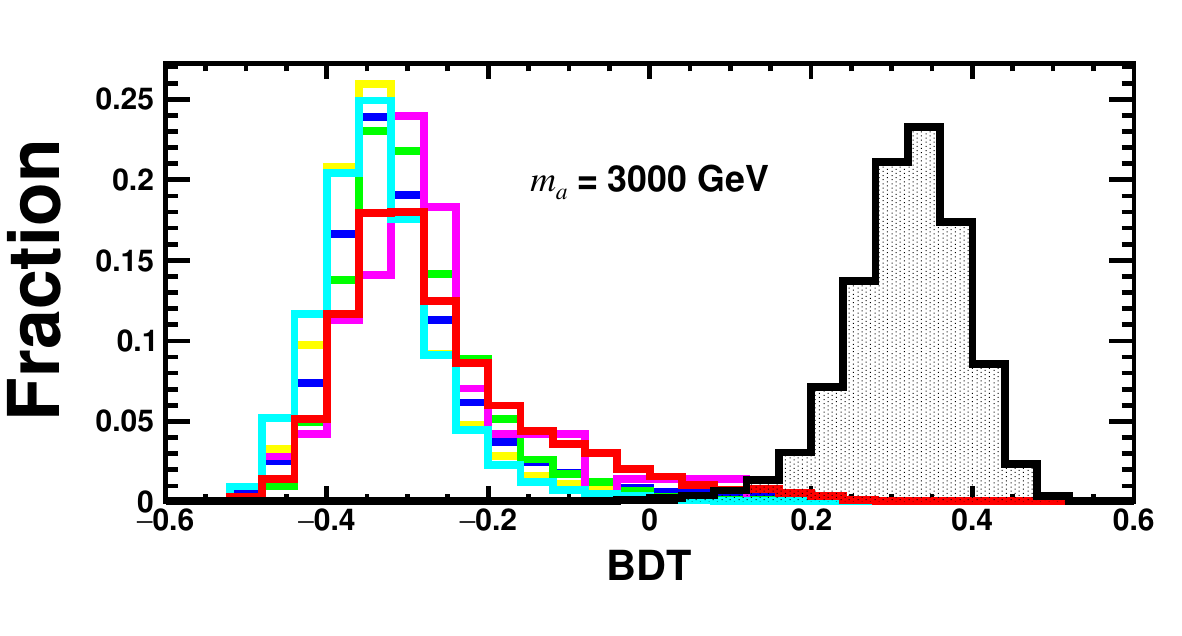}
}
\end{figure}
\vspace{-1.10cm}
\begin{figure}[H] 
\centering
%\addtocounter{figure}{-1}
\subfigure{
\includegraphics[width=7.3cm, height=4.7cm]{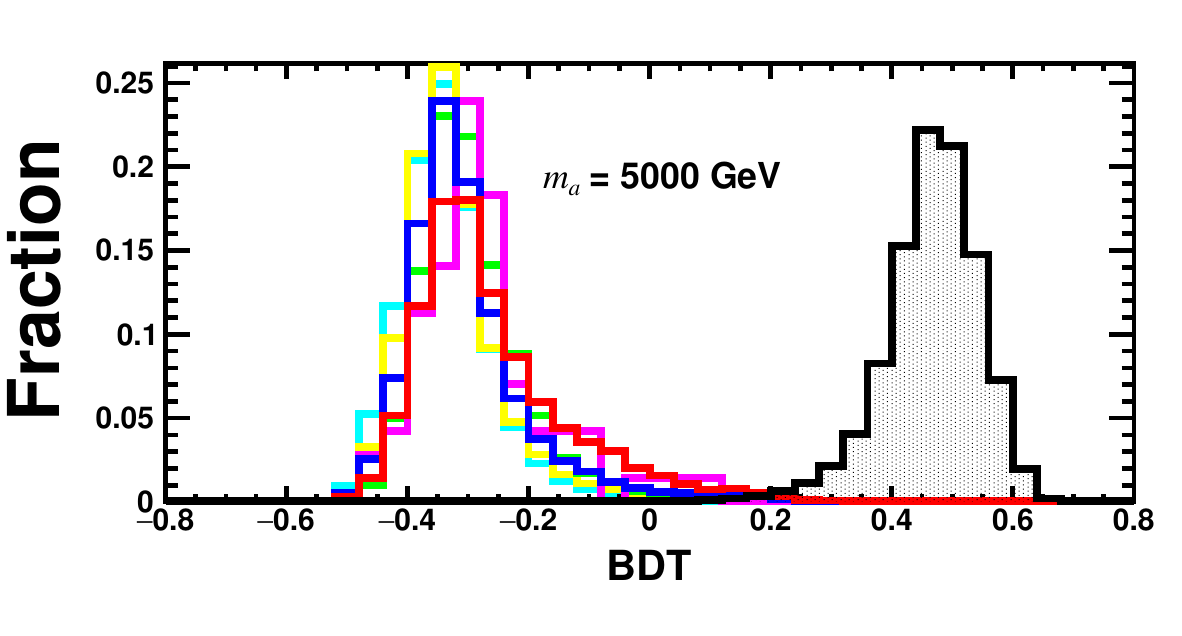}\,\,\,\,\,
\includegraphics[width=7.3cm, height=4.7cm]{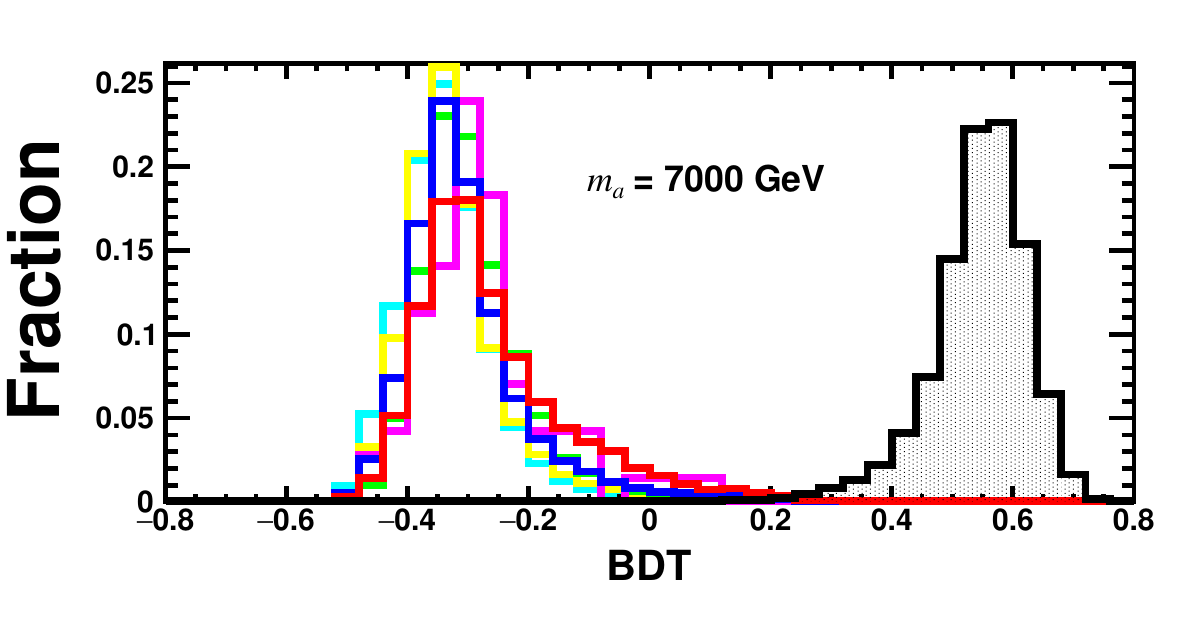}
}
\caption{
Distributions of BDT responses for the signal (black, shaded) and the six background processes at the SppC/FCC-hh with $\sqrt{s}=100$~TeV, assuming different $m_a$ values.
}
\label{fig:BDTall}
\end{figure}

\subsection{BDT selection efficiencies}
\label{app:gammaZjj_efficiency}

\begin{table*}[h]
\centering 
\scalebox{0.75}{
%\begin{ruledtabular}
\begin{tabular}{ccccccccc}
\hline
\hline
$m_a$ [GeV] & BDT & signal & $ZW\gamma$ & $Z(\to \ell^+\ell^-)\,\gamma jj$ & $ZZ\gamma$ & $W^+W^-\gamma$ & $W(\to \ell\nu)\,\gamma jj$ & $t\bar{t}\gamma$ \\
\hline
100  & 0.108 & $7.99\mltp10^{-1}$ & $3.89\mltp10^{-3}$ & $3.14\mltp10^{-3}$ & $4.64\mltp10^{-3}$ & $5.54\mltp10^{-4}$ & $-$               & $2.94\mltp10^{-4}$ \\
165  & 0.121 & $7.47\mltp10^{-1}$ & $3.44\mltp10^{-3}$ & $2.96\mltp10^{-3}$ & $5.74\mltp10^{-3}$ & $2.22\mltp10^{-3}$ & $-$               & $3.82\mltp10^{-3}$ \\
400  & 0.149 & $7.49\mltp10^{-1}$ & $1.95\mltp10^{-3}$ & $1.18\mltp10^{-3}$ & $2.36\mltp10^{-3}$ & $4.46\mltp10^{-3}$ & $-$               & $6.32\mltp10^{-3}$ \\  
700  & 0.140 & $8.88\mltp10^{-1}$ & $2.40\mltp10^{-3}$ & $1.10\mltp10^{-3}$ & $1.86\mltp10^{-3}$ & $8.04\mltp10^{-3}$ & $-$               & $4.71\mltp10^{-3}$ \\
1000 & 0.147 & $8.84\mltp10^{-1}$ & $3.05\mltp10^{-3}$ & $7.73\mltp10^{-4}$ & $1.56\mltp10^{-3}$ & $8.04\mltp10^{-3}$ & $-$               & $2.79\mltp10^{-3}$ \\
1500 & 0.200 & $8.35\mltp10^{-1}$ & $1.62\mltp10^{-3}$ & $3.24\mltp10^{-4}$ & $9.51\mltp10^{-4}$ & $7.48\mltp10^{-3}$ & $-$               & $1.03\mltp10^{-3}$ \\
2000 & 0.190 & $9.10\mltp10^{-1}$ & $2.72\mltp10^{-3}$ & $2.51\mltp10^{-4}$ & $1.18\mltp10^{-3}$ & $6.93\mltp10^{-3}$ & $-$               & $7.36\mltp10^{-4}$ \\
4000 & 0.200 & $9.88\mltp10^{-1}$ & $1.75\mltp10^{-3}$ & $1.25\mltp10^{-4}$ & $5.33\mltp10^{-4}$ & $3.88\mltp10^{-3}$ & $-$               & $-$               \\
5000 & 0.225 & $9.88\mltp10^{-1}$ & $1.04\mltp10^{-3}$ & $6.48\mltp10^{-5}$ & $3.42\mltp10^{-4}$ & $2.22\mltp10^{-3}$ & $-$               & $-$               \\
7000 & 0.199 & $9.88\mltp10^{-1}$ & $6.48\mltp10^{-4}$ & $5.26\mltp10^{-5}$ & $2.66\mltp10^{-4}$ & $1.38\mltp10^{-3}$ & $-$               & $-$               \\
\hline
\hline
\end{tabular}
%\end{ruledtabular}
}
\caption{
Selection efficiencies of the BDT cut for both signal and background processes at the SppC/FCC-hh with $\sqrt{s}=100$~TeV for different $m_a$ values. The second column gives the lower threshold of the BDT response; “$-$” indicates that the yield becomes negligible with $\mathcal{L}=20~\mathrm{ab}^{-1}$.
}
\label{tab:BDT}
\end{table*}

%\newpage
\section{Details of the $W^\pm W^\pm W^\mp$ analyses}

\subsection{Distributions of representative observables}
\label{app:WWW_observables}

\begin{figure}[H]
\centering
\subfigure{
\includegraphics[width=7.3cm, height=4.7cm]{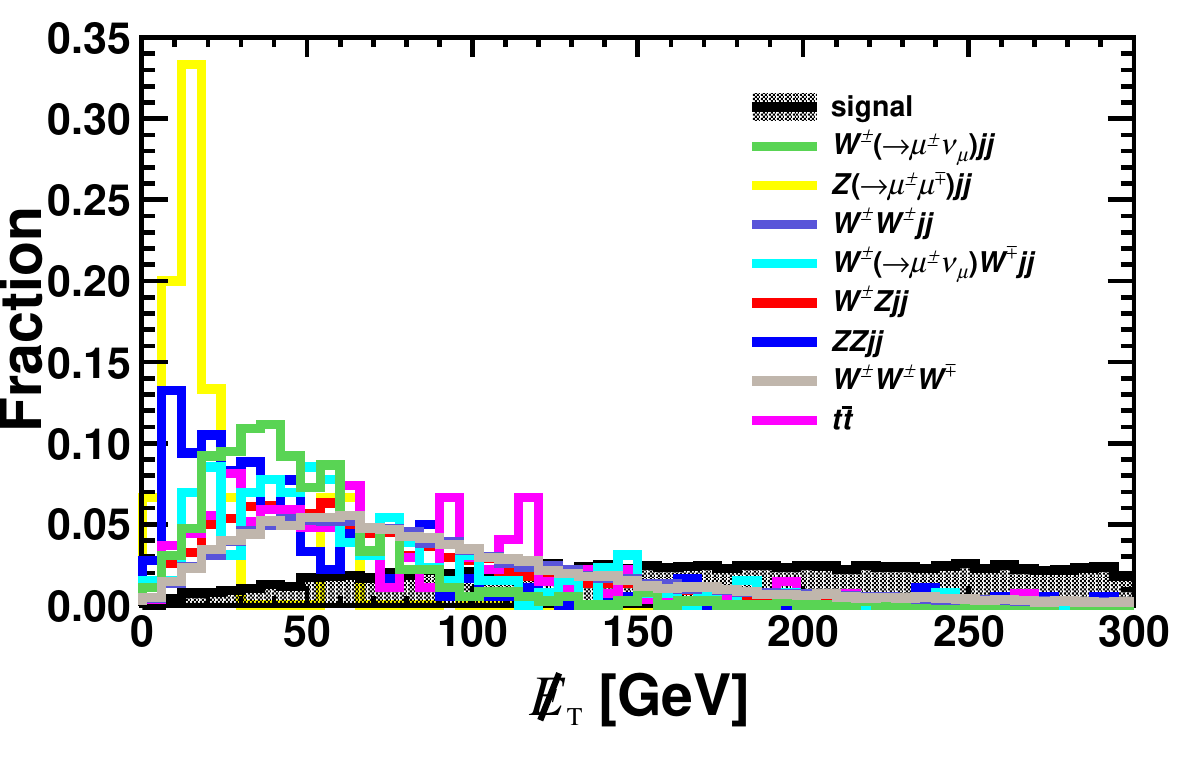}\,\,\,\,\,
\includegraphics[width=7.3cm, height=4.7cm]{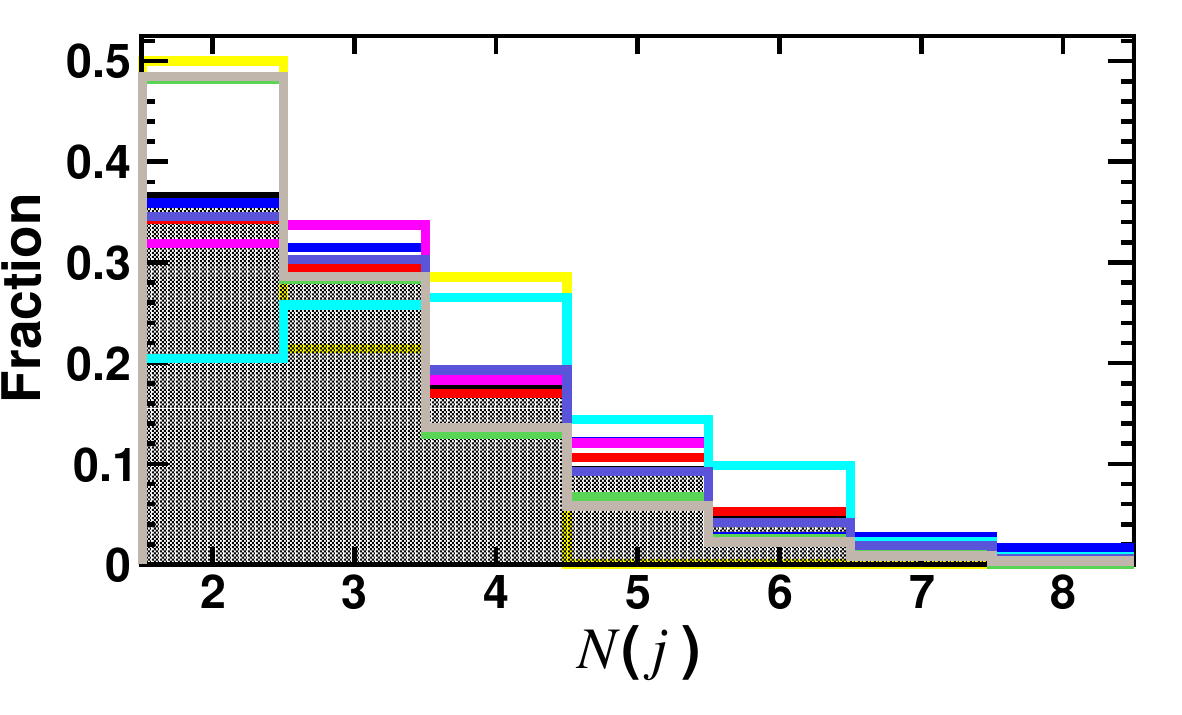}
}
\end{figure}
\addtocounter{figure}{-1}
\vspace{-1.10cm}
\begin{figure}[H]
\centering
\addtocounter{figure}{1}
\subfigure{
\includegraphics[width=7.3cm, height=4.7cm]{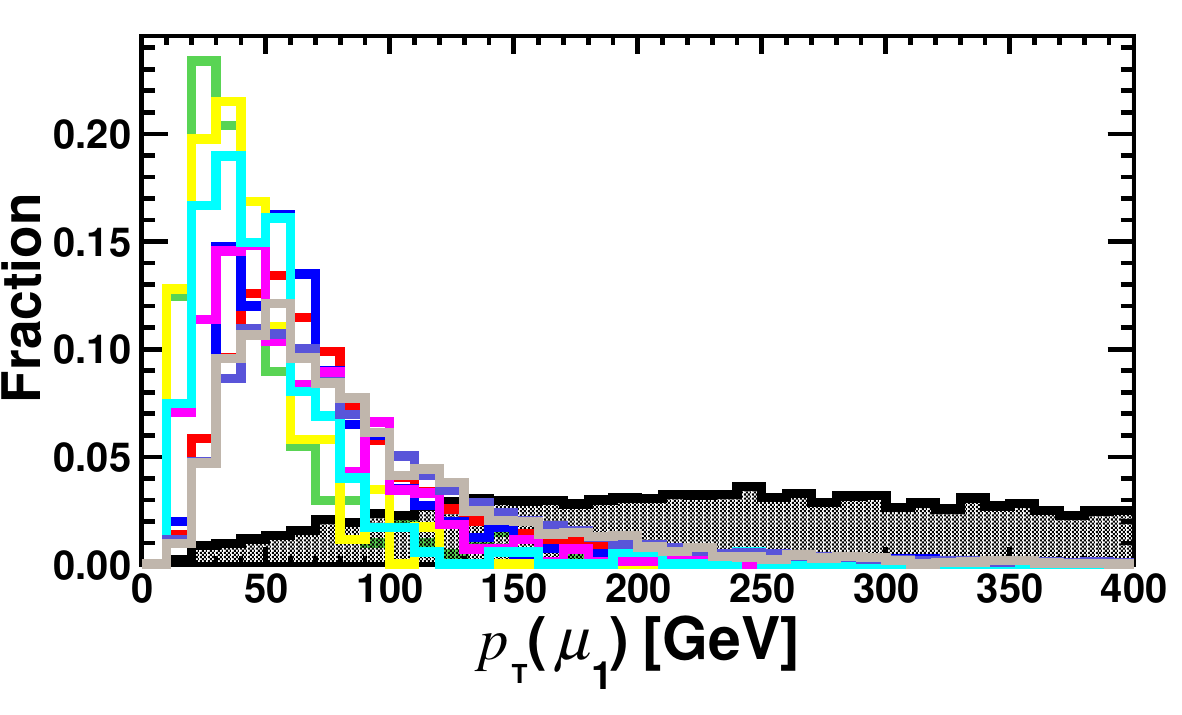}\,\,\,\,\,
\includegraphics[width=7.3cm, height=4.7cm]{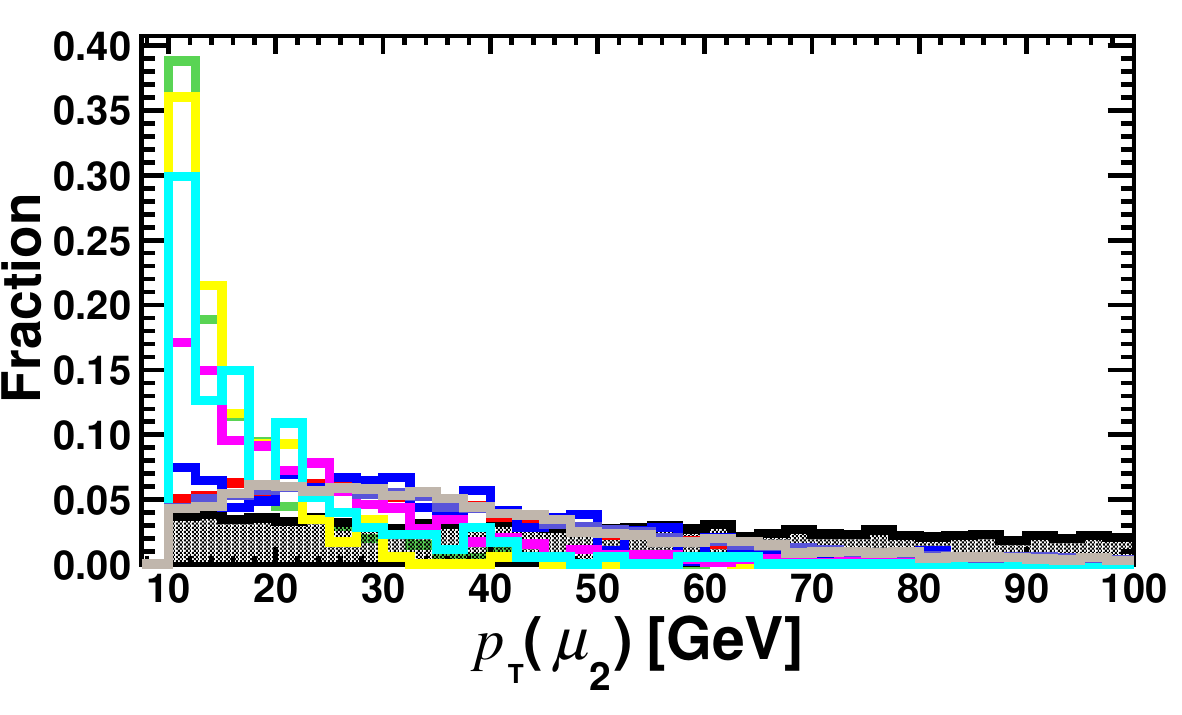}
}
\end{figure}
\vspace{-1.10cm}
\begin{figure}[H] 
\centering
\addtocounter{figure}{-1}
\subfigure{
\includegraphics[width=7.3cm, height=4.7cm]{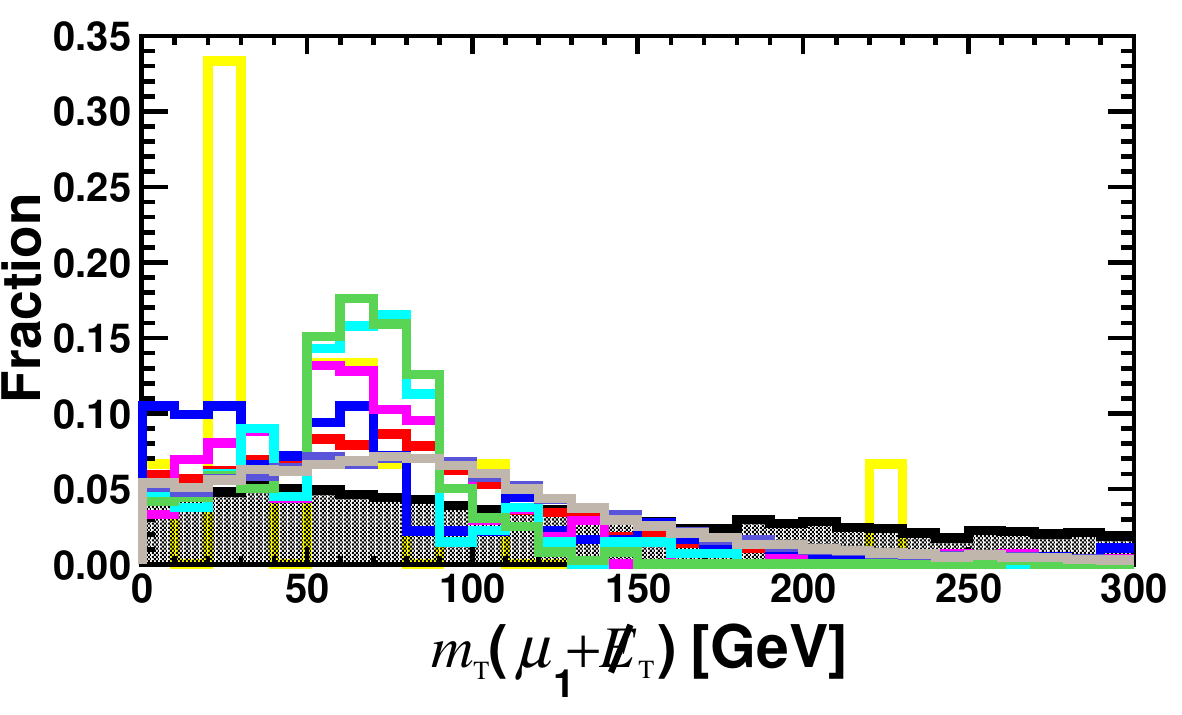}\,\,\,\,\,
\includegraphics[width=7.3cm, height=4.7cm]{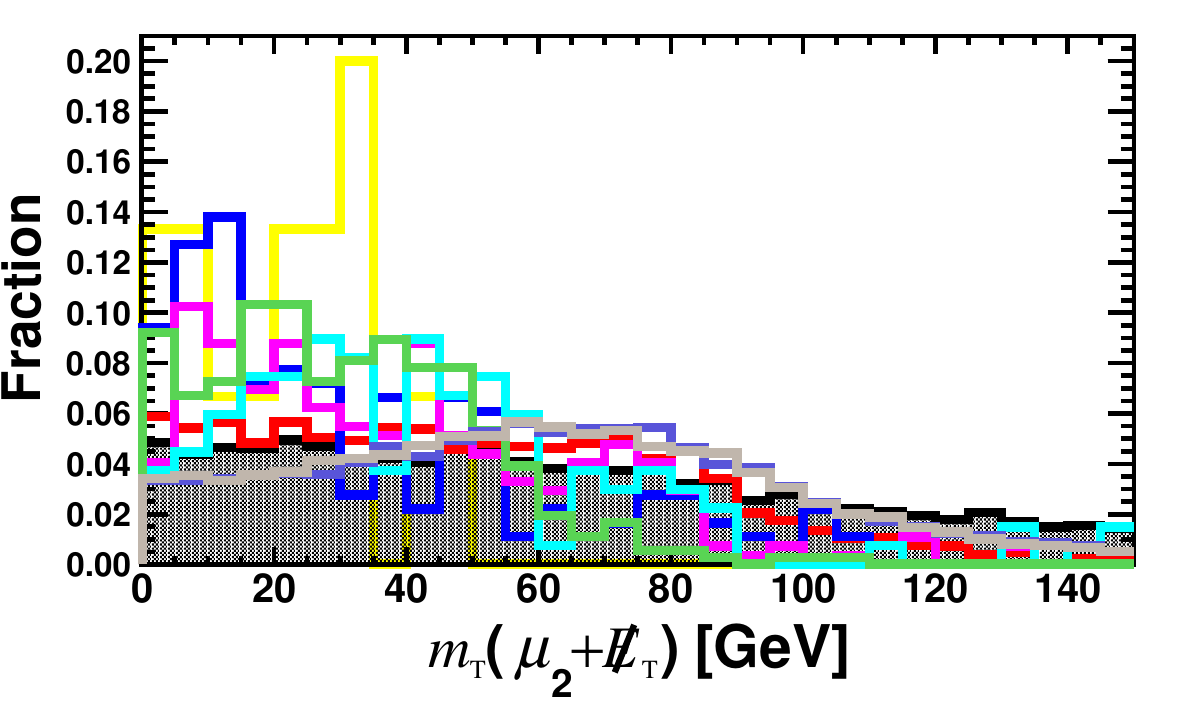}
}
\end{figure}
\vspace{-1.10cm}
\begin{figure}[H] 
\centering
\addtocounter{figure}{1}
\subfigure{
\includegraphics[width=7.3cm, height=4.7cm]{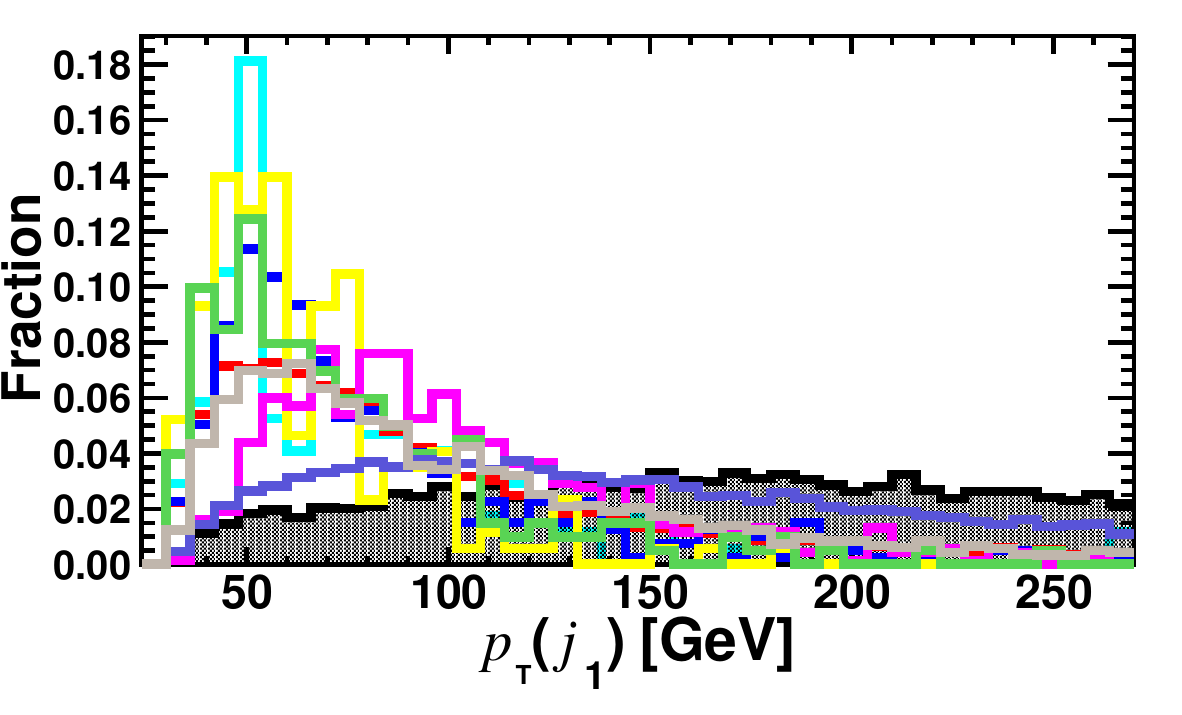}\,\,\,\,\,
\includegraphics[width=7.3cm, height=4.7cm]{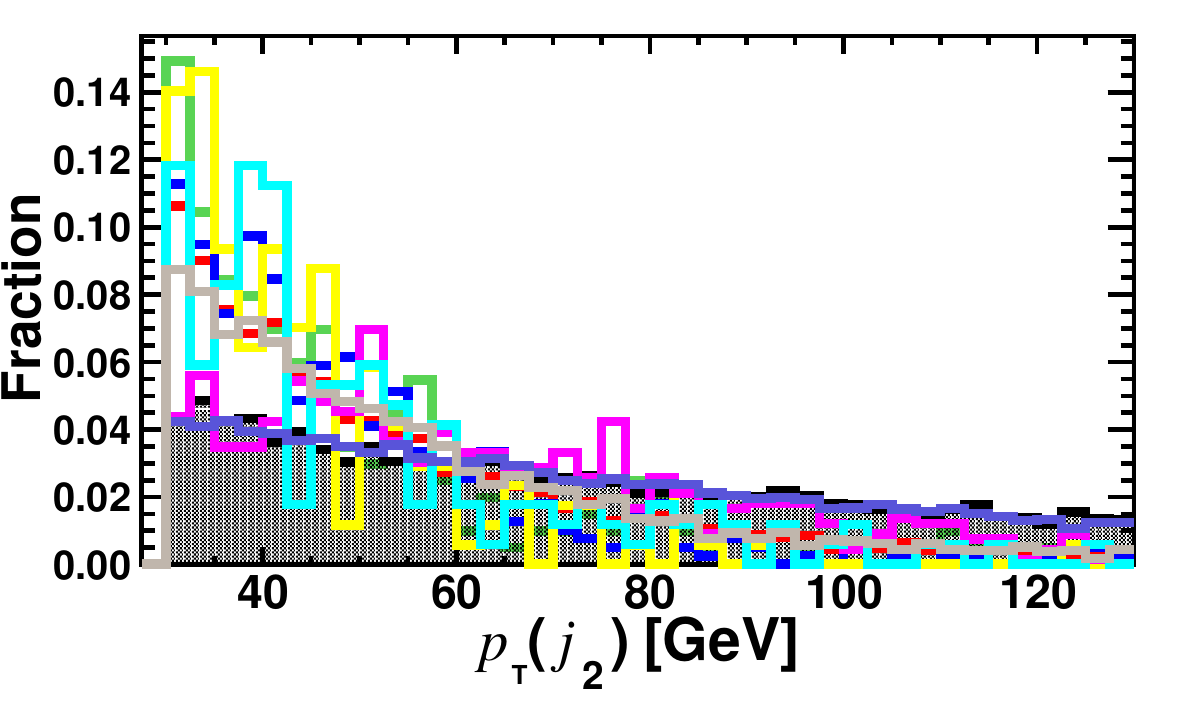}
}
\end{figure}
\vspace{-1.10cm}
\begin{figure}[H] 
\centering
%\addtocounter{figure}{-1}
\subfigure{
\includegraphics[width=7.3cm, height=4.7cm]{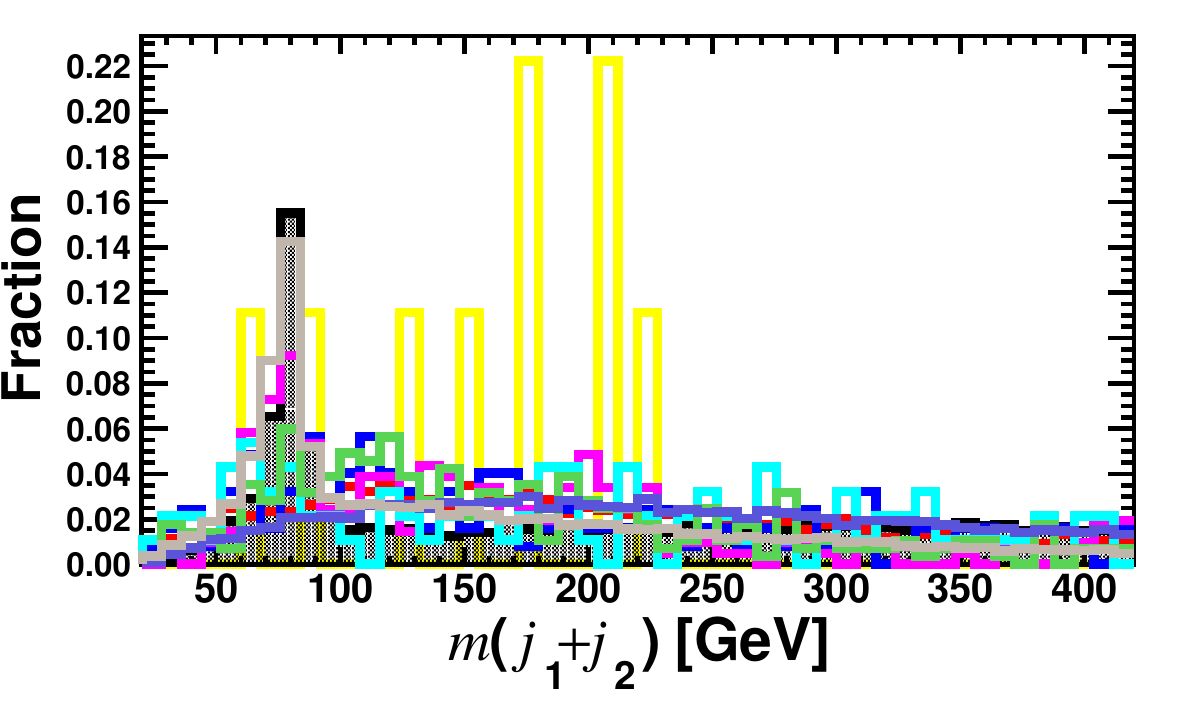}\,\,\,\,\,
\includegraphics[width=7.3cm, height=4.7cm]{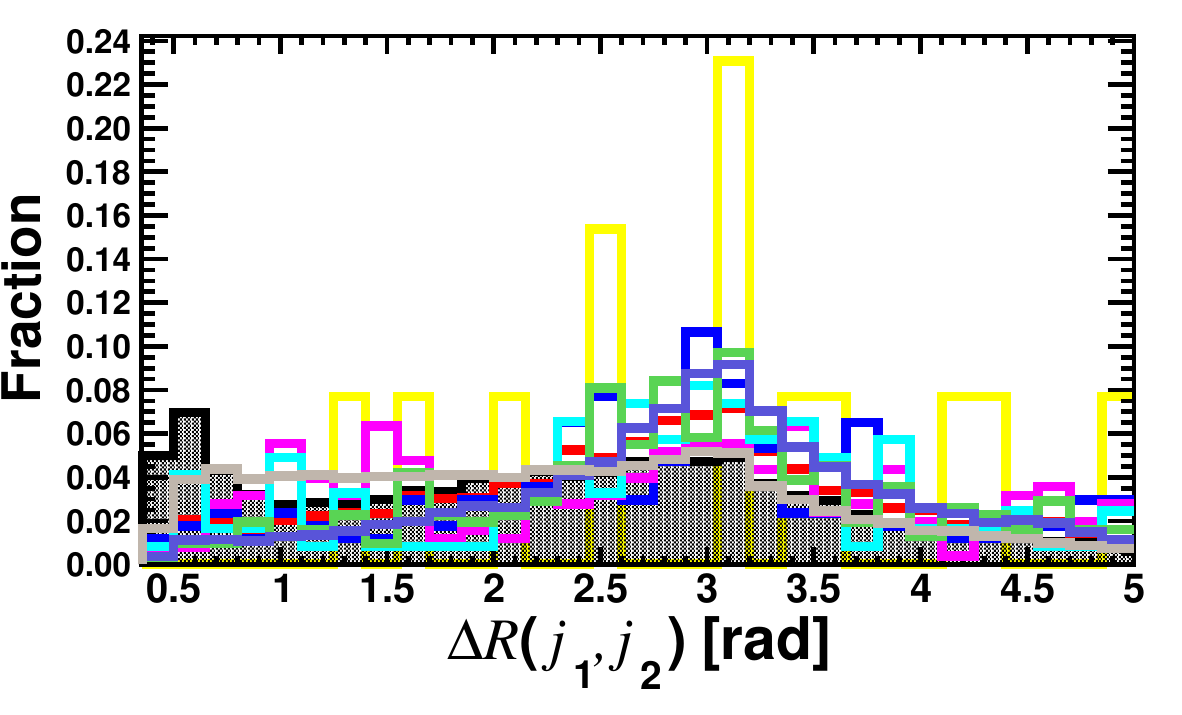}
}
\caption{
Distributions of kinematic observables for $\met$, $N(j)$, $\mu_{1}$, $\mu_{2}$, $j_{1}$, and $j_{2}$ after applying the preselection criteria for the signal (black, shaded) and background processes at the SppC/FCC-hh with $\sqrt{s}=100$~TeV, assuming the benchmark $m_{a}=900$~GeV.
}
\label{Representative_Muon}
\end{figure}

\begin{figure}[H]
\centering
\subfigure{
\includegraphics[width=7.3cm, height=4.7cm]{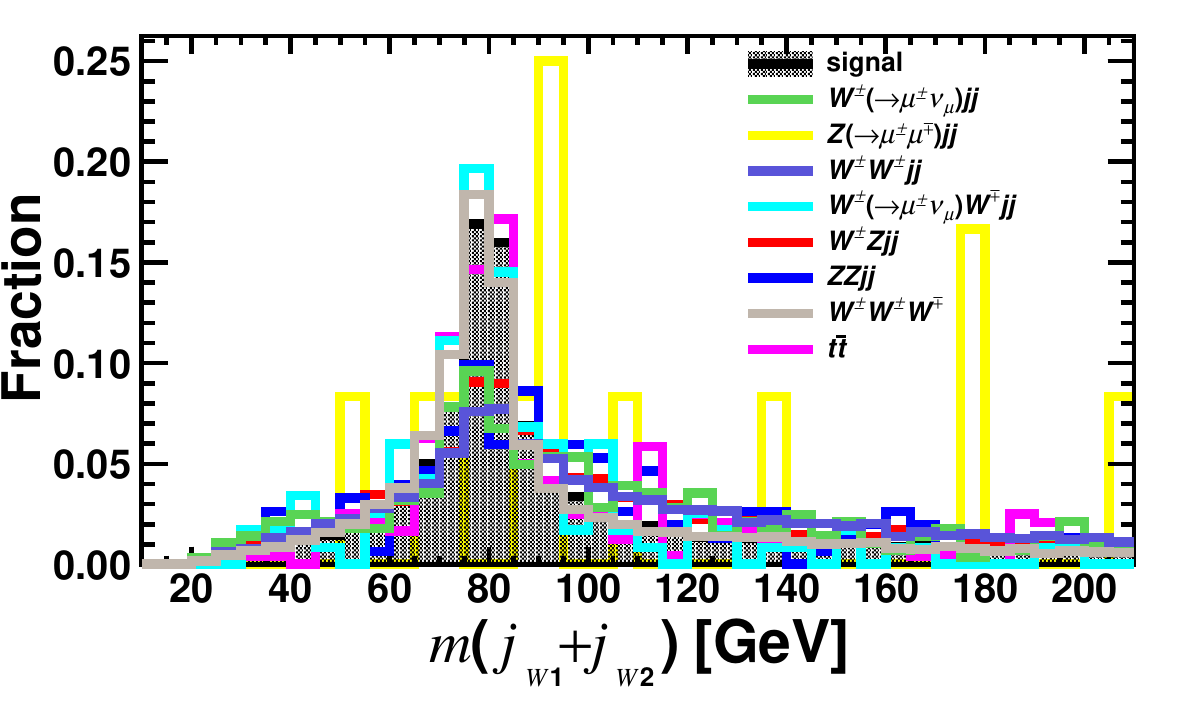}\,\,\,\,\,
\includegraphics[width=7.3cm, height=4.7cm]{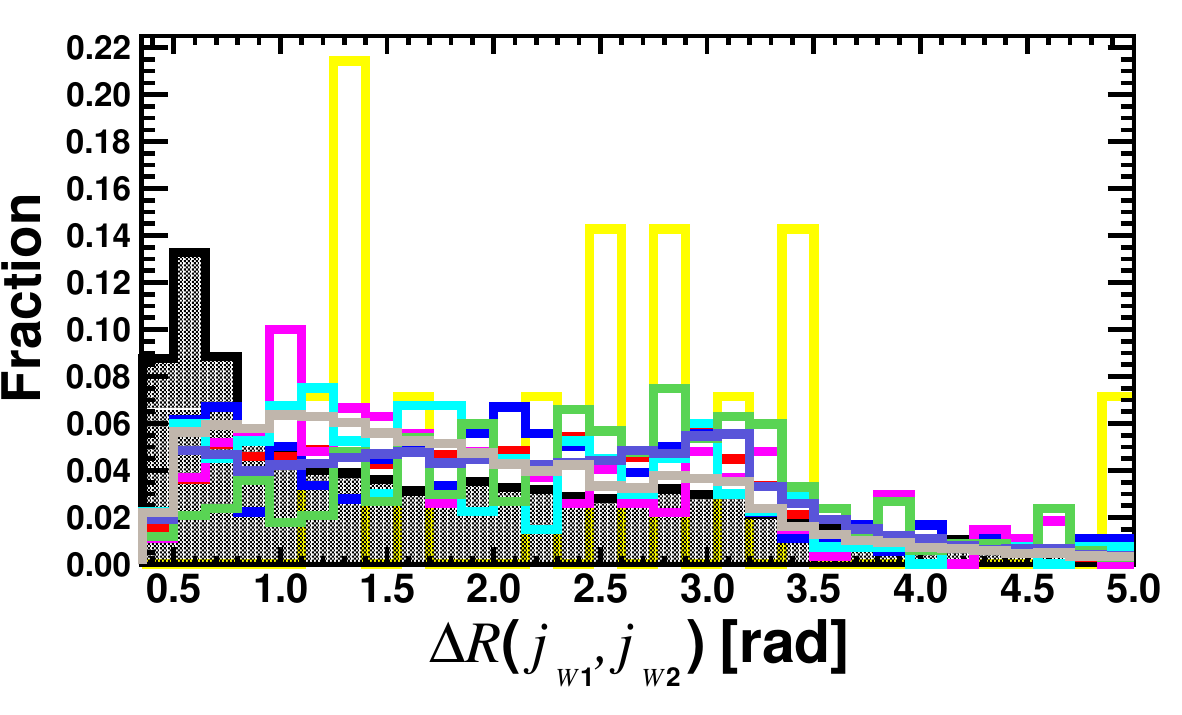}
}
\end{figure}
\addtocounter{figure}{-1}
\vspace{-1.10cm}
\begin{figure}[H]
\centering
\addtocounter{figure}{1}
\subfigure{
\includegraphics[width=7.3cm, height=4.7cm]{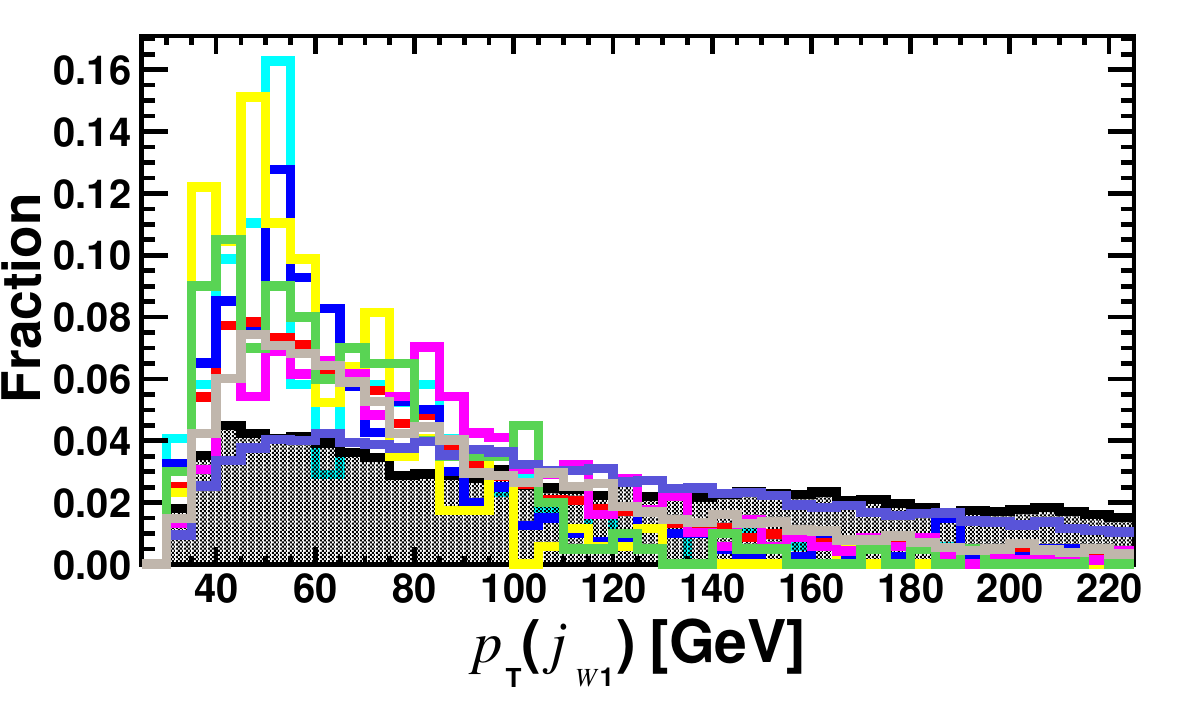}\,\,\,\,\,
\includegraphics[width=7.3cm, height=4.7cm]{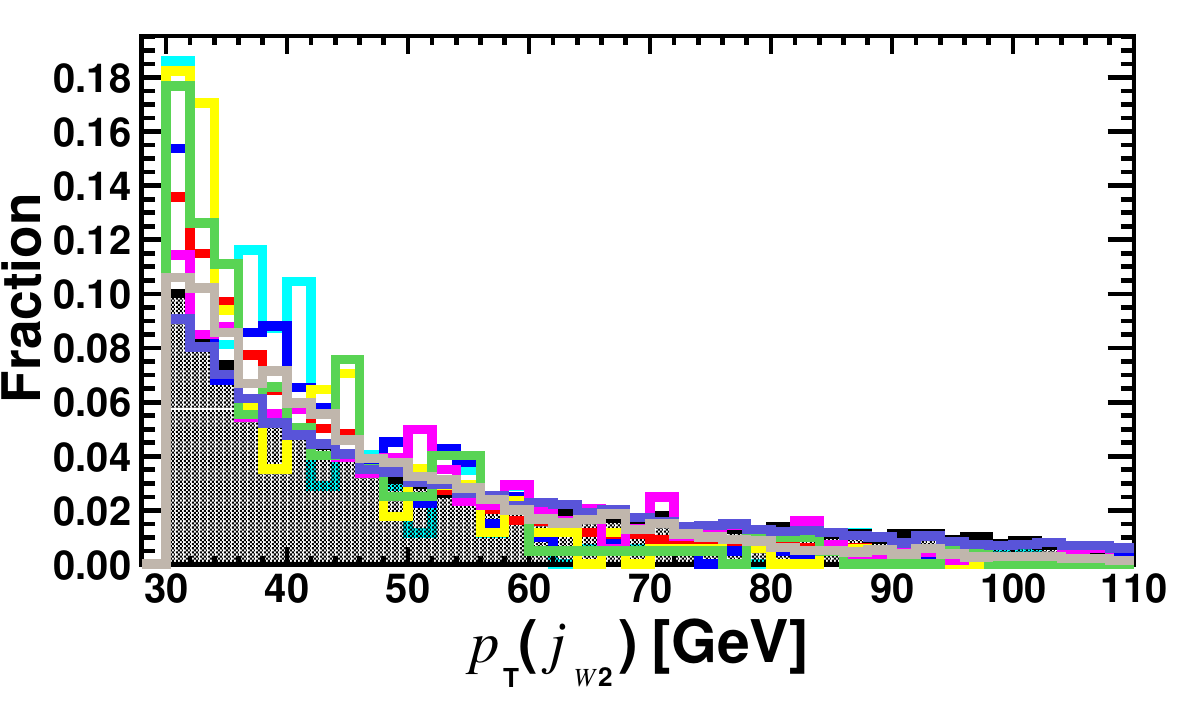}
}
\end{figure}
\vspace{-1.10cm}
\begin{figure}[H] 
\centering
\addtocounter{figure}{-1}
\subfigure{
\includegraphics[width=7.3cm, height=4.7cm]{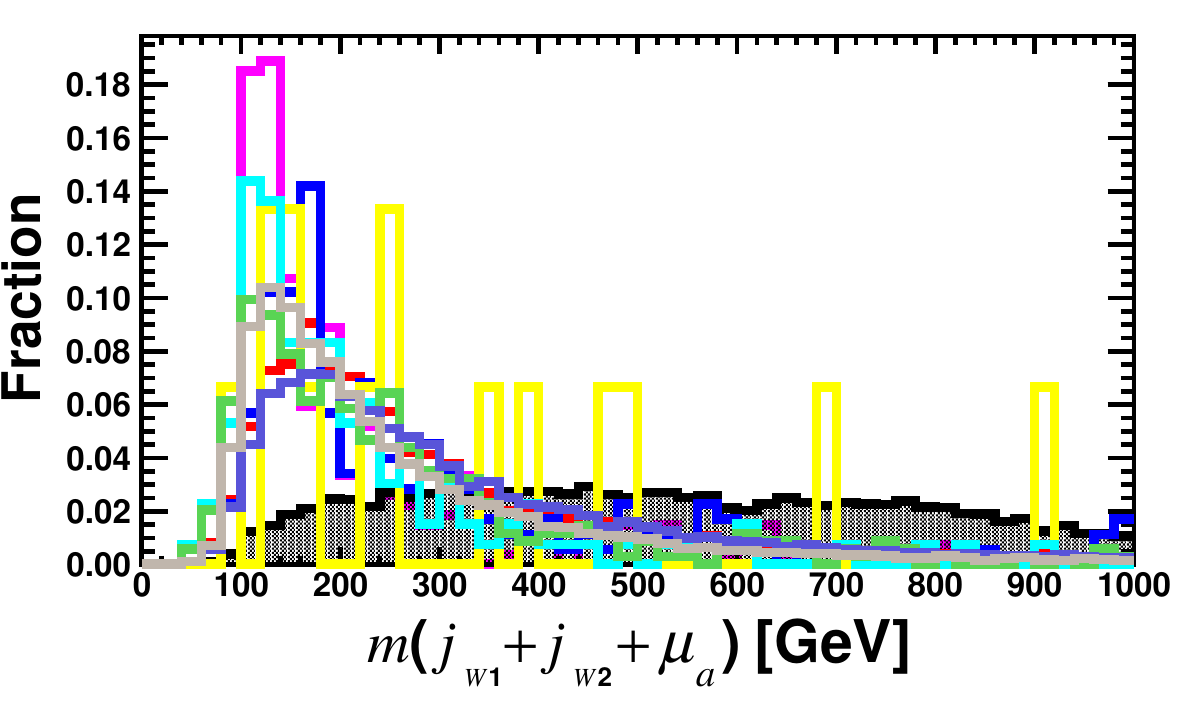}\,\,\,\,\,
\includegraphics[width=7.3cm, height=4.7cm]{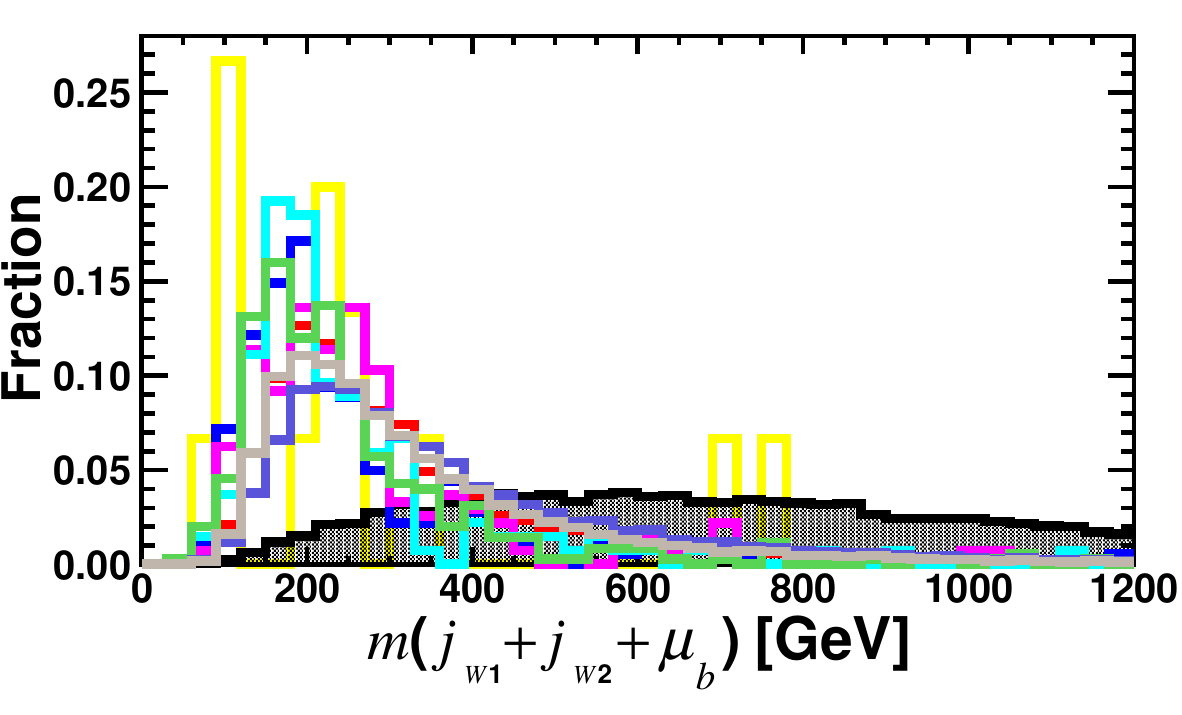}
}
\end{figure}
\vspace{-1.10cm}
\begin{figure}[H] 
\centering
\addtocounter{figure}{1}
\subfigure{
\includegraphics[width=7.3cm, height=4.7cm]{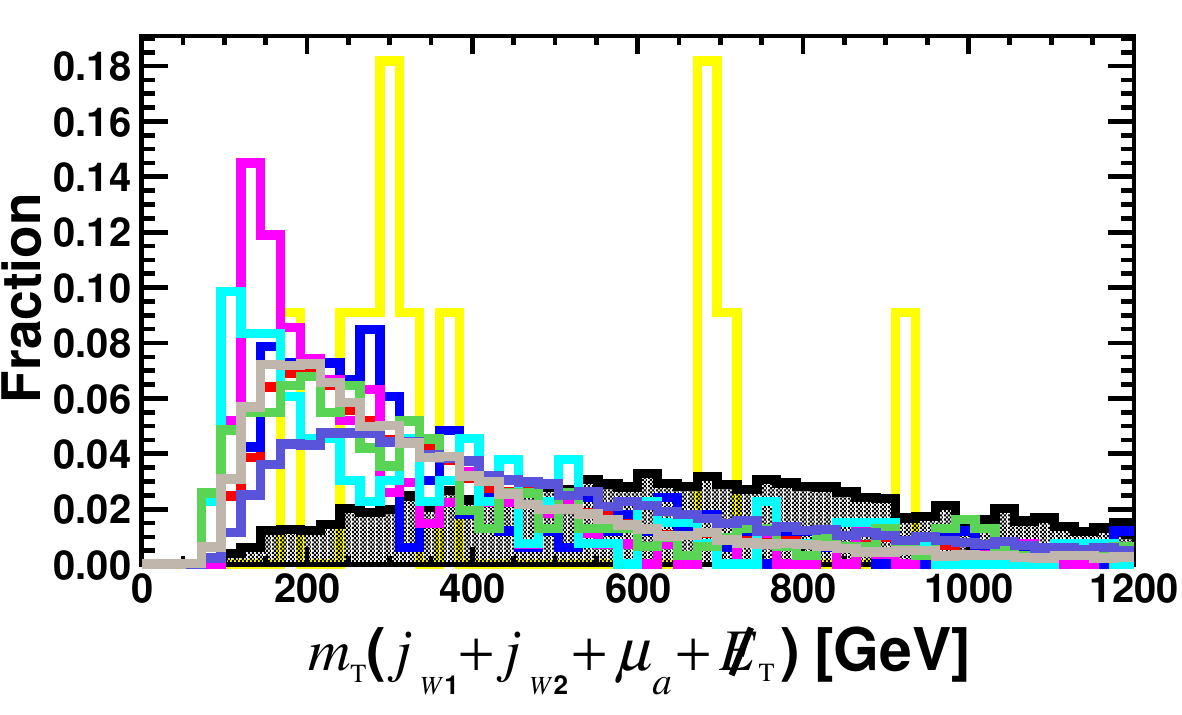}\,\,\,\,\,
\includegraphics[width=7.3cm, height=4.7cm]{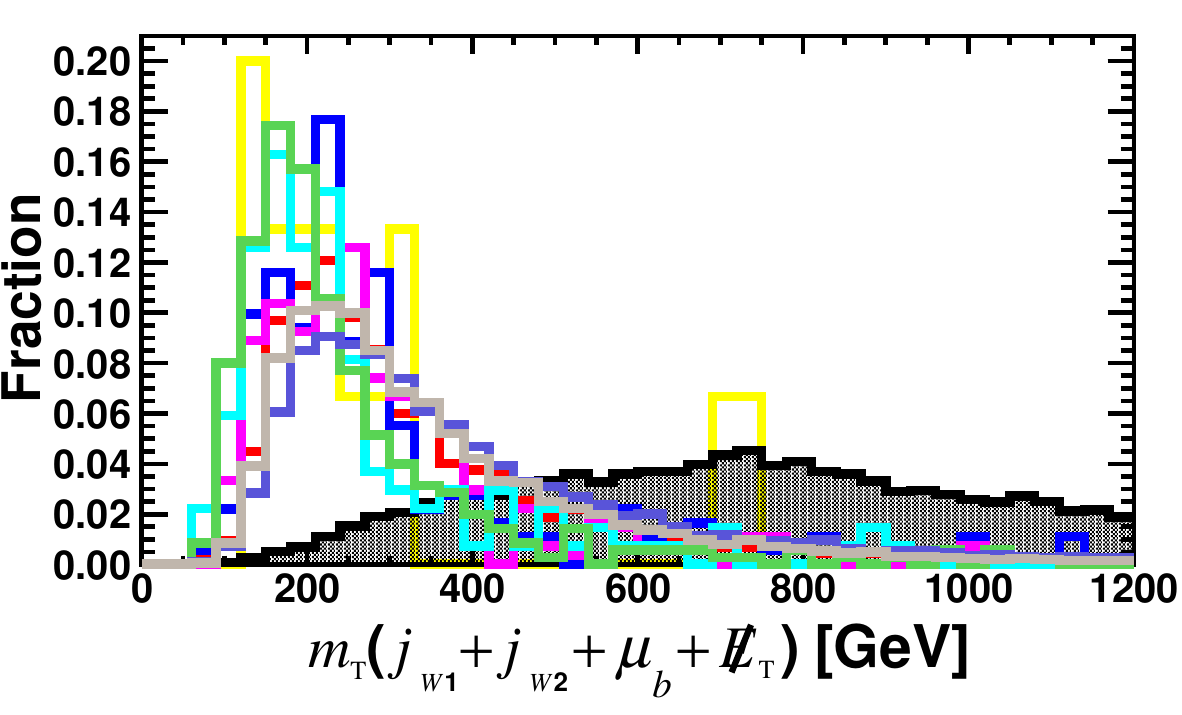}
}
\end{figure}
\vspace{-1.10cm}
\begin{figure}[H] 
\centering
%\addtocounter{figure}{-1}
\subfigure{
\includegraphics[width=7.3cm, height=4.7cm]{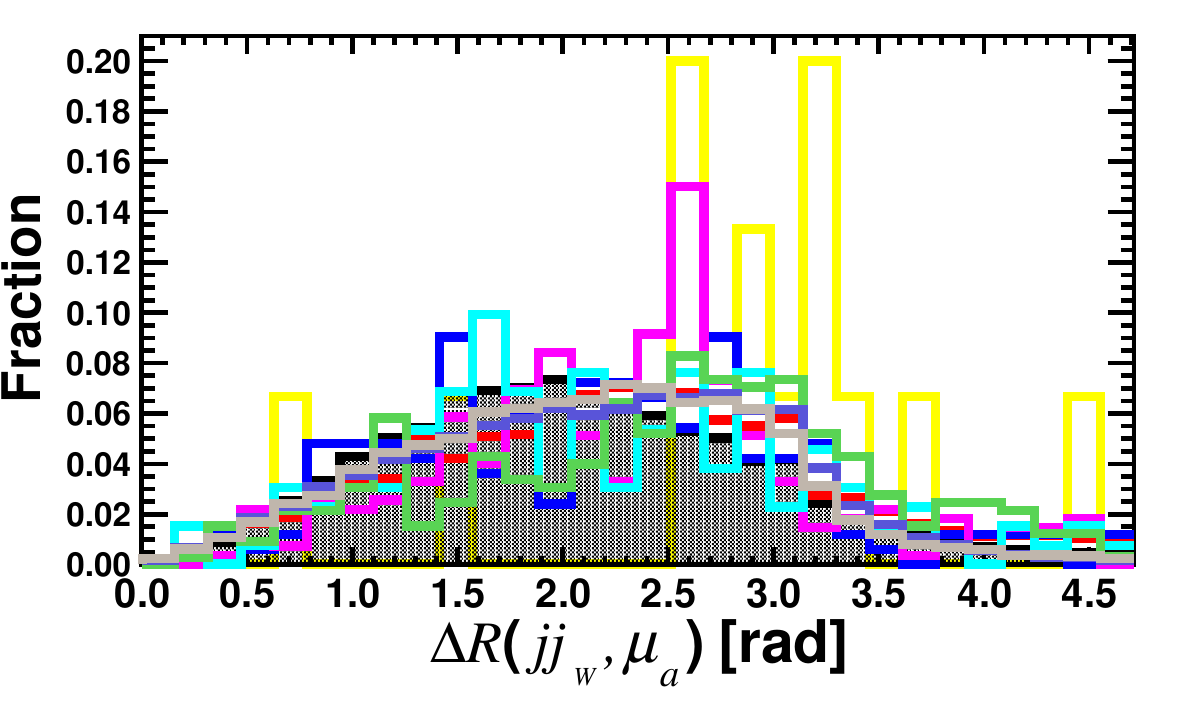}\,\,\,\,\,
\includegraphics[width=7.3cm, height=4.7cm]{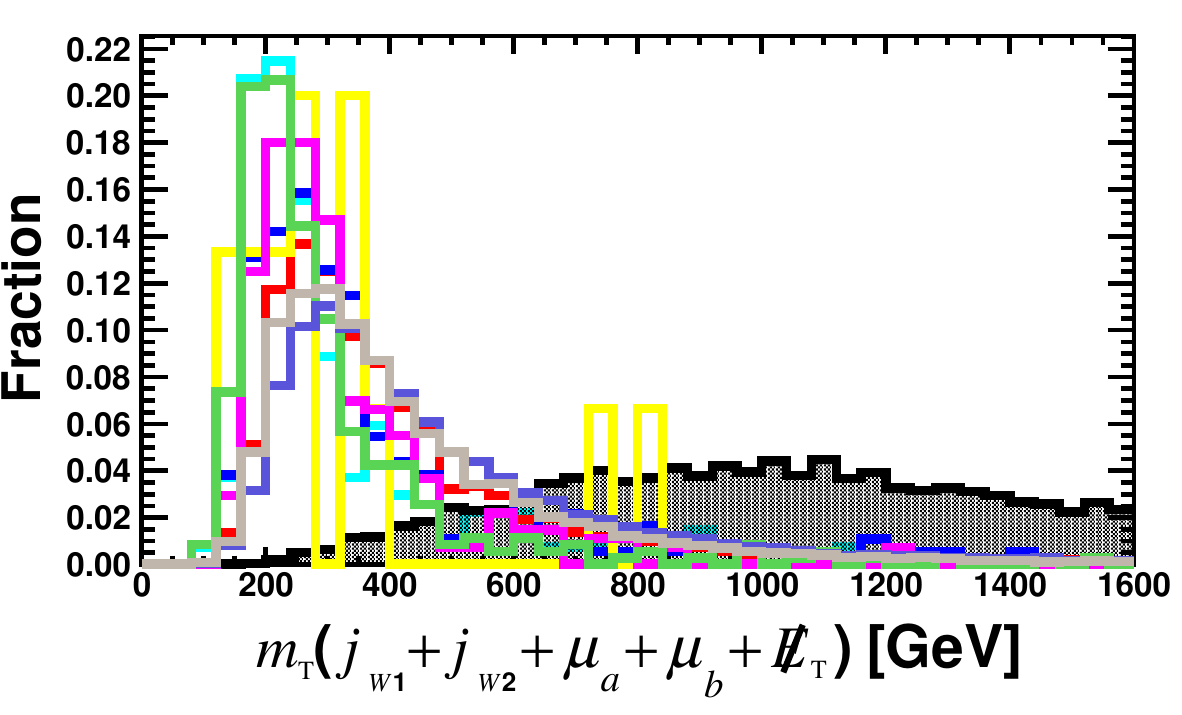}
}
\caption{
Same as Fig.~\ref{Representative_Muon}, but for $j_{_{W1}}$, $j_{_{W2}}$, and observables related to the reconstruction of the ALP mass and the off-shell $W^{(*)\pm}$ boson, assuming the benchmark $m_{a}=900$~GeV.
}
\label{fig:jjW}
\end{figure}

\begin{figure}[H]
\centering
\subfigure{
\includegraphics[width=7.3cm, height=4.7cm]{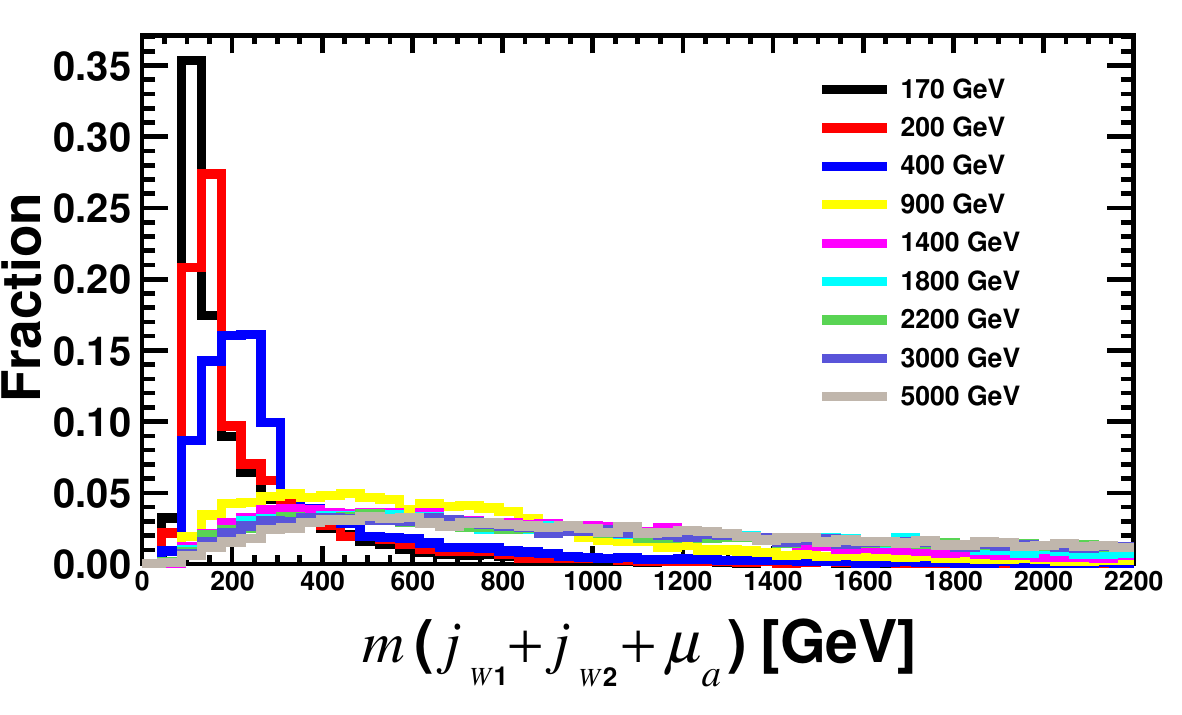}\,\,\,\,\,
\includegraphics[width=7.3cm, height=4.7cm]{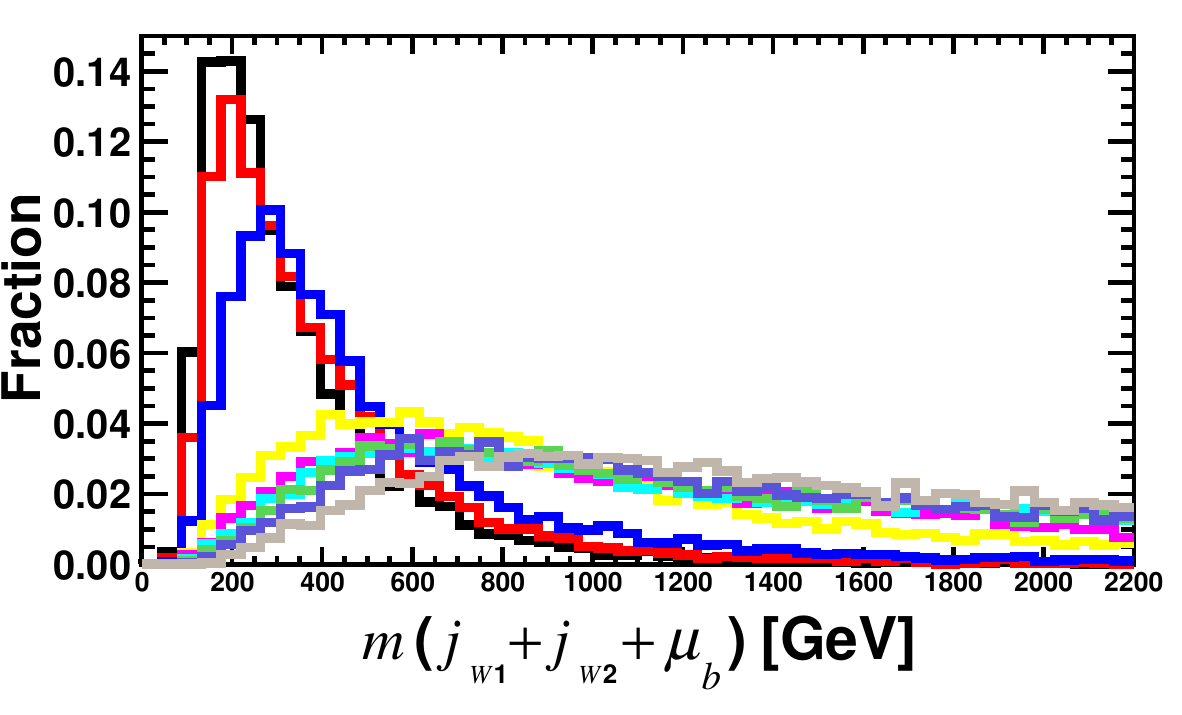}
}
\end{figure}
\addtocounter{figure}{-1}
\vspace{-1.10cm}
\begin{figure}[H]
\centering
\addtocounter{figure}{1}
\subfigure{
\includegraphics[width=7.3cm, height=4.7cm]{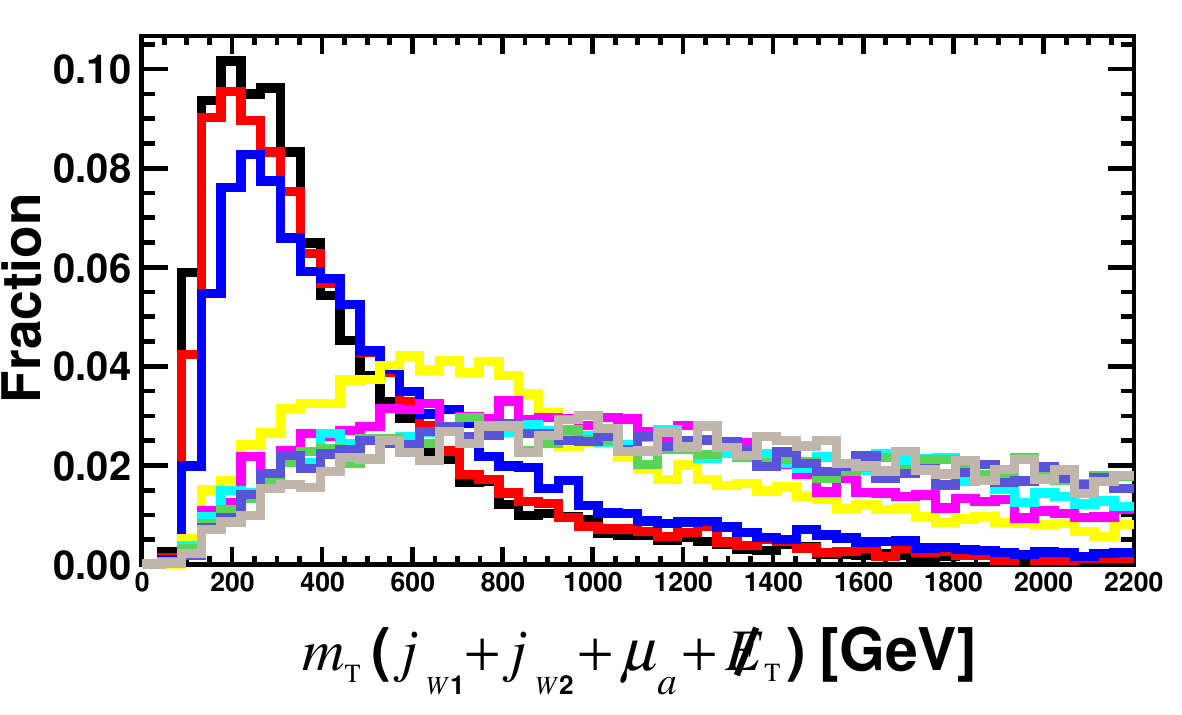}\,\,\,\,\,
\includegraphics[width=7.3cm, height=4.7cm]{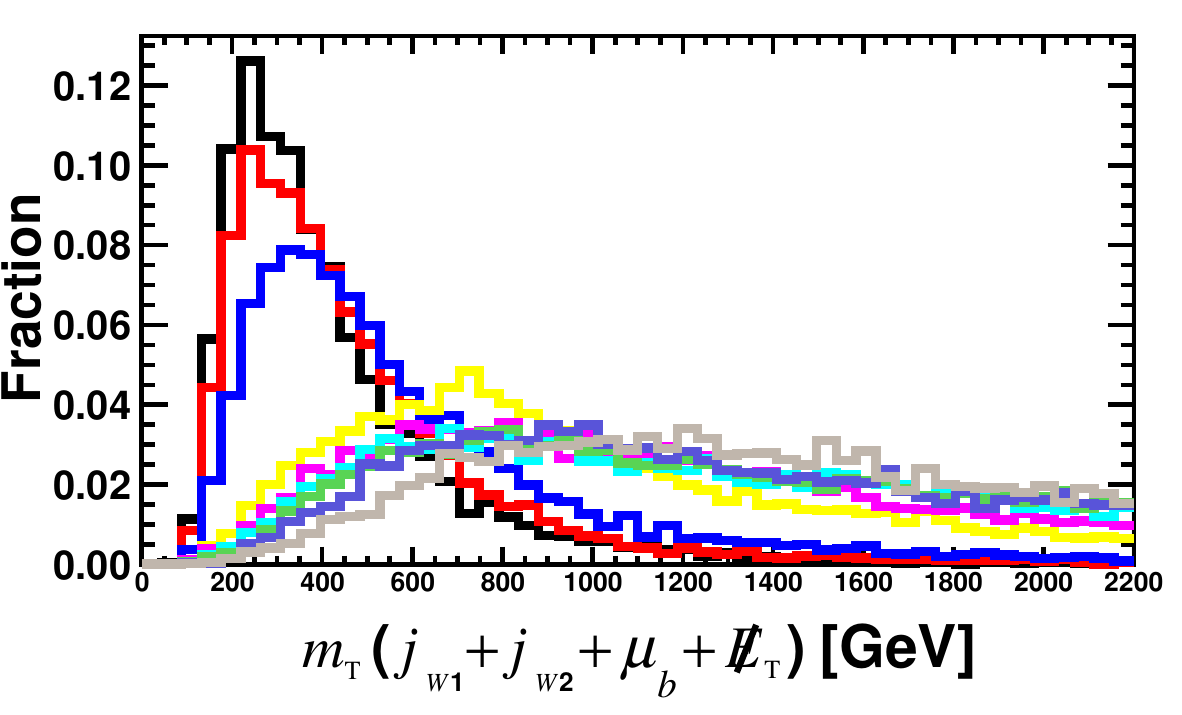}
}
\end{figure}
\vspace{-1.10cm}
\begin{figure}[H] 
\centering
%\addtocounter{figure}{-1}
\subfigure{
\includegraphics[width=7.3cm, height=4.7cm]{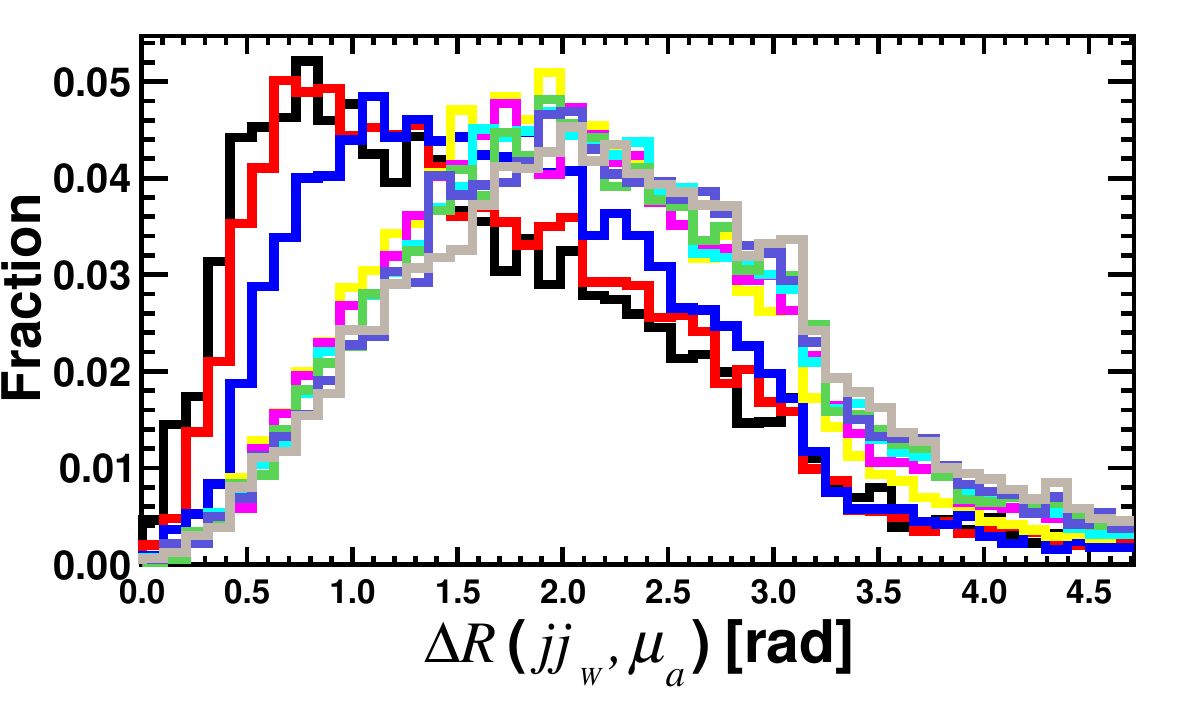}\,\,\,\,\,
\includegraphics[width=7.3cm, height=4.7cm]{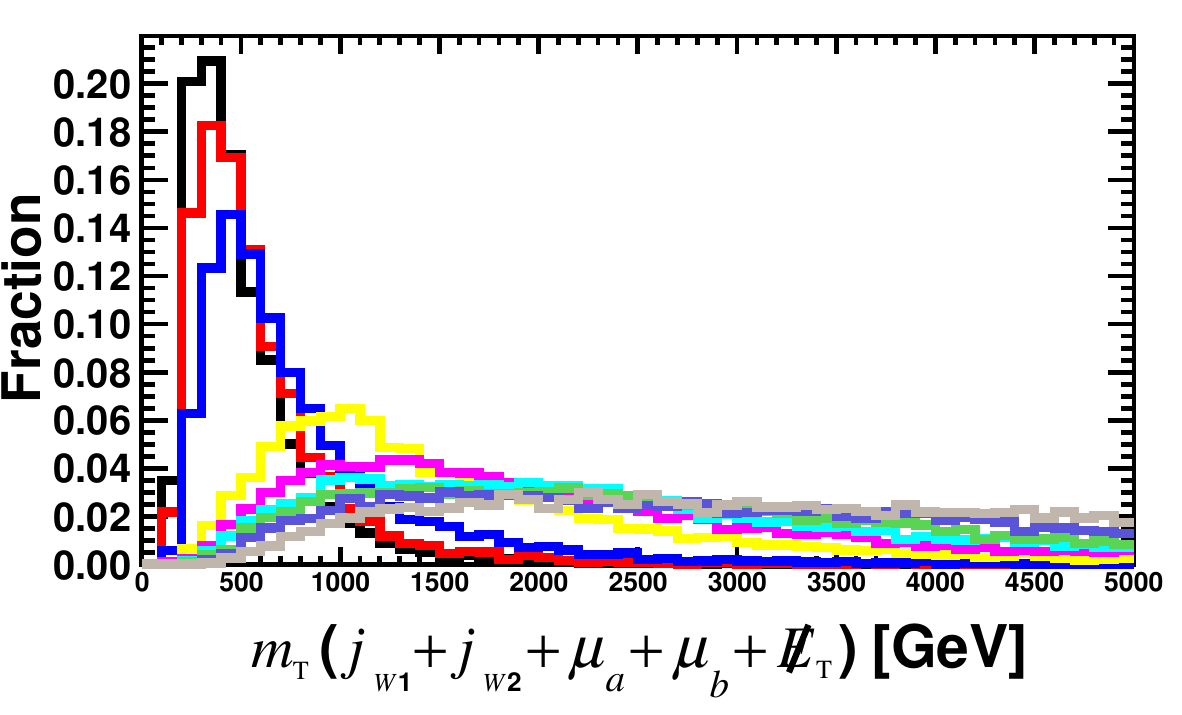}
}
\caption{
Distributions of the observables related to the reconstruction of the ALP mass and the off-shell $W^{(*)\pm}$ boson, for the signal only, assuming different $m_a$ values.
}
\label{fig:ALPandWstarMasses}
\end{figure}

%\vspace{+60cm}

%\newpage
\subsection{Distributions of BDT responses}
\label{app:WWW_BDT}

\begin{figure}[H]
\centering
\subfigure{
\includegraphics[width=7.3cm, height=4.7cm]{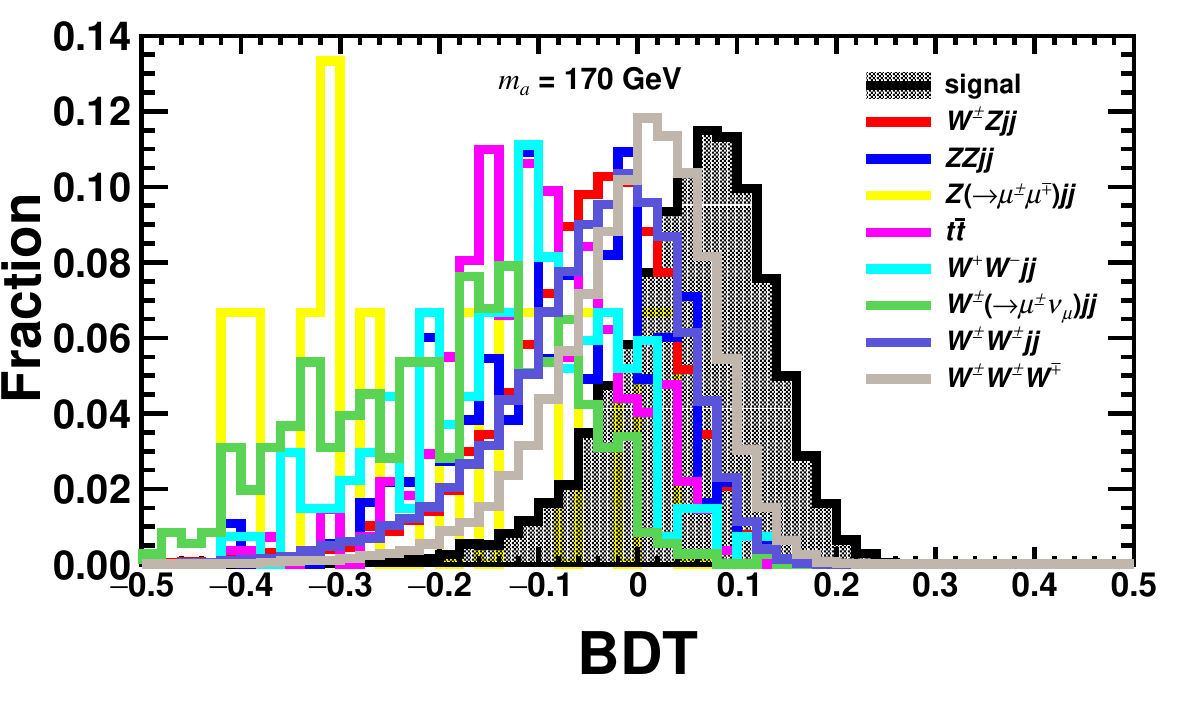}\,\,\,\,\,
\includegraphics[width=7.3cm, height=4.7cm]{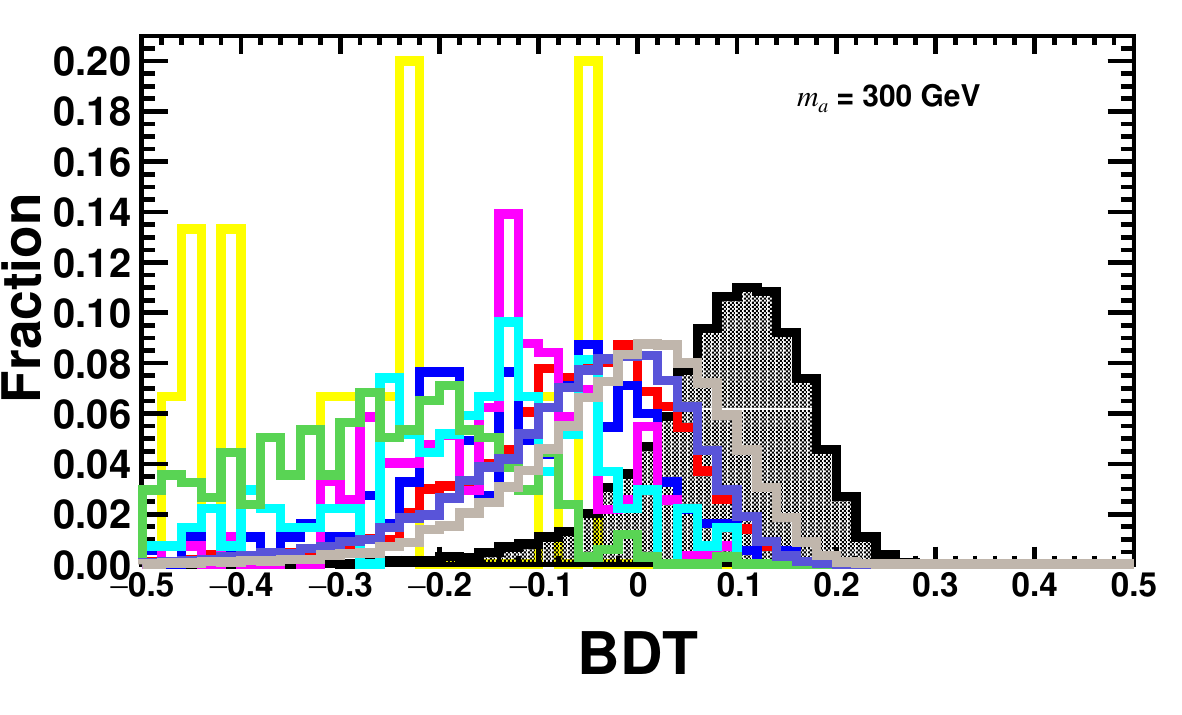}
}
\end{figure}
\addtocounter{figure}{-1}
\vspace{-1.10cm}
\begin{figure}[H]
\centering
\addtocounter{figure}{1}
\subfigure{
\includegraphics[width=7.3cm, height=4.7cm]{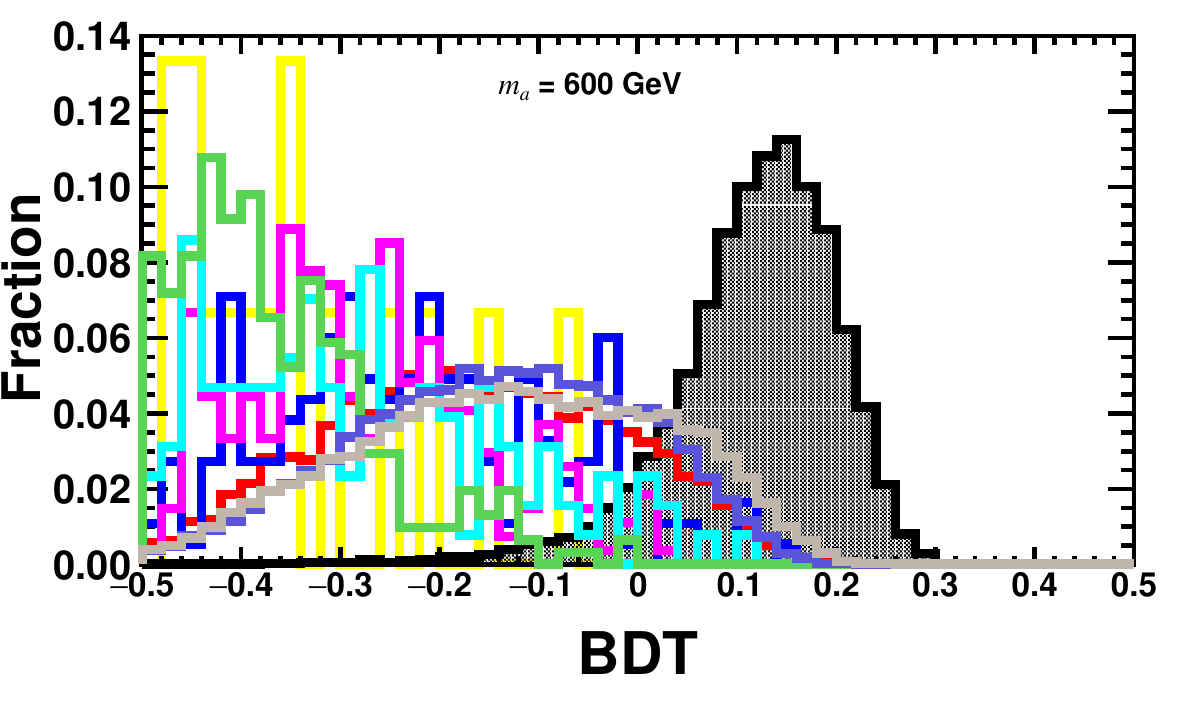}\,\,\,\,\,
\includegraphics[width=7.3cm, height=4.7cm]{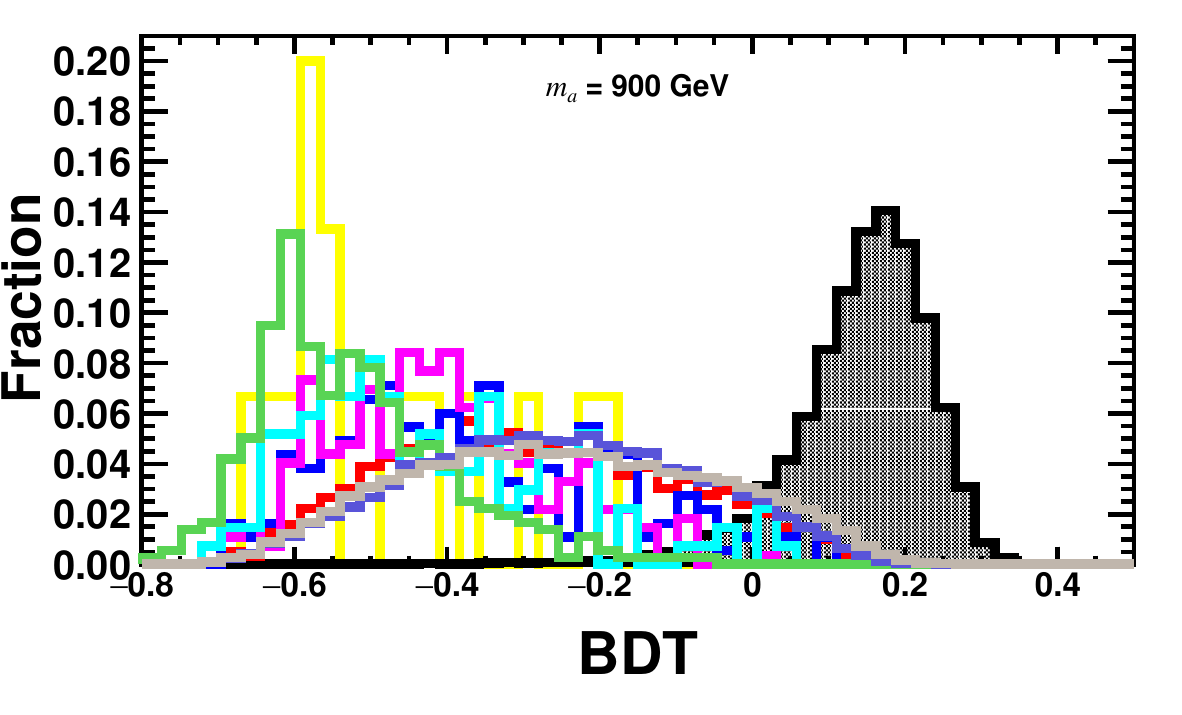}
}
\end{figure}
\vspace{-1.10cm}
\begin{figure}[H] 
\centering
\addtocounter{figure}{-1}
\subfigure{
\includegraphics[width=7.3cm, height=4.7cm]{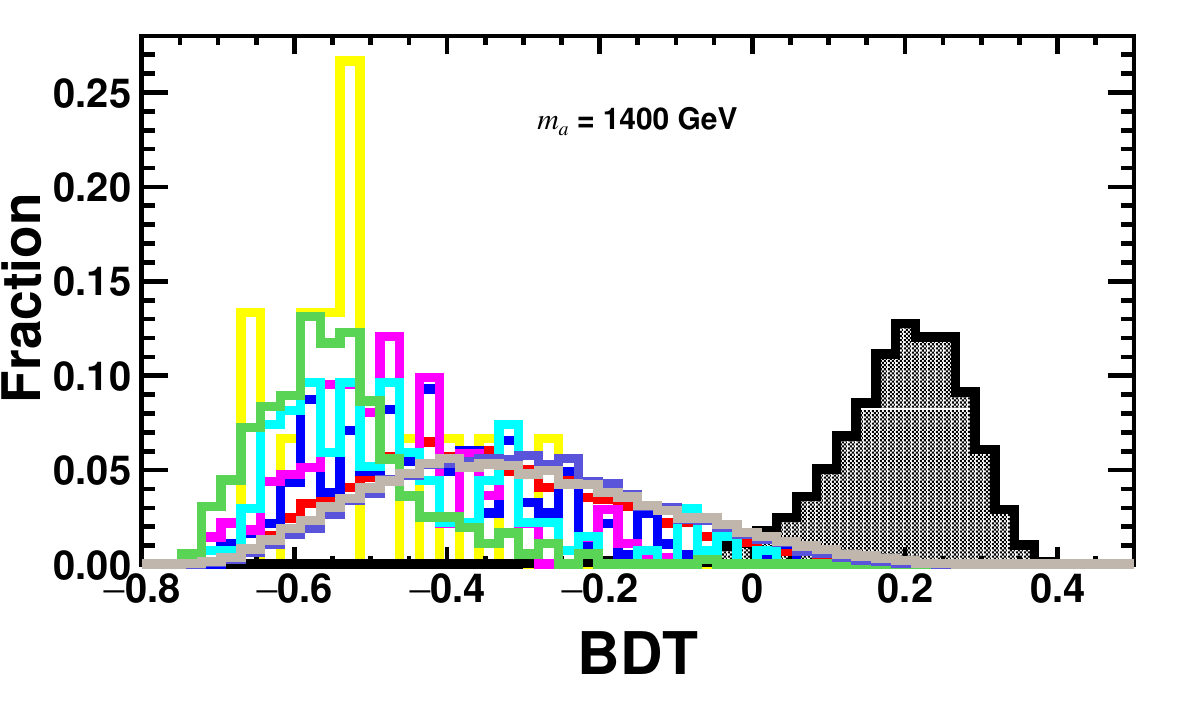}\,\,\,\,\,
\includegraphics[width=7.3cm, height=4.7cm]{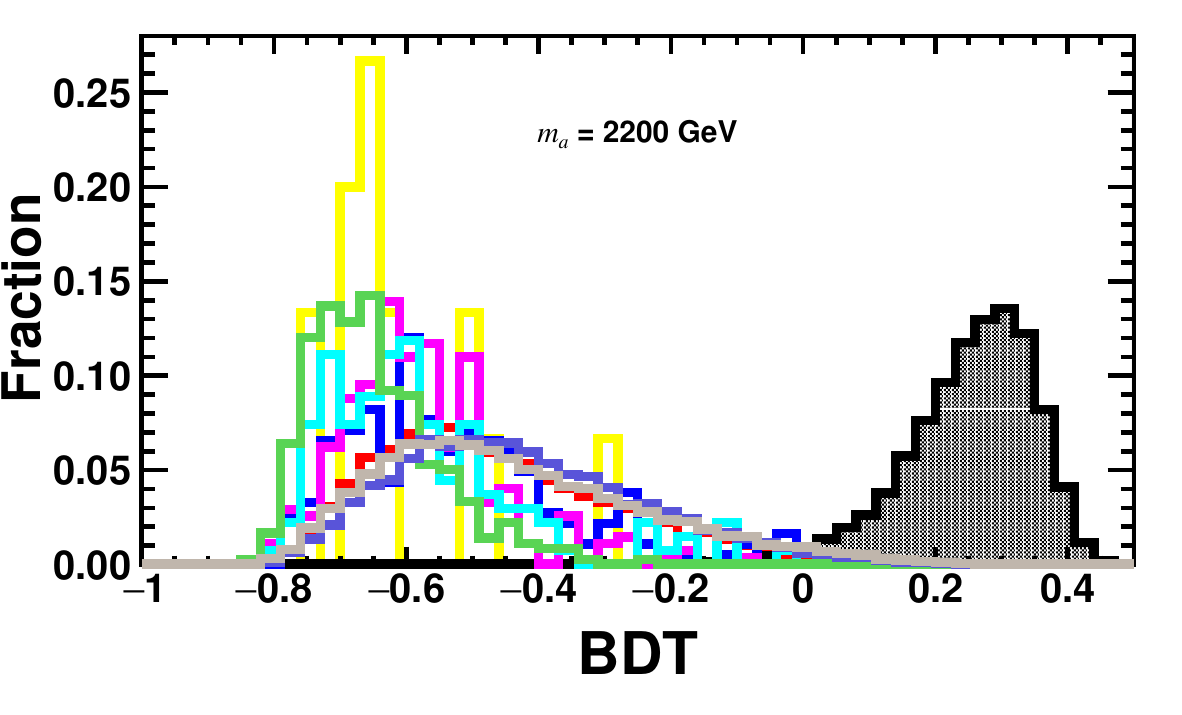}
}
\end{figure}
\vspace{-1.10cm}
\begin{figure}[H] 
\centering
\addtocounter{figure}{1}
\subfigure{
\includegraphics[width=7.3cm, height=4.7cm]{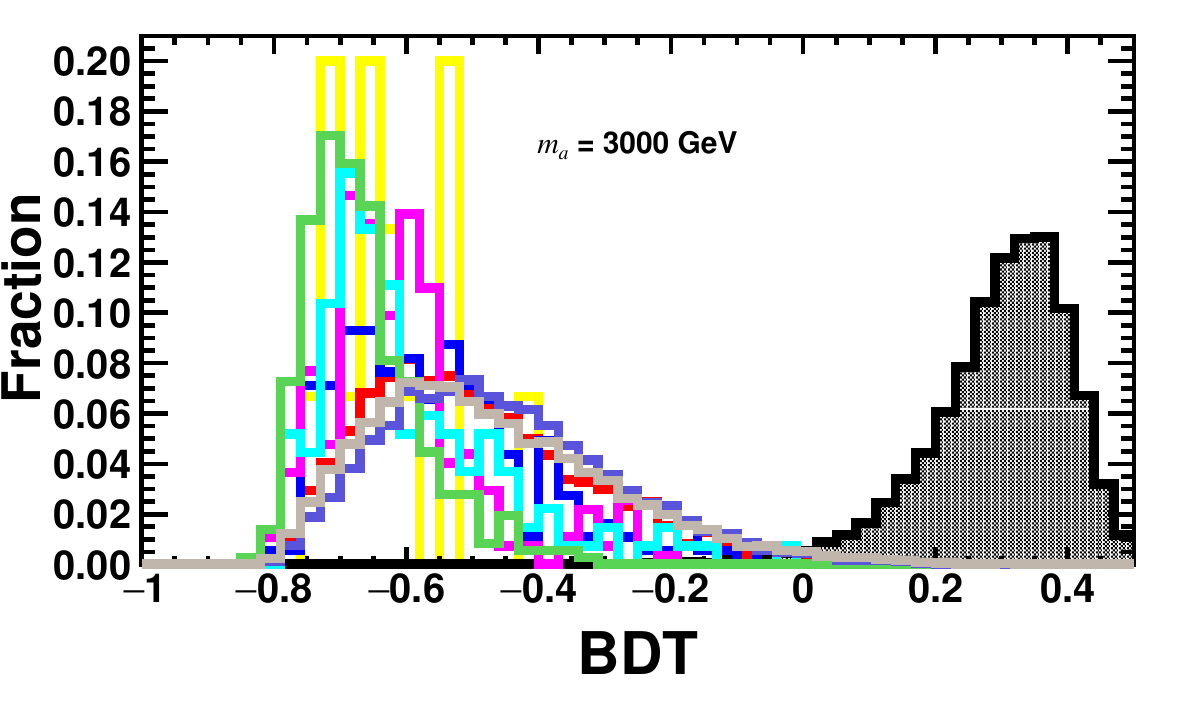}\,\,\,\,\,
\includegraphics[width=7.3cm, height=4.7cm]{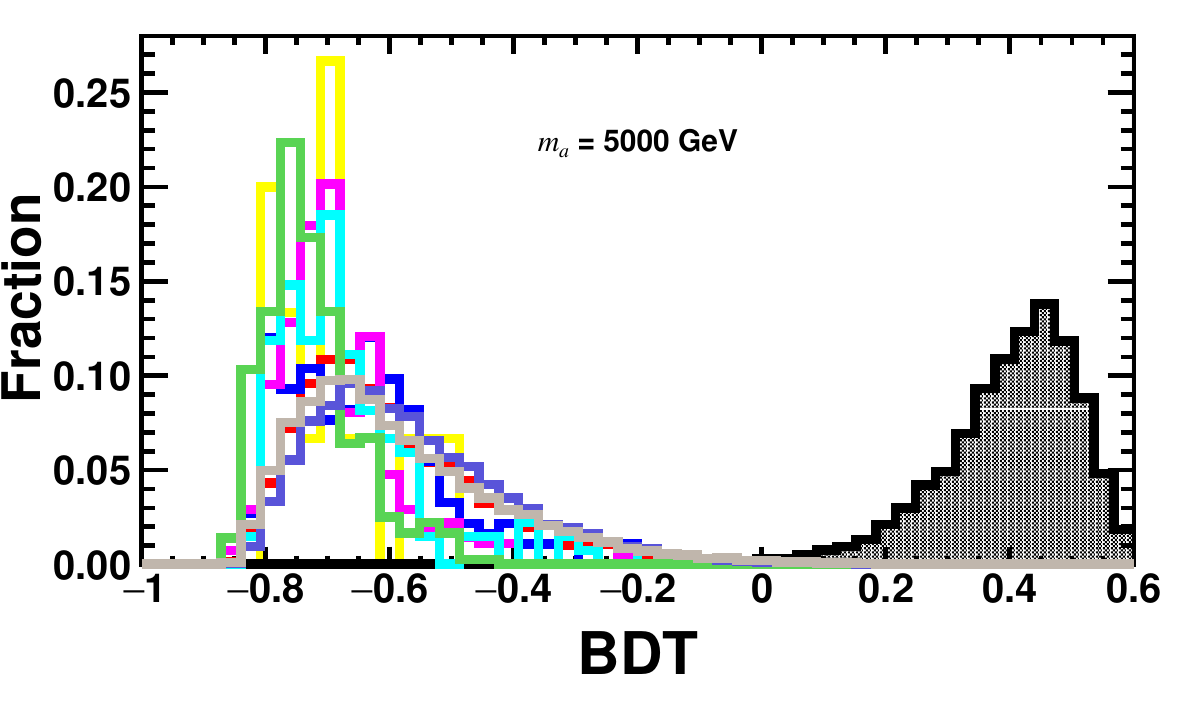}
}
\caption{
Distributions of BDT responses after applying the preselection criteria for the signal (black, shaded) and background processes at the SppC/FCC-hh with $\sqrt{s}=100$~TeV for representative $m_{a}$ values.
}
\label{fig:BDT_ALPmasses}
\end{figure}

%\newpage
\subsection{Selection efficiencies}
\label{app:WWW_Sel_Eff}

\begin{table*}[h]
\centering
\scalebox{0.55}{
%\begin{ruledtabular}
\begin{tabular}{c c c c c c c c c c c}
\hline
\hline
$m_a$ & selection & signal & $W^\pm Zjj$ & $ZZjj$ & $Z(\to\mu^{+}\mu^{-})jj$ & $t\bar{t}$ & $W^\pm(\to\mu^{\pm}\nu_{\mu})\,W^\mp jj$ & $W^{\pm}(\to\mu^{\pm}\nu_{\mu})jj$ & $W^\pm W^\pm jj$ & $W^\pm W^\pm W^\mp$\\
\hline
\multirow{2}{*}{170 GeV} & preselection & $2.94\mltp10^{-3}$&$3.15\mltp10^{-4}$&$3.66\mltp10^{-5}$&$8.57\mltp10^{-6}$&$2.57\mltp10^{-5}$&$5.40\mltp10^{-5}$ &$2.97\mltp10^{-5}$ &$6.19\mltp10^{-3}$ &$3.87\mltp10^{-3}$\\
&  BDT$>$0.069&$4.56\mltp10^{-1}$&$4.83\mltp10^{-2}$&$3.28\mltp10^{-2}$&$-$&$6.02\mltp10^{-3}$&$2.96\mltp10^{-2}$&$2.79\mltp10^{-3}$&$6.22\mltp10^{-2}$&$1.51\mltp10^{-1}$ \\
\hline
\multirow{2}{*}{200 GeV}  & preselection & $3.43\mltp10^{-3}$&$3.15\mltp10^{-4}$&$3.66\mltp10^{-5}$&$8.57\mltp10^{-6}$&$2.57\mltp10^{-5}$&$5.40\mltp10^{-5}$ &$2.97\mltp10^{-5}$ &$6.19\mltp10^{-3}$ &$3.87\mltp10^{-3}$\\
& BDT$>$0.049 &$6.16\mltp10^{-1}$&$1.10\mltp10^{-1}$&$8.20\mltp10^{-2}$&$-$&$4.22\mltp10^{-2}$&$2.96\mltp10^{-2}$&$2.79\mltp10^{-3}$&$1.36\mltp10^{-1}$&$3.00\mltp10^{-1}$ \\
\hline
\multirow{2}{*}{300 GeV}  & preselection & $4.44\mltp10^{-3}$&$3.15\mltp10^{-4}$&$3.66\mltp10^{-5}$&$8.57\mltp10^{-6}$&$2.57\mltp10^{-5}$&$5.40\mltp10^{-5}$ &$2.97\mltp10^{-5}$ &$6.19\mltp10^{-3}$ &$3.87\mltp10^{-3}$\\
& BDT$>$0.086  &$5.48\mltp10^{-1}$&$4.12\mltp10^{-2}$&$2.19\mltp10^{-2}$&$-$&$-$&$7.41\mltp10^{-3}$&$2.79\mltp10^{-3}$&$5.27\mltp10^{-2}$&$1.49\mltp10^{-1}$ \\
\hline
\multirow{2}{*}{400 GeV} & preselection & $4.41\mltp10^{-3}$&$3.15\mltp10^{-4}$&$3.66\mltp10^{-5}$&$8.57\mltp10^{-6}$&$2.57\mltp10^{-5}$&$5.40\mltp10^{-5}$ &$2.97\mltp10^{-5}$ &$6.19\mltp10^{-3}$ &$3.87\mltp10^{-3}$\\
& BDT$>$0.083  &$5.29\mltp10^{-1}$&$3.70\mltp10^{-2}$&$3.28\mltp10^{-2}$&$-$&$6.02\mltp10^{-3}$&$-$&$2.79\mltp10^{-3}$&$4.36\mltp10^{-2}$&$1.35\mltp10^{-1}$ \\  
\hline
\multirow{2}{*}{600 GeV} & preselection & $5.09\mltp10^{-3}$&$3.15\mltp10^{-4}$&$3.66\mltp10^{-5}$&$8.57\mltp10^{-6}$&$2.57\mltp10^{-5}$&$5.40\mltp10^{-5}$ &$2.97\mltp10^{-5}$ &$6.19\mltp10^{-3}$ &$3.87\mltp10^{-3}$\\
& BDT$>$0.099 &$6.50\mltp10^{-1}$&$1.92\mltp10^{-2}$&$2.19\mltp10^{-2}$&$-$&$-$&$7.41\mltp10^{-3}$&$-$&$2.47\mltp10^{-2}$&$6.10\mltp10^{-2}$ \\
\hline
\multirow{2}{*}{900 GeV}  & preselection & $5.07\mltp10^{-3}$&$3.15\mltp10^{-4}$&$3.66\mltp10^{-5}$&$8.57\mltp10^{-6}$&$2.57\mltp10^{-5}$&$5.40\mltp10^{-5}$ &$2.97\mltp10^{-5}$ &$6.19\mltp10^{-3}$ &$3.87\mltp10^{-3}$\\
& BDT$>$0.121 &$6.70\mltp10^{-1}$&$5.08\mltp10^{-3}$&$-$&$-$&$-$&$-$&$-$&$9.56\mltp10^{-3}$&$2.06\mltp10^{-2}$ \\
\hline
\multirow{2}{*}{1400 GeV}  & preselection & $5.13\mltp10^{-3}$&$3.15\mltp10^{-4}$&$3.66\mltp10^{-5}$&$8.57\mltp10^{-6}$&$2.57\mltp10^{-5}$&$5.40\mltp10^{-5}$ &$2.97\mltp10^{-5}$ &$6.19\mltp10^{-3}$ &$3.87\mltp10^{-3}$\\
& BDT$>$0.120&$8.05\mltp10^{-1}$&$2.54\mltp10^{-3}$&$-$&$-$&$-$&$7.41\mltp10^{-3}$&$-$&$6.78\mltp10^{-3}$&$1.38\mltp10^{-2}$ \\                                       
\hline
\multirow{2}{*}{1800 GeV} & preselection & $5.18\mltp10^{-3}$&$3.15\mltp10^{-4}$&$3.66\mltp10^{-5}$&$8.57\mltp10^{-6}$&$2.57\mltp10^{-5}$&$5.40\mltp10^{-5}$ &$2.97\mltp10^{-5}$ &$6.19\mltp10^{-3}$ &$3.87\mltp10^{-3}$\\
& BDT$>$0.152&$7.80\mltp10^{-1}$&$1.13\mltp10^{-3}$&$-$&$-$&$-$&$-$&$-$&$2.52\mltp10^{-3}$&$5.49\mltp10^{-3}$ \\                                        
\hline
\multirow{2}{*}{2200 GeV} & preselection & $5.17\mltp10^{-3}$&$3.15\mltp10^{-4}$&$3.66\mltp10^{-5}$&$8.57\mltp10^{-6}$&$2.57\mltp10^{-5}$&$5.40\mltp10^{-5}$ &$2.97\mltp10^{-5}$ &$6.19\mltp10^{-3}$ &$3.87\mltp10^{-3}$\\
& BDT$>$0.120 & $8.98\mltp10^{-1}$&$1.98\mltp10^{-3}$&$-$&$-$&$-$&$-$&$-$&$4.52\mltp10^{-3}$&$7.62\mltp10^{-3}$ \\                                           
\hline
\multirow{2}{*}{3000 GeV} & preselection & $5.20\mltp10^{-3}$&$3.15\mltp10^{-4}$&$3.66\mltp10^{-5}$&$8.57\mltp10^{-6}$&$2.57\mltp10^{-5}$&$5.40\mltp10^{-5}$ &$2.97\mltp10^{-5}$ &$6.19\mltp10^{-3}$ &$3.87\mltp10^{-3}$\\
&  BDT$>$0.080  	&$9.57\mltp10^{-1}$&$1.69\mltp10^{-3}$&$-$&$-$&$-$&$7.41\mltp10^{-3}$&$-$&$5.75\mltp10^{-3}$&$8.66\mltp10^{-3}$ \\                                            
\hline
\multirow{2}{*}{5000 GeV} & preselection & $5.42\mltp10^{-3}$&$3.15\mltp10^{-4}$&$3.66\mltp10^{-5}$&$8.57\mltp10^{-6}$&$2.57\mltp10^{-5}$&$5.40\mltp10^{-5}$ &$2.97\mltp10^{-5}$ &$6.19\mltp10^{-3}$ &$3.87\mltp10^{-3}$\\
&  BDT$>$0.080&$9.80\mltp10^{-1}$&$1.13\mltp10^{-3}$&$-$&$-$&$-$&$-$&$-$&$2.45\mltp10^{-3}$&$3.49\mltp10^{-3}$ \\                                            
\hline
\multirow{2}{*}{10000 GeV} & preselection & $5.30\mltp10^{-3}$&$3.15\mltp10^{-4}$&$3.66\mltp10^{-5}$&$8.57\mltp10^{-6}$&$2.57\mltp10^{-5}$&$5.40\mltp10^{-5}$ &$2.97\mltp10^{-5}$ &$6.19\mltp10^{-3}$ &$3.87\mltp10^{-3}$\\
&  BDT$>$0.030 &$9.93\mltp10^{-1}$&$1.13\mltp10^{-3}$&$-$&$-$&$-$&$-$&$-$&$7.11\mltp10^{-4}$&$2.07\mltp10^{-3}$ \\                                         
\hline
\multirow{2}{*}{20000 GeV} & preselection & $5.43\mltp10^{-3}$&$3.15\mltp10^{-4}$&$3.66\mltp10^{-5}$&$8.57\mltp10^{-6}$&$2.57\mltp10^{-5}$&$5.40\mltp10^{-5}$ &$2.97\mltp10^{-5}$ &$6.19\mltp10^{-3}$ &$3.87\mltp10^{-3}$\\
&  BDT$>$0.050  &$9.97\mltp10^{-1}$&$8.47\mltp10^{-4}$&$-$&$-$&$-$&$-$&$-$&$1.29\mltp10^{-4}$&$5.17\mltp10^{-4}$ \\                                            
\hline
\hline
\end{tabular}
%\end{ruledtabular}
}
\caption{
Selection efficiencies of the preselection and BDT criteria for the signal and background processes at the SppC/FCC-hh with $\sqrt{s}=100$~TeV, assuming different ALP masses. Here, “$-$” indicates that the yield becomes negligible with $\mathcal{L}=20~\iab$.
}
\label{tab:allEfficiencies}
\end{table*}

\newpage
\section{Details of the $W^+W^- jj$ analyses}

\subsection{Distributions of representative observables}
\label{app:WWjj_observables}

\begin{figure}[H]
\centering
\subfigure{
\includegraphics[width=7.3cm, height=4.7cm]{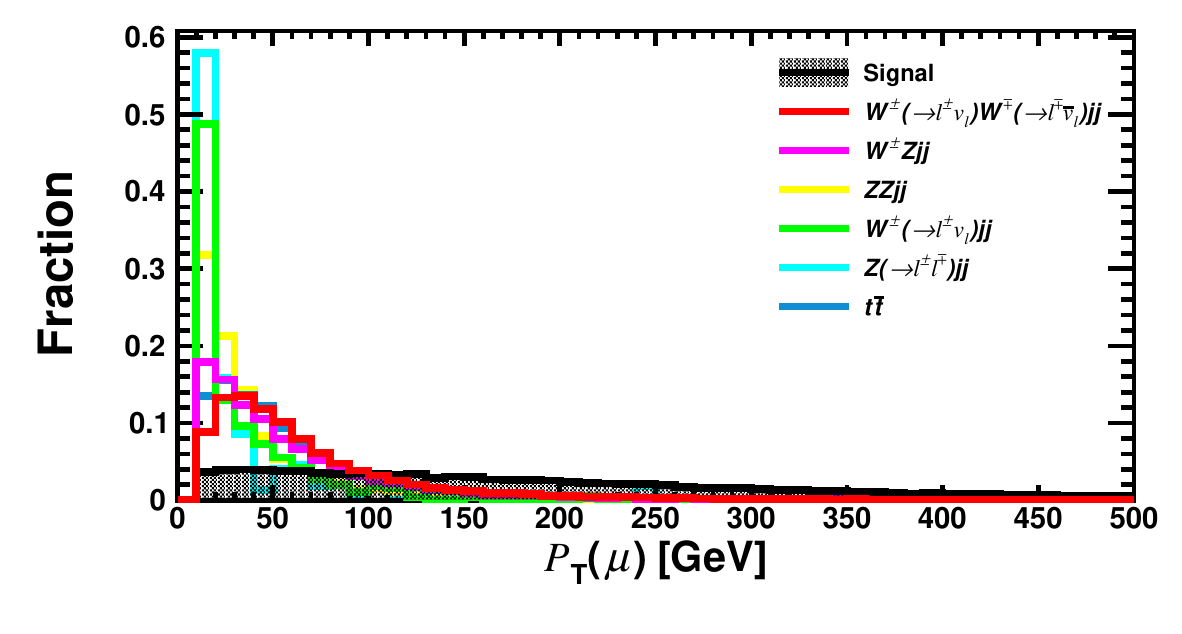}\,\,\,\,\,
\includegraphics[width=7.3cm, height=4.7cm]{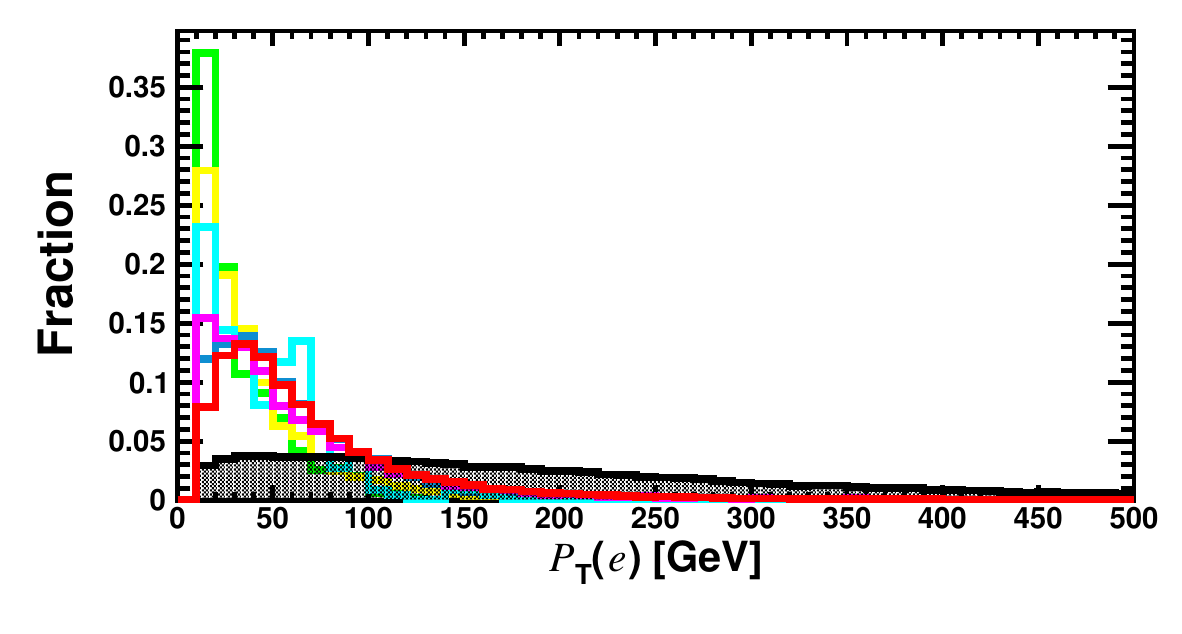}
}
\end{figure}
\addtocounter{figure}{-1}
\vspace{-1.10cm}
\begin{figure}[H]
\centering
\addtocounter{figure}{1}
\subfigure{
\includegraphics[width=7.3cm, height=4.7cm]{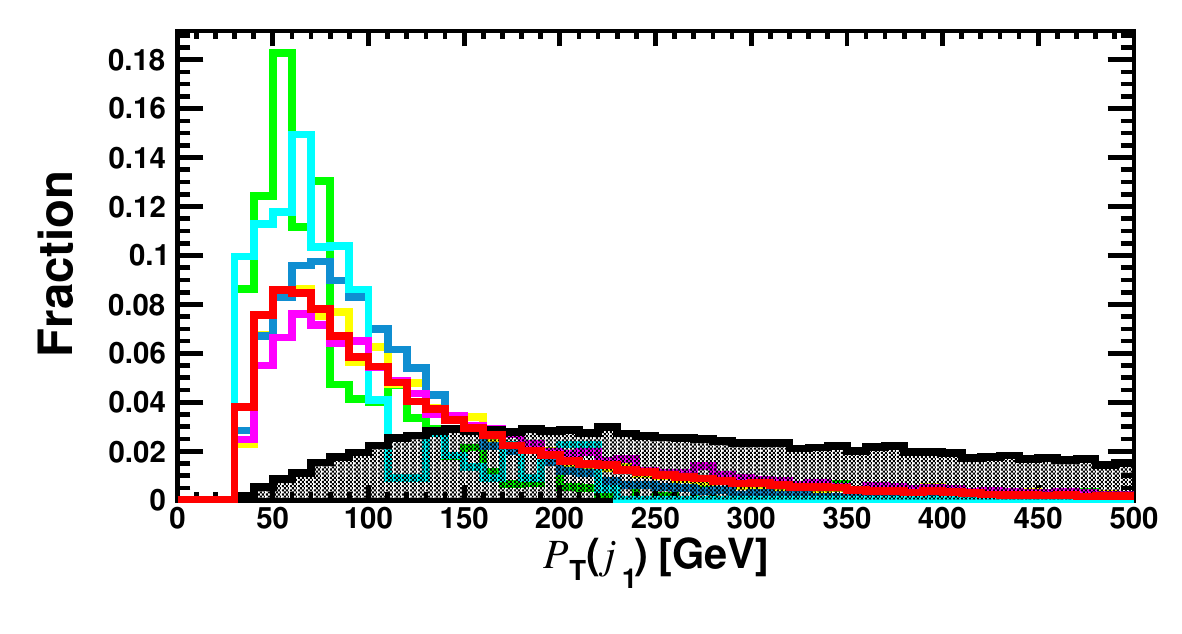}\,\,\,\,\,
\includegraphics[width=7.3cm, height=4.7cm]{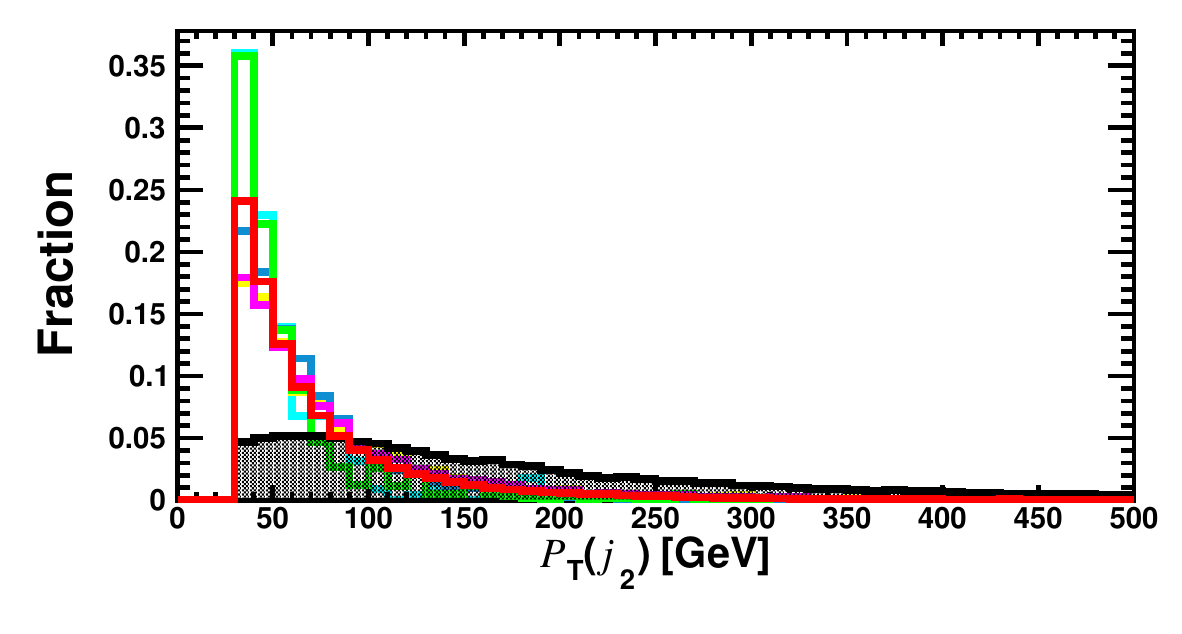}
}
\end{figure}
\addtocounter{figure}{-1}
\vspace{-1.10cm}
\begin{figure}[H]
\centering
\addtocounter{figure}{1}
\subfigure{
\includegraphics[width=7.3cm, height=4.7cm]{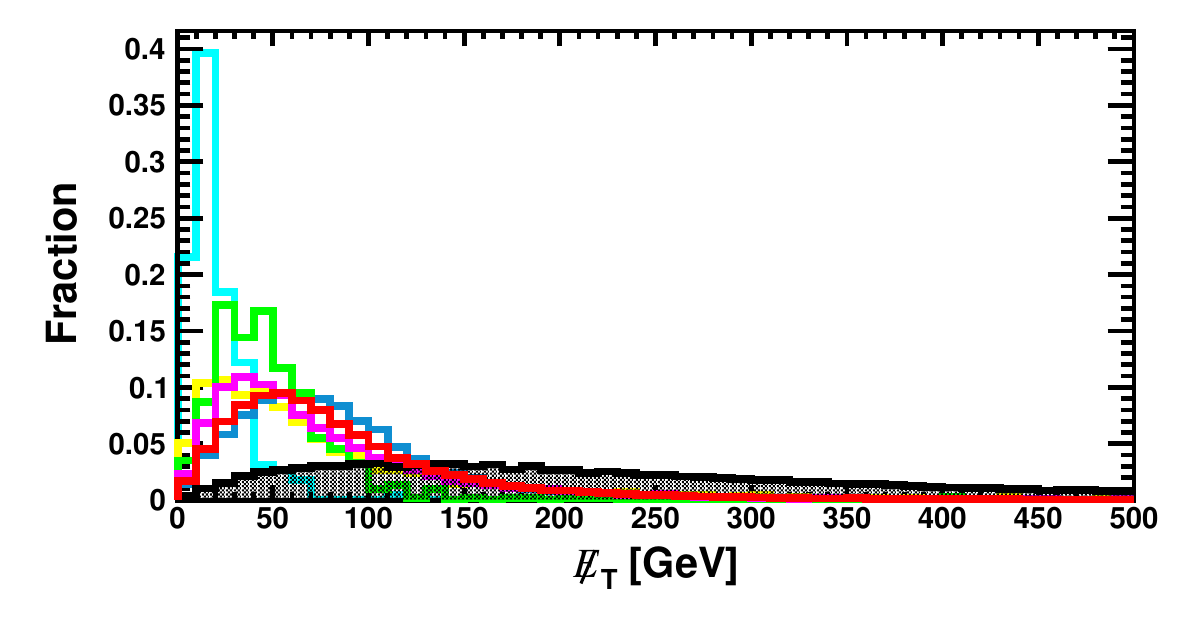}\,\,\,\,\,
\includegraphics[width=7.3cm, height=4.7cm]{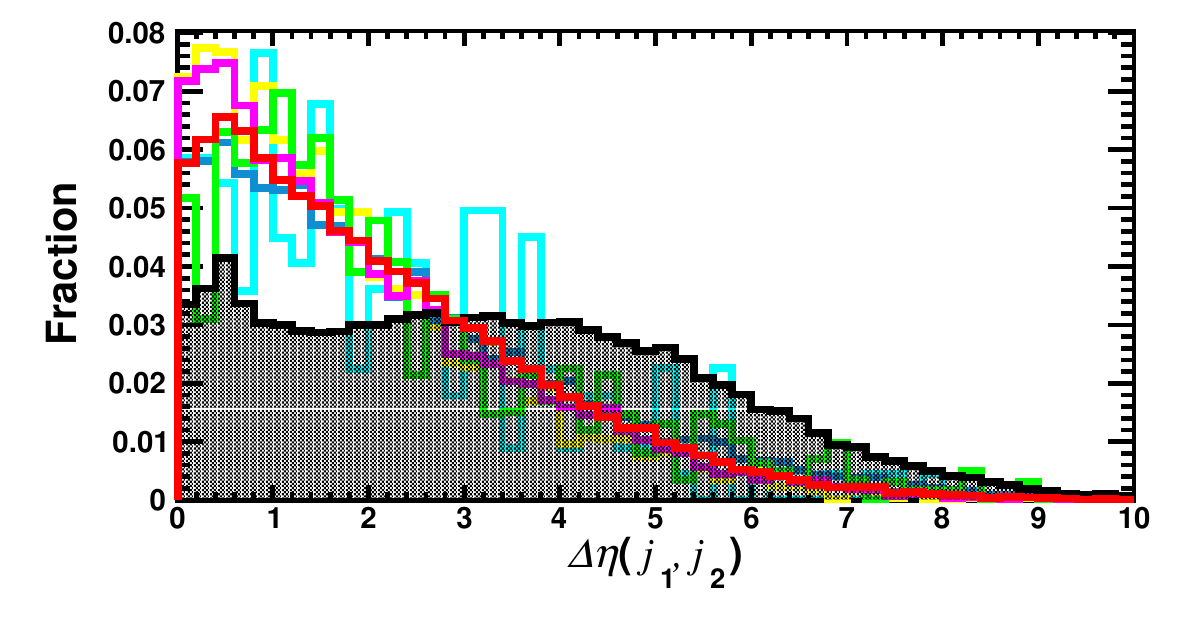}
}
\end{figure}
\addtocounter{figure}{-1}
\vspace{-1.10cm}
\begin{figure}[H]
\centering
\addtocounter{figure}{1}
\subfigure{
\includegraphics[width=7.3cm, height=4.7cm]{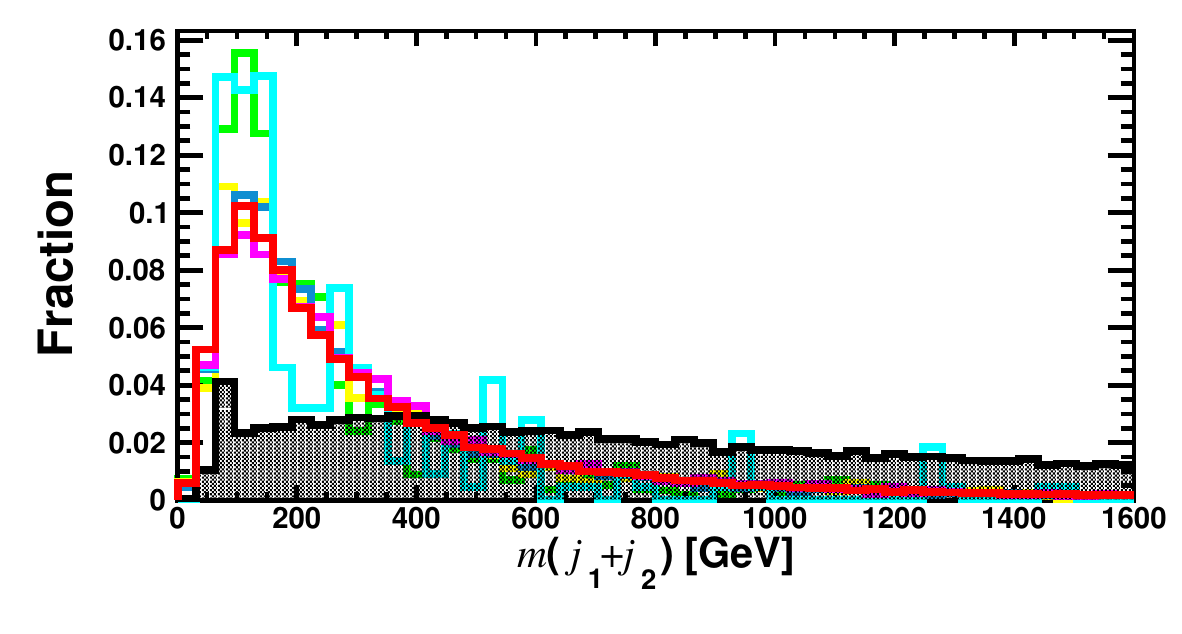}\,\,\,\,\,
\includegraphics[width=7.3cm, height=4.7cm]{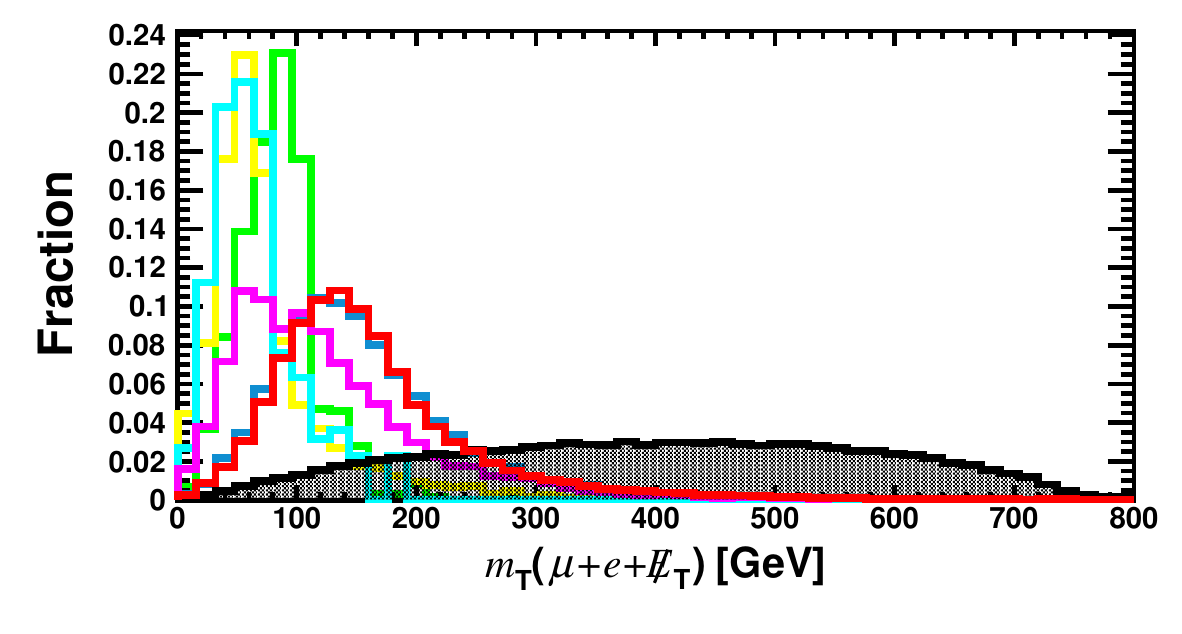}
}
\caption{
Distributions of representative observables of the signal (black, dashed) and the six background processes at the SppC / FCC-hh with $\sqrt{s}=100$~TeV, assuming $m_a=750$~GeV.
}
\label{fig:obs100TeV1}
\end{figure}

%\newpage
\subsection{Distributions of BDT responses}
\label{app:WWjj_BDT}

\begin{figure}[H]
\centering
\subfigure{
\includegraphics[width=7.3cm, height=4.7cm]{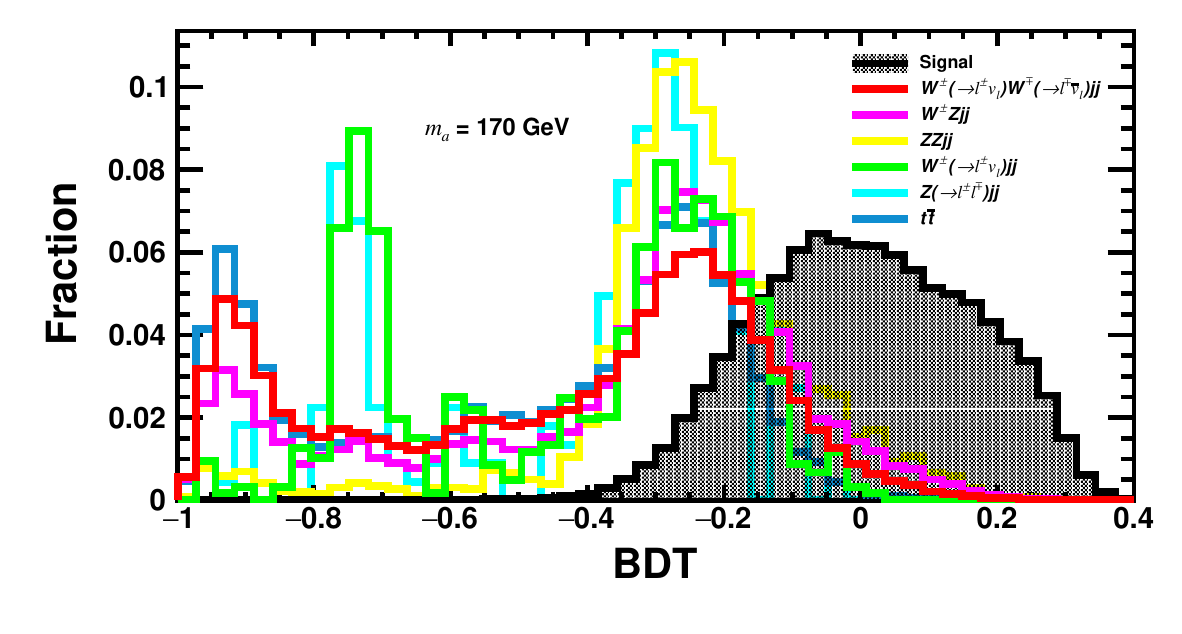}\,\,\,\,\,
\includegraphics[width=7.3cm, height=4.7cm]{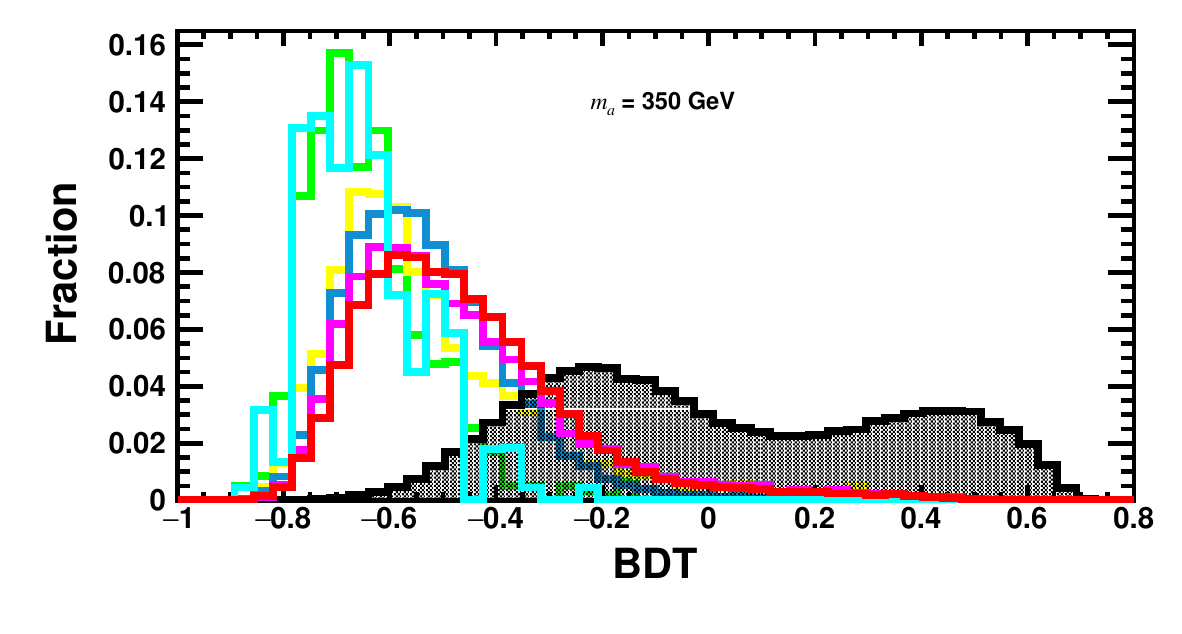}
}
\end{figure}
\addtocounter{figure}{-1}
\vspace{-1.10cm}
\begin{figure}[H]
\centering
\addtocounter{figure}{1}
\subfigure{
\includegraphics[width=7.3cm, height=4.7cm]{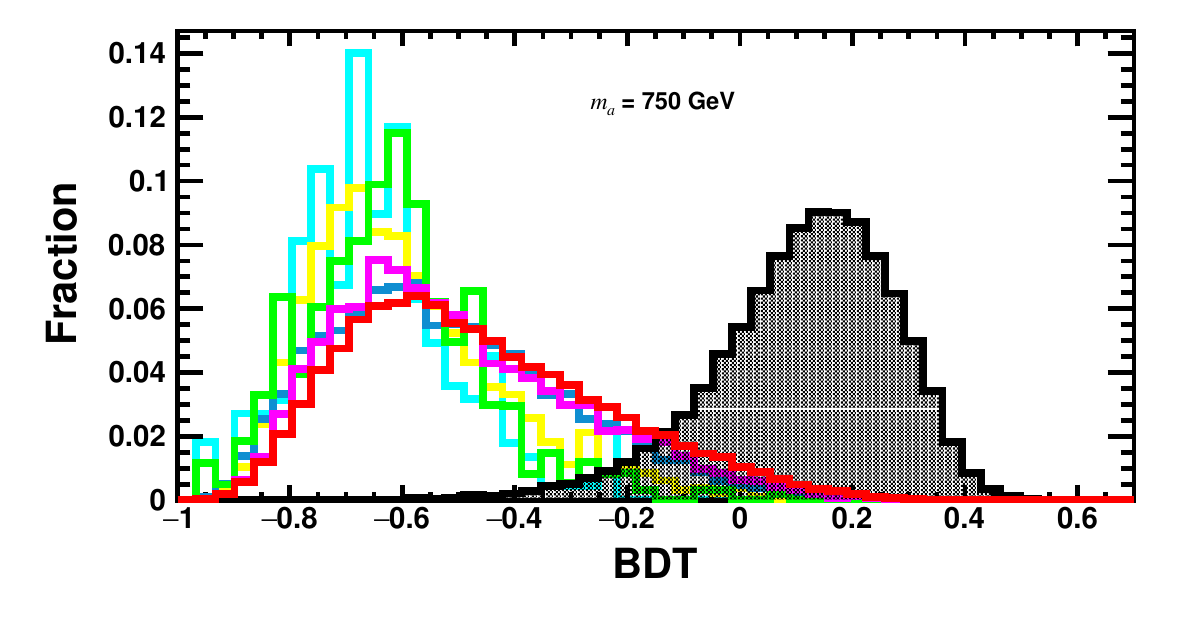}\,\,\,\,\,
\includegraphics[width=7.3cm, height=4.7cm]{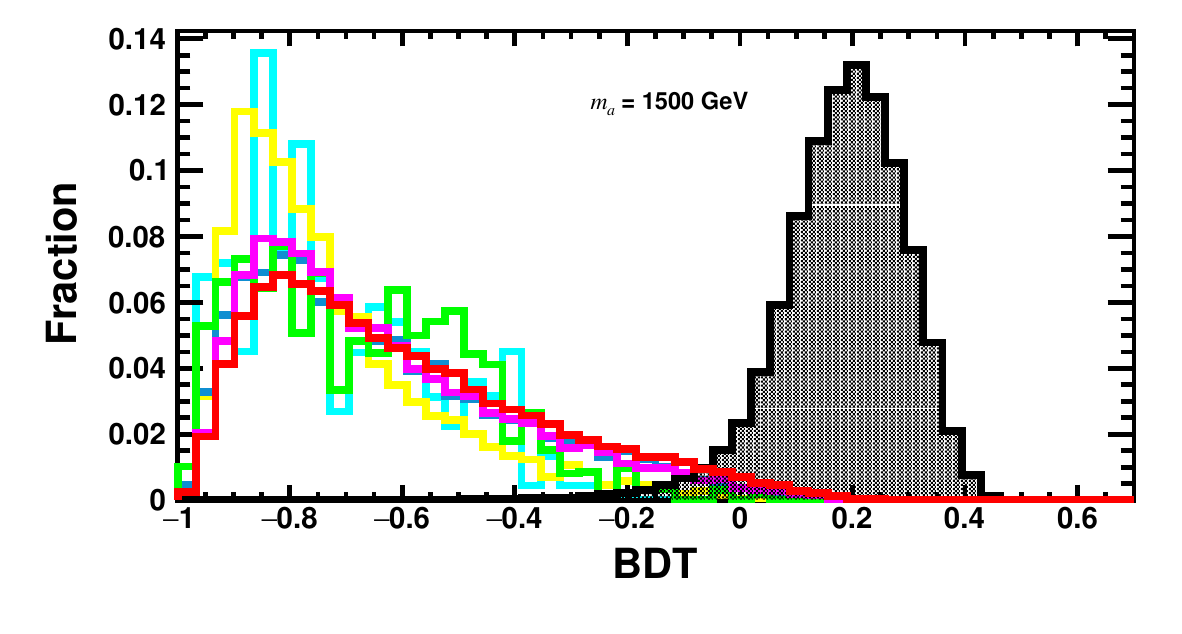}
}
\end{figure}
\addtocounter{figure}{-1}
\vspace{-1.10cm}
\begin{figure}[H]
\centering
\addtocounter{figure}{1}
\subfigure{
\includegraphics[width=7.3cm, height=4.7cm]{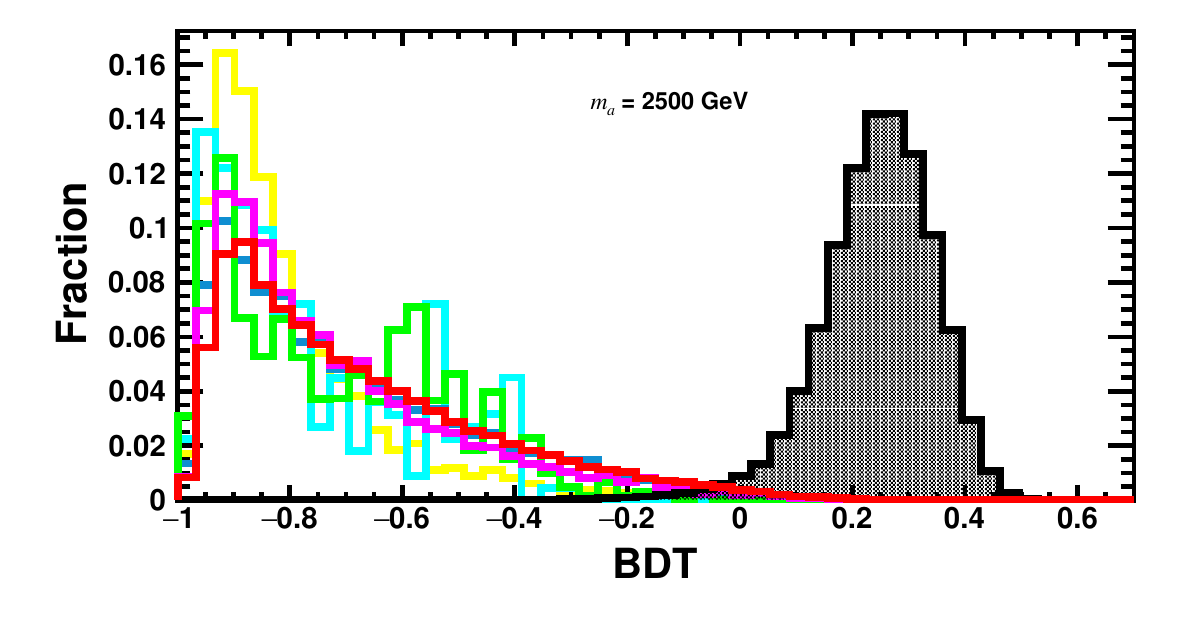}\,\,\,\,\,
\includegraphics[width=7.3cm, height=4.7cm]{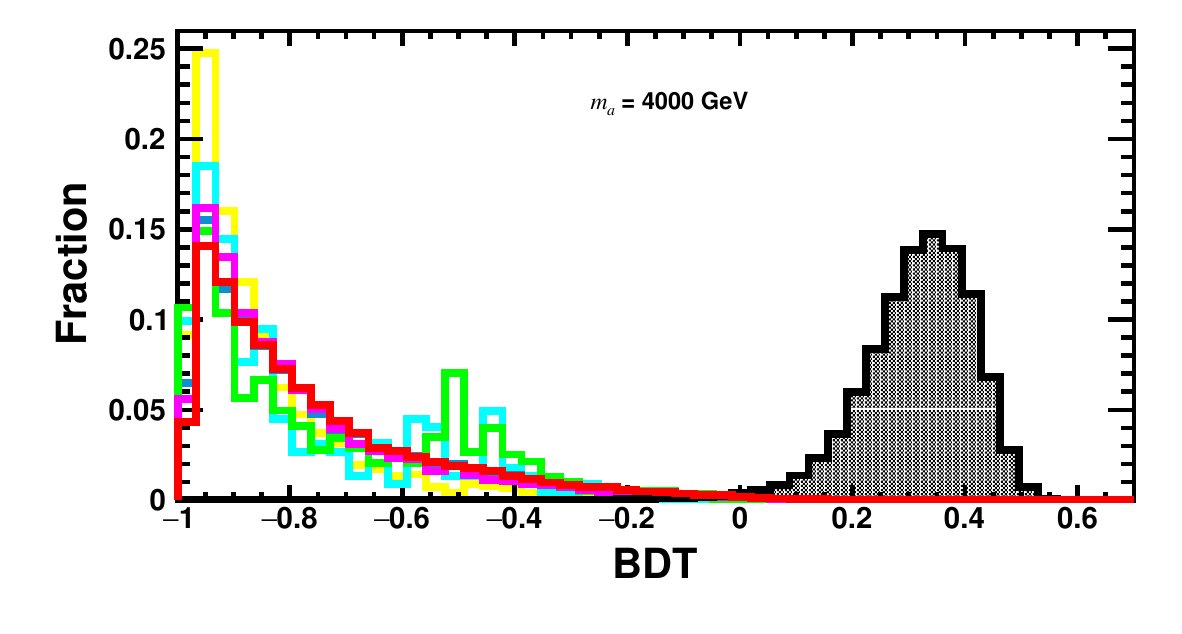}
}
\end{figure}
\addtocounter{figure}{-1}
\vspace{-1.10cm}
\begin{figure}[H]
\centering
\addtocounter{figure}{1}
\subfigure{
\includegraphics[width=7.3cm, height=4.7cm]{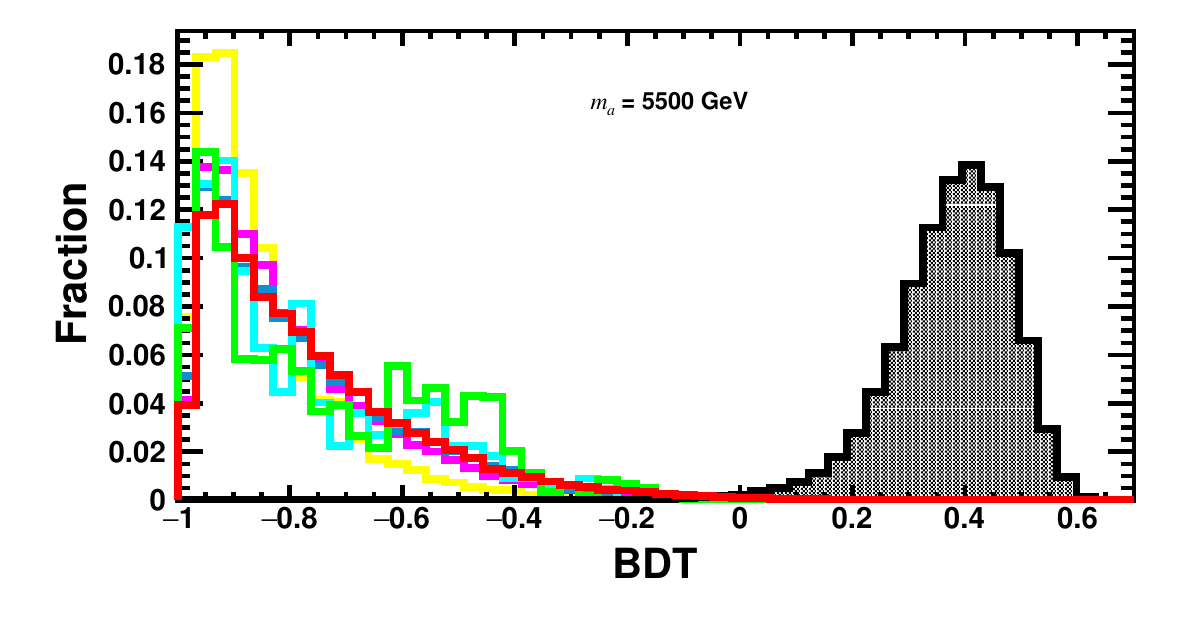}\,\,\,\,\,
\includegraphics[width=7.3cm, height=4.7cm]{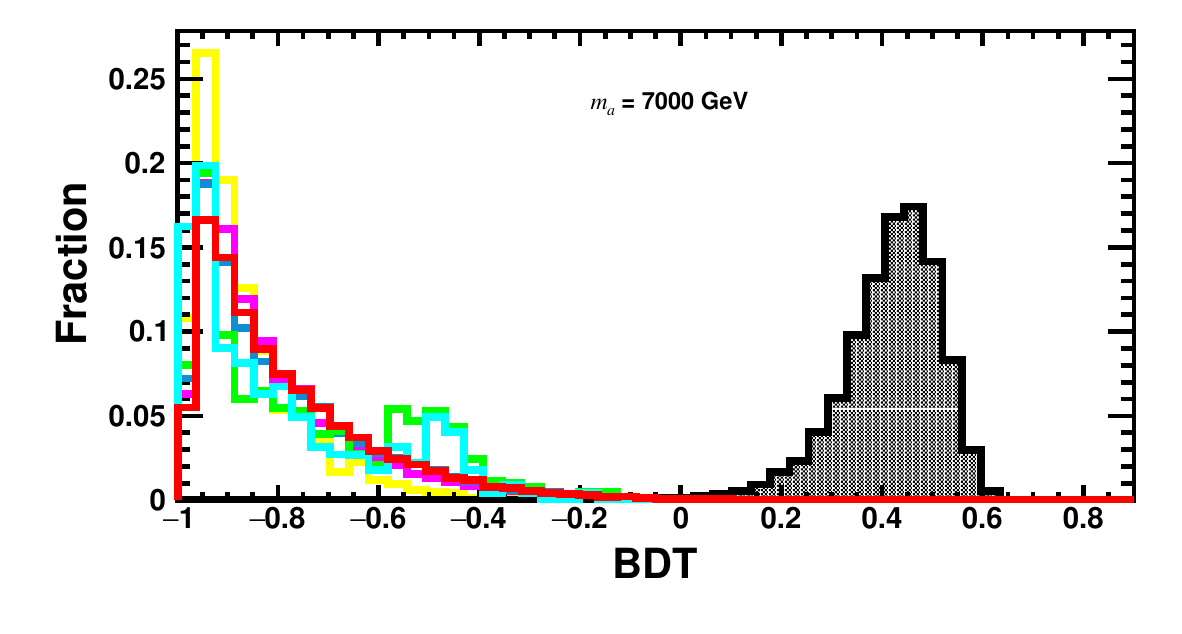}
}
\caption{
Distributions of BDT responses after applying the preselection criteria for the signal (black, shaded) and background processes at the SppC / FCC-hh with $\sqrt{s}=100$~TeV for representative $m_a$ values.
}
\label{fig:BDT_ALPmasses}
\end{figure}

%\newpage
\subsection{Selection efficiencies}
\label{app:WWjj_Sel_Eff}

\begin{table*}[h]
\centering 
\scalebox{0.9}{
%\begin{ruledtabular}
\begin{tabular}{ccccccccc}
\hline
\hline
$m_a$ [GeV] & BDT & signal & $W^+ (\to \ell^+ \nu)\,W^-(\to \ell^- \bar{\nu})jj$ & $WZjj$ & $ZZjj$ & $t\bar{t}$ \\
\hline
170 &  0.047 
&$4.05\mltp10^{-1}$&$1.20\mltp10^{-2}$&$2.62\mltp10^{-2}$&$3.43\mltp10^{-2}$&$1.99\mltp10^{-3}$  \\
185 &  0.126 
&$3.45\mltp10^{-1}$&$8.88\mltp10^{-3}$&$1.93\mltp10^{-2}$&$2.66\mltp10^{-2}$&$1.39\mltp10^{-3}$  \\
200 &  0.038  
&$4.22\mltp10^{-1}$&$2.10\mltp10^{-2}$&$3.72\mltp10^{-2}$&$4.70\mltp10^{-2}$&$3.90\mltp10^{-3}$ \\
225 &  -0.086  
&$5.07\mltp10^{-1}$&$4.01\mltp10^{-2}$&$5.77\mltp10^{-2}$&$6.36\mltp10^{-2}$&$8.48\mltp10^{-2}$ \\
250 &  -0.132 
&$5.40\mltp10^{-1}$&$5.39\mltp10^{-2}$&$7.06\mltp10^{-2}$&$7.09\mltp10^{-2}$&$1.30\mltp10^{-3}$  \\
350 & 0.256  
&$2.85\mltp10^{-1}$&$1.06\mltp10^{-2}$&$9.57\mltp10^{-3}$&$1.19\mltp10^{-2}$&$1.65\mltp10^{-3}$\\
750 & 0.096 
&$5.91\mltp10^{-1}$&$1.43\mltp10^{-2}$&$7.35\mltp10^{-3}$&$1.16\mltp10^{-3}$&$5.02\mltp10^{-3}$\\ 
1500 & 0.162 
&$6.14\mltp10^{-1}$&$2.03\mltp10^{-3}$&$1.16\mltp10^{-3}$&$-$&$5.19\mltp10^{-4}$\\ 
2500 & 0.165
&$8.08\mltp10^{-1}$&$1.09\mltp10^{-3}$&$4.83\mltp10^{-4}$&$-$&$3.46\mltp10^{-4}$\\     
4000 &  0.199 
&$8.86\mltp10^{-1}$&$5.00\mltp10^{-4}$&$3.87\mltp10^{-4}$&$-$&$8.66\mltp10^{-5}$\\ 
5500 & 0.200
&$9.39\mltp10^{-1}$&$2.93\mltp10^{-4}$&$9.67\mltp10^{-5}$&$-$&$8.66\mltp10^{-5}$\\     
7000 &  0.205 
&$9.62\mltp10^{-1}$&$2.59\mltp10^{-4}$&$-$&$-$&$8.66\mltp10^{-5}$\\
\hline
\hline
\end{tabular}
%\end{ruledtabular}
}
\caption{
Selection efficiencies of the BDT cut for both signal and background processes at the SppC/FCC-hh with $\sqrt{s}=100$~TeV, assuming different ALP masses. The second column gives the lower threshold of the BDT response, and “$-$” indicates that the number of events becomes negligible with $\mathcal{L}=20~\mathrm{ab}^{-1}$.
}
\label{tab:BDT_WWjj}
\end{table*}

%\acknowledgments
\begin{acknowledgments}
%\noindent
We thank Bin Diao and Ye Lu for helpful discussions. 
Z.D., J.F. and K.W. are supported by the National Natural Science Foundation of China under grant no. 11905162, the Excellent Young Talents Program of the Wuhan University of Technology under grant no. 40122102, and the research program of the Wuhan University of Technology under grants no. 2020IB024 and 104972025KFYjc0101. 
Y.N.M. and Y.X. are supported by the National Natural Science Foundation of China under grant no. 12205227. 
The simulation and analysis work of this article was completed with the computational cluster provided by the Theoretical Physics Group at the Department of Physics, School of Physics and Mechanics, Wuhan University of Technology.
\end{acknowledgments}

\paragraph{Note added.}
Following standard practice in high-energy physics, authors are listed in strict alphabetical order by surname. 
This ordering should not be interpreted as indicating any ranking of contribution, seniority, or leadership.
All authors contributed equally to this work.

%%%%%%%%%%%%%%%%%%%%%%%%%%%%%%%%%%%%%%%%%

\bibliography{Refs.bib}
\bibliographystyle{JHEP.bst}

\end{document}